\documentclass[aps,prd,onecolumn,nofootinbib,preprintnumbers,superscriptaddress,longbibliography,nobibnotes,10pt]{revtex4-2}
\usepackage{style}

\begin{document}

\preprint{FERMILAB-PUB-24-0356-T}

\title{The Non-Relativistic Effective Field Theory Of Dark Matter-Electron Interactions}
\author{Gordan Krnjaic\,\orcidlink{0000-0001-7420-9577}} 
\email{krnjaicg@fnal.gov}
\affiliation{Theoretical Physics Division, Fermi National Accelerator Laboratory, Batavia, IL 60510}
\affiliation{Department of Astronomy and Astrophysics, University of Chicago, Chicago, IL 60637}
\affiliation{Kavli Institute for Cosmological Physics, University of Chicago, Chicago, IL 60637}
\author{Duncan Rocha\,\orcidlink{0000-0002-8263-7982}} 
\email{drocha@uchicago.edu}
\affiliation{Theoretical Physics Division, Fermi National Accelerator Laboratory, Batavia, IL 60510}
\affiliation{Kavli Institute for Cosmological Physics, University of Chicago, Chicago, IL 60637}
\affiliation{Enrico Fermi Institute, Physics Department, University of Chicago, Chicago, IL 60637}
\author{Tanner Trickle\,\orcidlink{0000-0003-1371-4988}}
\email{ttrickle@fnal.gov}
\affiliation{Theoretical Physics Division, Fermi National Accelerator Laboratory, Batavia, IL 60510}

\date{\today}

\begin{abstract}
    Electronic excitations in atomic, molecular, and crystal targets are at the forefront of the ongoing search for light, sub-GeV dark matter (DM).
    In many light DM-electron interactions the energy and momentum deposited is much smaller than the electron mass, motivating a non-relativistic (NR) description of the electron.
    Thus, for any target, light DM-electron phenomenology relies on understanding the interactions between the DM and electron in the NR limit. 
    In this work we derive the NR effective field theory (EFT) of general DM-electron interactions from a top-down perspective, starting from general high-energy DM-electron interaction Lagrangians. 
    This provides an explicit connection between high-energy theories and their low-energy phenomenology in electron excitation based experiments.
    Furthermore, we derive Feynman rules for the DM-electron NR EFT, allowing observables to be computed diagrammatically, which can systematically explain the presence of in-medium screening effects in general DM models. 
    We use these Feynman rules to compute absorption, scattering, and dark Thomson scattering rates for a wide variety of high-energy DM models.
\end{abstract}

\maketitle
\newpage
\tableofcontents
\clearpage 

\section{Introduction}
\label{sec:intro}

Recent years have seen an explosion of ideas motivating light dark matter (DM) candidates with sub-GeV mass. In addition to familiar thermal freeze-out,  this mass range accommodates multiple new production mechanisms, for example, modified freeze-out~\cite{Pospelov:2007mp,Pospelov:2008jd,Hochberg:2014kqa,Hochberg:2014dra,Kuflik:2015isi,Kuflik:2017iqs}, asymmetric DM~\cite{Kaplan:2009ag,Zurek:2013wia,Lin:2011gj}, freeze-in~\cite{Dvorkin:2019zdi,Krnjaic:2022wor,Dodelson:1993je,Hall:2009bx}, misalignment~\cite{Dine:1982ah,Abbott:1982af,Turner:1985si,Turner:1983he} and inflationary production~\cite{Graham:2015rva}. Such light DM evades conventional nuclear-recoil direct detection searches since the energy deposited rapidly vanishes for DM lighter than the target nuclei. This limitation has inspired the development of next generation direct detection experiments, many at the interface of condensed matter and particle physics -- see Refs.~\cite{Lin:2019uvt,Kahn:2021ttr,Zurek:2024qfm} for recent reviews.

To efficiently probe light DM in a direct detection experiment, the target and DM must be kinematically matched, such that the target responds to the energy deposited during an interaction with DM. This energy deposition $\omega$ varies dramatically depending on the interaction: 
\begin{itemize}
    \item {\bf Absorption:}
    if a DM particle of mass $m$ is absorbed into the target, the energy deposition is $\omega \sim m$.
    \item{\bf Elastic Scatter:}
    if DM scatters elastically with the target, $\omega \lesssim m v^2$, where $v \sim 10^{-3}$ is the local DM velocity. 
    \item{\bf Inelastic Process:} if the incoming and outgoing dark states differ in mass, e.g., inelastic DM  \cite{Tucker-Smith:2001myb} or ``dark-Compton" scattering \cite{Hochberg:2021zrf}, $\omega$ may lie between the absorption and elastic scattering regimes, $m v^2 \lesssim \omega \lesssim m$.
\end{itemize}
Therefore, to extend direct detection sensitivity down to MeV scale masses via scattering, or eV scale masses via absorption, the target must have $\sim$ eV energy levels. Since this is the typical energy scale of electronic excitations in most materials, these systems are kinematically well-matched for sub-GeV DM direct detection.

There is currently a large experimental effort devoted to probing DM induced electronic excitations in a variety of targets. Electrons in liquid Xe or Ar can be ionized in the XENON~\cite{XENON:2021qze,XENON:2022ltv} and DarkSide~\cite{DarkSide:2022knj} experiments, respectively, if DM deposits $\omega \gtrsim 10 \; \text{eV}$. Electrons can also be excited across the band gap in semiconducting Si or Ge targets with eV scale energy deposits in the CDEX~\cite{CDEX:2019exx,CDEX:2022kcd}, DAMIC~\cite{DAMIC-M:2023gxo}, EDELWEISS~\cite{EDELWEISS:2020fxc}, SENSEI~\cite{SENSEI:2023zdf}, and SuperCDMS~\cite{SuperCDMS:2019jxx,SuperCDMS:2020ymb} experiments. 

Furthermore, there are several new ideas for exploiting electronic transitions in various other materials. In the $\sim$ few-eV energy range,   scintillators~\cite{Derenzo:2016fse,Zema:2024epe}, dielectrics~\cite{Hochberg:2021pkt,Lasenby:2021wsc,Knapen:2021run}, carbon nanotubes~\cite{Cavoto:2017otc,Cavoto:2019flp}, molecular targets~\cite{Arvanitaki:2017nhi,Blanco:2019lrf,Blanco:2021hlm}, and quantum dots~\cite{Blanco:2022cel} have all been proposed as promising targets. 
There are also many ideas for exploiting meV-scale electronic excitations to extend sensitivity to even lower DM masses. Targets such as graphene~\cite{Hochberg:2016ntt,Catena:2023awl,Catena:2023qkj}, Dirac materials~\cite{Hochberg:2017wce,Coskuner:2019odd,Geilhufe:2019ndy}, spin-orbit coupled materials~\cite{Inzani:2020szg,Chen:2022pyd}, narrow gap semiconductors~\cite{SuperCDMS:2022kse}, superconductors~\cite{Hochberg:2015fth,Hochberg:2015pha,Hochberg:2016ajh,Mitridate:2021ctr,Hochberg:2021ymx}, and doped semiconductors~\cite{Du:2022dxf} have been recently studied.

Given the unknown nature of DM, and the rapidly expanding experimental program, it is crucial to understand how a general DM candidate can interact with a given target, and several aspects of this problem have been studied in the recent literature. Elastic electron scattering in atomic~\cite{Bernabei:2007gr,Kopp:2009et,Essig:2011nj} and crystal~\cite{Essig:2011nj,Graham:2012su,Lee:2015qva,Essig:2015cda,Griffin:2019mvc,Griffin:2021znd} targets has been well studied in the context of DM models with a dark photon-like mediator. Elastic electron scattering in more general DM models has been studied using a ``bottom-up" approach, which enumerates possible interactions in the non-relativistic (NR) interaction Hamiltonian of the DM-electron system~\cite{Catena:2019gfa,Catena:2021qsr,Catena:2022fnk,Catena:2023awl,Catena:2023qkj,Liang:2024lkk,Liang:2024ecw}. In complementary analyses, it has also been shown that some DM-electron scattering rates can be related to the target dielectric function~\cite{Hochberg:2021pkt,Knapen:2021run}, and other generalized susceptibilities using linear response theory~\cite{Catena:2024rym}.

Additionally there has been similar progress towards understanding DM absorption into electronic excitations, where initial work primarily studied axion and dark photon absorption in atomic and isotropic crystal targets ~\cite{Pospelov:2008jk,Dzuba:2010cw,Hochberg:2016sqx,Bloch:2016sjj,Hochberg:2016ajh,An:2013yua,An:2014twa,Arvanitaki:2017nhi,Hochberg:2017wce,Coskuner:2019odd,Geilhufe:2019ndy,Mitridate:2021ctr}. It has also been shown that the DM absorption rate, in some DM models, can be related to the target dielectric function~\cite{Pospelov:2008jk,Hochberg:2016sqx}. More recent work has extended the absorption calculation to scalar DM models in crystal targets~\cite{Mitridate:2021ctr}, electron-coupled axion DM in spin-polarized targets~\cite{Berlin:2023ubt}, photon-coupled axion DM in magnetized targets~\cite{Berlin:2023ppd}, and electric and magnetic dipole-coupled DM in both atomic and crystal targets~\cite{Krnjaic:2023nxe}. First principles calculations have also been performed for axion and dark photon DM absorption into Dirac materials~\cite{Hochberg:2017wce,Coskuner:2019odd,Geilhufe:2019ndy}. 
In addition to scattering and absorption, a dark Compton-like process was shown to be important for heavier axion and dark photon DM candidates~\cite{Hochberg:2021zrf}.

In this paper we develop a framework for calculating DM-electron interaction rates in {\it any} target material, given {\it any} \he DM-electron interaction.\footnote{To be more specific, the NR EFT developed in Secs.~\ref{sec:NRQED} -~\ref{sec:NR_DM_electron_interaction} applies for any target and \he DM-electron interaction. The Feynman rules and interaction rate calculations in Secs.~\ref{sec:feynman_rules} -~\ref{sec:dark_thomson_scattering} rely on an independent electron, or single-particle, approximation~\cite{Girvin_2019}, discussed in detail in Sec.~\ref{subsec:nr_electron_field_quantization}, which can fail to be a good description for some electronic states in some targets, e.g., the Cooper-paired electrons in superconductors.} The key to this generalization is the NR effective field theory (EFT) describing the DM-electron interactions at energies and momenta below the electron mass. While the kinematics of different interaction processes (absorption, scattering, inelastic processes $\dots$) can vary dramatically, they all deposit momentum $\bm q$ and energy $\omega$ well below the electron mass ($|{\bm q}|, \omega \ll m_e)$, and can therefore be described with the NR EFT of DM-electron interactions. Moreover, all of these processes induce transitions between the electronic states of the target, so all observables of interest can be related to matrix elements of the form
\be
    {\cal M} \sim \int d^3{\bm x} \, e^{i \bm{q} \cdot {\bm x}} \, \psi_F^\dagger({\bm x}) \, \ONRGeneral \, \psi_I({\bm x})~,
\label{eq:intro_example_transition_matrix_element}
\ee
where $\psi_I({\bm x}), \psi_F({\bm x})$ are initial and final state NR electronic wavefunctions, and $\ONRGeneral$ is an NR interaction operator. The NR EFT Lagrangian then determines the operators that appear in the transition matrix elements (and their relative weights), while the target electronic structure determines the wavefunctions. Therefore, interaction rates written in terms of the matrix elements from Eq.~\eqref{eq:intro_example_transition_matrix_element} can be used for any target, and the NR EFT Lagrangian specifies which NR interaction operators, $\ONRGeneral$, contribute to the NR matrix element.

An additional benefit of the Lagrangian formalism for the DM-electron NR EFT is it allows observables to be computed using Feynman diagrams. Given the DM-electron NR EFT Lagrangian, we derive Feynman rules (Sec.~\ref{sec:feynman_rules}) for the NR electron propagator, three-point and four-point vertices, and some commonly appearing loop diagrams (built from the previous vertex and propagator Feynman rules). These Feynman rules depend on the transition matrix elements in Eq.~\eqref{eq:intro_example_transition_matrix_element}, and include the target-dependent wavefunctions for the relevant observable. This diagrammatic approach is useful because it allows complicated results to be built from simple primitive rules, and can be used to identify new observables. Furthermore, it clarifies when screening effects are present, which manifests as the cancellation between diagrams which can be easily overlooked in a tree level calculation. 

To determine the NR EFT of DM-electron interactions we adopt a ``top-down" approach: starting from a \he theory defined at energies above $m_e$, we match on to the low-energy NR EFT by finding a map between the \he and low-energy electron fields. This is sometimes referred to as ``integrating out the positron," since the low-energy theory will only contain two components of the high-energy, four-component electron field. This top-down approach is beneficial because it provides an explicit connection between 
\he model building and low-energy DM-electron phenomenology. This approach can be contrasted with the other EFTs of electron~\cite{Catena:2019gfa,Catena:2021qsr,Catena:2022fnk}, nuclear recoil~\cite{Fan:2010gt,Fitzpatrick:2012ix,Fitzpatrick:2012ib,Cirelli:2013ufw,Gresham:2014vja}, and collective excitation scattering~\cite{Trickle:2020oki} because it is only an EFT of the electron degrees of freedom. This EFT is valid when the initial and final dimensionful scales of the electronic system (e.g., the electron momentum, energy, or binding potential) are much less than $m_e$. We do \textit{not} assume that the DM kinematics are non-relativistic. The EFT can be applied whenever $\omega, |{\bm q}| \ll m_e$, no further approximations are required on the DM side of the calculation.

This paper is organized as follows: in Sec.~\ref{sec:NRQED} we derive the NR QED Lagrangian to order $m_e^{-2}$ using the \FWspell method, which provides the mapping between the \he electron field, involved in \he DM-electron interactions, and the two component, low-energy electron field, involved in the low-energy DM-electron interactions. In Sec.~\ref{sec:NR_DM_electron_interaction} we use this mapping to connect a general \he DM-electron interaction Lagrangian to the low-energy, NR EFT of DM-electron interactions. In Sec.~\ref{sec:feynman_rules} we derive the Feynman rules for this NR EFT, allowing any observable to be computed diagrammatically. In Secs.~\ref{sec:absorption},~\ref{sec:scattering}, and~\ref{sec:dark_thomson_scattering} we use the Feynman rules derived in Sec.~\ref{sec:feynman_rules} to compute absorption, elastic scattering, and dark Thomson scattering (the low-energy limit of the dark Compton scattering process in Ref.~\cite{Hochberg:2021zrf}), respectively, for a wide variety of DM models. App.~\ref{app:summary_tables} contains a summary of the NR EFT interaction Lagrangians and their corresponding Feynman rules. App.~\ref{app:in_medium_photon_propagator} contains a detailed derivation of the in-medium photon propagator which is needed whenever screening effects are relevant.


\section{Non-Relativistic QED}
\label{sec:NRQED}

Our main goal is to derive the NR EFT of DM-electron interactions. That is, we want to find an interaction Lagrangian coupling the DM fields to the electron field $\psi$ in the NR limit. To do this we must first understand how $\psi$ is related to the \he electron spinor $\Psi$, which appears in the \he DM-electron interaction Lagrangian. To leading order in the DM-electron coupling, the mapping of $\Psi \rightarrow \psi$ is independent of the \he DM-electron interaction, and is found by taking the NR limit of QED. Therefore, we begin by deriving the NR QED Lagrangian, $\LNR_\text{QED}$, which describes the dynamics of $\psi$. While this has been developed before in a variety of different contexts (see, for example, Refs.~\cite{Paz:2015uga,Mitridate:2021ctr,Krnjaic:2023nxe}), here we present a pedagogical derivation for completeness, focusing on the method best suited for the calculations in Sec.~\ref{sec:NR_DM_electron_interaction}. This approach is ``top-down," as our starting point is the Dirac Lagrangian and we derive NR QED by taking its low energy limit to a fixed order in $m_e$. This approach contrasts with previous ``bottom-up" calculations (see for example Refs.~\cite{Paz:2015uga,Gunawardana:2017zix,Kobach:2017xkw}), which focus on symmetry principles to build a basis of operators in the NR limit.


\subsection{Framing the Problem: Diagonalizing QED}
\label{subsec:warmup}

To motivate our approach consider the Dirac Lagrangian
\begin{align}
    \label{eq:DiracLagrangian}
    \LUV_\text{QED} = \bar{\Psi} \left( i \gamma^\mu D_\mu - m_e \right) \Psi \, ,~~~ \Psi  =
        \begin{pmatrix}
            \psi_1 \\ \psi_2
        \end{pmatrix},
\end{align}
where $\Psi$ is the usual four-component electron Dirac spinor, $\psi_1$, $\psi_2$ are two-component spinors, $D_\mu = \partial_\mu + i e A_\mu$ is the gauge covariant derivative, $e = - |e|$ is the electron charge, and $A_\mu$ is the 
photon field. We will find it convenient to work in the Dirac basis for the gamma matrices
\begin{align}
    \gamma^0 = \begin{pmatrix} 1 & 0 \\ 0 & -1 \end{pmatrix}~~~,~~~\gamma^i = \begin{pmatrix} 0 & \sigma^i \\ - \sigma^i & 0 \end{pmatrix}~~~,~~~\gamma^5 = \begin{pmatrix} 0 & 1 \\ 1 & 0 \end{pmatrix} \, ,
    \label{eq:dirac_basis}
\end{align}
where $\sigma^i$ are the Pauli matrices. In the zero-momentum ($\partial_i \rightarrow 0$), non-interacting ($A_\mu \to 0$) limit, the QED Lagrangian in Eq.~\eqref{eq:DiracLagrangian} can be written
\begin{align}
    \label{eq:QED-simple}
    \LUV_\text{QED} = \begin{pmatrix}
        \psi_1^\dagger & \psi^\dagger_2
    \end{pmatrix} \begin{pmatrix}
        i \partial_t - m_e & 0 \\ 0 & i \partial_t + m_e
    \end{pmatrix} \begin{pmatrix}
        \psi_1\\ \psi_2
    \end{pmatrix} \, ,
\end{align}
where $\psi_1$ can now be identified as the two-component electron field within $\Psi$, since its equation of motion in this limit is
\begin{align}
\frac{ \delta \LUV_\text{QED} }{\delta \psi_1^\dagger} =  
 \left( i \partial_t - m_e \right) \psi_1 = 0 ~~\implies ~~ \psi_1 \propto e^{-im_et},
\end{align}
and we see that $\psi_1$ has positive frequency $m_e$. Furthermore, by re-phasing the four-component field according to
\begin{align}
    \label{eq:phase_me}
    \Psi \to e^{-i m_e t}\Psi \, , 
\end{align}
the QED Lagrangian in the zero-momentum, non-interacting limit, Eq.~\eqref{eq:QED-simple}, can be simplified to yield
\begin{align}
    \LUV_\text{QED} = \begin{pmatrix}
        \psi_1^\dagger & \psi^\dagger_2
    \end{pmatrix} \begin{pmatrix}
        i \partial_t  & 0 \\ 0 & i \partial_t + 2 m_e
    \end{pmatrix} \begin{pmatrix}
        \psi_1 \\ \psi_2
    \end{pmatrix} \, ,
\end{align}
explicitly demonstrating that the electron field $\psi_1$ satisfies the free NR Schr\"odinger equation, $i \partial_t \psi_1 = 0$. Since $\psi_1, \psi_2$ have no interactions, $\psi_2$ decouples from the theory in this zero-momentum, non-interacting limit. 

However, once momentum dependence and interactions are restored, the full QED Lagrangian in Eq.~\eqref{eq:DiracLagrangian} contains off-diagonal terms that couple $\psi_1$ to $\psi_2$
\begin{align}
\label{eq:dirac-off-diag}
    \LUV_\text{QED} = \begin{pmatrix}
        \psi_1^\dagger & \psi^\dagger_2
    \end{pmatrix} \begin{pmatrix}
        i D_t & i \sigma^i D_i \\ i \sigma^i D_i & i D_t + 2 m_e
    \end{pmatrix} \begin{pmatrix}
         \psi_1 \\  \psi_2
    \end{pmatrix} \, ,
\end{align}
so developing the NR EFT of QED at energy scales below $m_e$  reduces to the task of integrating out $\psi_2$ (which is considered ``heavy") to remove these off-diagonal interactions. A common approach to eliminating these interactions is to substitute for $\psi_2$ using its equation of motion
\begin{align}
    \psi_2 = -\frac{1}{ i D_t + 2m_e }  (i \sigma^i D_i \, \psi_1) \, ,
\end{align}
and then expand in powers of $1/m_e$ to derive NR QED at a given order. While this approach is conceptually simple, it introduces technical complexity since at every order in $1/m_e$ terms with additional time derivatives appear, which alter the kinetic term for $\psi_1$. These additional terms can be removed with further field redefinitions, but determining the appropriate form for such transformations is generically non-trivial and calls for a more systematic treatment, which we describe below.


\subsection{\FWspell Transformations in QED}
\label{subsec:FW}

An alternative technique for removing the $\psi_1, \psi_2$ interactions in Eq.~\eqref{eq:dirac-off-diag} is known as the \FWspell (FW) method~\cite{PhysRev.78.29,PhysRev.87.688,drell,Gardestig:2007mk,Smith:2023htu}, which perturbatively diagonalizes the Lagrangian with successive field redefinitions. This approach is tantamount to determining a set of $n$ Hermitian operators $\{ \FWX_0, \dots, \FWX_{n - 1} \}$ for which the field redefinition
\begin{align}
\label{eq:main-trans}
    \Psi \rightarrow \FWU_{n} \Psi \equiv \left[ \exp\left( -i \frac{ \FWX_0 }{m_e} \right) \dots \exp\left( -i \frac{\FWX_{n-1}}{m_e^n} \right) \right] \begin{pmatrix}
        \psi \\ \psi_H
    \end{pmatrix},
\end{align}
diagonalizes the QED Lagrangian in Eq.~\eqref{eq:dirac-off-diag} to order $m_e^{-n + 1}$, where $\FWU_n$ is understood to be the operator in the square brackets, and the post-field redefinition $\Psi$ is written in terms of the electron field $\psi$ and the heavy field $\psi_H$. To be explicit, after the field redefinition in Eq.~\eqref{eq:main-trans} there will be no terms in the QED Lagrangian which contract $\psi$ with $\psi_H$ at order $m_e^{-n + 1}$ or lower; they may only exist at order $m_e^{-n}$ or higher.

For future convenience in determining the $\FWX_i$ diagonalization operators, we introduce the following operator classification:
\begin{itemize}
    \item{\bf Even Operators:} An \textit{even} operator is diagonal in the Dirac basis, and commutes with $\gamma^0$ in any basis; if $[E, \gamma^0] = 0$ then $E$ is even. Upon contraction with Dirac spinors an even operator does not mix components in the Dirac basis, so $\Psi^\dagger E \Psi \rightarrow \psi^\dagger (\cdots) \psi + \psi_H^\dagger (\cdots) \psi_H$, where $\Psi = (\psi \; \psi_H)^T$. As an example, $\gamma^0$ is an even operator.
    \item{\bf Odd Operators:} An \textit{odd} operator is off-diagonal in the Dirac basis, and anti-commutes with $\gamma^0$ in any basis, so if $\{ O, \gamma^0 \} = 0$ then $O$ is an odd operator. Upon contraction with Dirac spinors, an odd operator mixes components in the Dirac basis, so that  $\Psi^\dagger O \Psi \rightarrow \psi^\dagger (\cdots) \psi_H + \psi_H^\dagger (\cdots) \psi$. As an example, $\gamma^i$ is an odd operator.
\end{itemize}
These definitions follow the standard rules of even and odd quantities: an even operator multiplied by an even (odd) operator is even (odd), and an odd operator multiplied by an odd operator is even. In this language diagonalizing the QED Lagrangian to order $m_e^{-n}$ amounts to removing all terms containing an odd operator to order $m_e^{-n}$. Additionally we define the projection operators
\begin{align}
    P_+ = \frac{1 + \gamma^0}{2} = \begin{pmatrix}
        1 & 0 \\ 0 & 0 
    \end{pmatrix}~~~,~~~P_- = \frac{1 - \gamma^0}{2} = \begin{pmatrix}
        0 & 0 \\ 0 & 1
    \end{pmatrix}~~,
\end{align}
which have simple interpretations in the Dirac basis: $P_+$ projects out the upper component of a spinor, and $P_-$ projects out the lower component. Note that both $P_\pm$ are even operators. Lastly, since they will appear frequently, we define special even and odd operators
\begin{align}
\label{eq:evenodd:dirac}
    \FWE \equiv i D_t ~~~,~~~\FWO \equiv i \gamma^0 \gamma^i D_i \, ,
\end{align}
so that the QED Lagrangian in Eq.~\eqref{eq:dirac-off-diag} is
\begin{align}
    \label{eq:Dirac_lagrangian_2}
    \LUV_\text{QED} = \Psi^\dagger \left( \FWE + \FWO + 2 m_e P_-  \right) \Psi \, .
\end{align}
To illustrate the procedure we will explicitly find $\FWX_0$ and $\FWX_1$, and then in Sec.~\ref{subsec:FW_summary} provide the algorithm for finding the others.


\vspace{1em}
\begin{center}
    \textit{ \normalsize Expanding QED to First Order }
\end{center}
\vspace{1em}

We begin by diagonalizing the QED Lagrangian in Eq.~\eqref{eq:Dirac_lagrangian_2} to order $m_e^0$ by removing the odd operator, $\FWO$, present at order $m_e^0$. Substituting the $n = 1$ FW transformation from \Eq{eq:main-trans} in to Eq.~\eqref{eq:Dirac_lagrangian_2} 
\begin{align}
    \LUV_\text{QED} & \rightarrow \Psi^\dagger \left[ \FWU_1^\dagger \left( \FWE + \FWO + 2 m_e P_- \right) \FWU_1 \right] \Psi \nonumber \\
     & = \Psi^\dagger \left[ e^{i \FWX_0 / m_e} \left( \FWE + \FWO + 2 m_e P_- \right) e^{-i \FWX_0 / m_e} \right] \Psi  \nonumber \\
     &\approx \Psi^\dagger \left( \FWE + \FWO + [ i \FWX_0, 2 P_- ] \right) \Psi  \, ,
    \label{eq:Dirac_lagrangian_expanded}
\end{align}
where in the last line we have only kept terms to order $m_e^0$. The operator $\FWX_0$ is defined by requiring that it removes all the odd terms at order $m_e^0$, which means removing $\FWO$ in Eq.~\eqref{eq:Dirac_lagrangian_expanded}. Since $\FWO$ is odd and $P_-$ is even, $\FWX_0$ must be odd, so its commutator with $P_-$ can be computed as $[i \FWX_0, 2 P_- ] = -i [\FWX_0, \gamma^0] = 2i \gamma^0 \FWX_0$. Thus, demanding that $\FWX_0$ cancels $\FWO$ implies
\begin{align}
    \label{eq:X0_QED}
    [ i \FWX_0, 2 P_- ] = 2i \gamma^0 \FWX_0 = -\FWO~~~\implies~~~\FWX_0 = -\frac{1}{2} \gamma^i D_i \, .
\end{align}
The NR QED Lagrangian, which contains only the $\psi$ degrees of freedom, can then be extracted to order $m_e^0$ by inserting $P_+$ projection operators, $\Psi \to P_+ \Psi$ after the $n = 1$ FW transformation has been applied, which eliminates the lower two components of $\Psi$ in the Dirac basis
\begin{align}
    \LNR_\text{QED} \approx \Psi^\dagger P_+ \left[ \hat \FWU_1^\dagger \left( \FWE + \FWO + 2 m_e P_- \right) \FWU_1  \right] P_+ \Psi 
     \approx \psi^\dagger  \, i D_t \, \psi \, ~.
\end{align}
Note that we did not need to know an explicit form for $\FWX_0$ to find $\LNR_\text{QED}$ to order $m_e^0$; we just needed to remove the odd operators from the Lagrangian at order $m_e^0$. For example, if we had simply removed $\FWO$ from the Lagrangian in Eq.~\eqref{eq:Dirac_lagrangian_2} we would have found the NR QED Lagrangian to order $m_e^0$. By repeating this procedure with additional 
operators in \Eq{eq:main-trans} we can systematically obtain a diagonal Lagrangian to any order in $1/m_e$.

\vspace{1em}
\begin{center}
    \textit{ \normalsize Expanding QED to Second Order}
\end{center}
\vspace{1em}

We now repeat the steps above using $\FWX_0$ from \Eq{eq:X0_QED} to determine both $\FWX_1$ and the NR QED Lagrangian to order $m_e^{-1}$. Applying the $n = 2$ FW transformation from \Eq{eq:main-trans} yields
\begin{align}  
    \Psi \rightarrow \FWU_2 \Psi   = \left[ \exp\left( -i \frac{\FWX_0}{m_e} \right)  \exp\left( - i\frac{\FWX_{1}}{m_e^2} \right) \right] \Psi,
\end{align}
and we insert this into Eq.~\eqref{eq:Dirac_lagrangian_2} to obtain
\begin{align}
    \LUV_\text{QED} \rightarrow \Psi^\dagger \left[ e^{i \FWX_1 / m_e^2} e^{i \FWX_0 / m_e} \left( \FWE + \FWO + 2 m_e P_- \right) e^{-i \FWX_0 / m_e} e^{-i \FWX_1 / m_e^2} \right] \Psi .
\end{align}
Expanding the exponentials to order $m_e^{-1}$ with the standard Baker-Campbell-Hausdorf (BCH) formula yields
\begin{align}
     \LUV_\text{QED} \approx \Psi^\dagger \Bigg\{ 2 m_e P_- + \FWE 
    + \frac{1}{m_e} \left( \left[ i \FWX_0, \FWE \right] + \left[ i \FWX_0, \FWO \right] + \frac{1}{2} \left[ i \FWX_0 , \left[ i \FWX_0, 2 P_- \right] \right] + \left[ i \FWX_1, 2 P_- \right]  \right) \Bigg\} \Psi \, ,~~\label{eq:QED_lagrangian_expanded_2}
\end{align}
and $\FWX_1$ is determined by requiring that there are no odd terms at order $m_e^{-1}$ in Eq.~\eqref{eq:QED_lagrangian_expanded_2}. Since $P_\pm$ are even and $\FWX_0$ is odd, the only odd term that needs to be eliminated from Eq.~\eqref{eq:QED_lagrangian_expanded_2} is $[ i \FWX_0, \FWE ] / m_e$. Therefore we must choose $\FWX_1$ such that $i [\FWX_1, 2 P_-] = - [i \FWX_0, \FWE ]$. Again using the fact that $\FWX_1$ is odd, $i [\FWX_1, 2 P_-] = 2 i \gamma^0 \FWX_1$, and therefore
\begin{align}
\label{eq:X1_QED}
    \FWX_1 = - \frac{i}{2} \gamma^0 \left( - [i \FWX_0, \FWE ] \right) = \frac{e}{4} \gamma^0 \gamma^i F_{0i} \, , 
\end{align}
where $F_{\mu \nu}$ is the electromagnetic field strength tensor satisfying $\left[ D_\mu, D_\nu \right] = i e F_{\mu \nu}$.

To identify the NR QED Lagrangian to order $m_e^{-1}$ we again explicitly project out the upper component of $\Psi$ in Eq.~\eqref{eq:QED_lagrangian_expanded_2} with the substitution $\Psi \rightarrow P_+ \Psi$ yielding
\begin{align}
    \LNR_\text{QED} \approx \Psi^\dagger P_+ \left\{ 2 m_e P_- + \FWE + \frac{1}{m_e} \left( \left[ i \FWX_0, \FWO \right] + \frac{1}{2} \left[ i \FWX_0 , \left[ i \FWX_0, 2 P_- \right] \right]  \right) \right\} P_+ \Psi  \, ,
\end{align}
which can be further simplified using $X_0$ from \Eq{eq:X0_QED} to yield
\begin{align}
    \LNR_\text{QED} \approx \psi^\dagger \left[ i D_t + \frac{1}{2 m_e} \left( \sigma^i \sigma^j D_i D_j \right) \right] \psi  \, .
    \label{eq:NR_QED_lagrangian_1}
\end{align}
As expected, the NR QED Lagrangian to order $m_e^{-1}$ renders an equation of motion for $\psi$ which is simply the Schr\"odinger equation for a spin-$\frac{1}{2}$ charged particle. 


\subsection{Summary: Expanding QED to Arbitrary Order}
\label{subsec:FW_summary}

The procedure for finding $\FWX_{n-1}$ and the NR QED Lagrangian to order $m_e^{-(n-1)}$ should now be clear: 
\begin{enumerate}
	\item{\bf Rephase:} 
            Using the known expressions for $\{\FWX_0, \dots, \FWX_{n-2}\}$, determined in the previous iterations of this process, rephase $\Psi$ with the $n^\text{th}$ order FW transformation
            \begin{align}
                \label{eq:main-trans3}
                    \Psi \rightarrow \FWU_n \Psi \equiv \left[ \exp\left( -i \frac{\FWX_0}{m_e} \right) \dots \exp\left( -i \frac{\FWX_{n-1}}{m_e^{n}} \right) \right] \Psi
            \end{align}
            where the $\FWX_{n-1}$ operator is initially treated as an unknown quantity. As we found above, the first two $\FWX$ operators can be written
            \begin{align}
                \label{eq:X0_X1_definitions}
                \FWX_0 =  -\frac{1}{2} \gamma^i D_i ~~,~~~ \FWX_1 =  
                  \frac{e}{4} \gamma^0 \gamma^i F_{0i} \, .
            \end{align}
	\item{\bf Expand:} 
            Insert \Eq{eq:main-trans3} into the QED Lagrangian 
    	\begin{align}
                \LUV_\text{QED} \rightarrow \Psi^\dagger \left[ \FWU_n^\dagger \left( \FWE + \FWO + 2 m_e P_- \right) \FWU_n \right] \Psi ,
                \label{eq:step_2}
    	\end{align}
            and use the BCH formula to expand the exponentials to order $m_e^{-(n - 1)}$. Identify all odd operators at order $m_e^{-(n - 1)}$ and remove them with a suitable definition for $\FWX_{n-1}$.
	\item{\bf Project:} 
            Replace $\Psi \to P_+ \Psi$, to extract only the electron degrees of freedom, $\psi$, from the full Dirac spinor $\Psi$, and expand to order $m_e^{-n + 1}$
    	\begin{align}
                \LNR_\text{QED} & =  \Psi^\dagger P_+ U_n^\dagger \left( \FWE + \FWO + 2 m_e P_- \right) U_n P_+ \Psi \approx \psi^\dagger (\cdots) \psi  , 
    	\end{align}
            which results in the desired expression: the NR QED Lagrangian in terms of $\psi$ to order $m_e^{-n + 1}$.
\end{enumerate}
Using this procedure we can find the NR QED Lagrangian to order $m_e^{-2}$. Note that we do not need to explicitly solve for $\FWX_2$ as long as we require that it remove all odd operators at order $m_e^{-2}$. Following this procedure we obtain
\begin{align}
    \mathcal{L}^{\rm NR}_\text{QED}  \approx \psi^\dagger \left\{ i D_t - \frac{1}{2m_e} \left( D^i D_i + e \sigma^i B_i \right) + \frac{e}{8 m_e^2} \left[ \sigma^i \sigma^j \left( E_i D_j - E_j D_i - \left( \partial_i E_j \right) \right) \right] \right\} \psi 
 \label{eq:NR_QED_lagrangian_2}
\end{align}
where $E^i = -F^{0i}$, and $B^i = - \epsilon^{ijk} F_{jk} / 2 \; (F_{ij} = \epsilon_{ijk} B^k)$ are the electric and magnetic fields, respectively.


\section{Non-Relativistic Dark Matter-Electron Interactions}
\label{sec:NR_DM_electron_interaction}

We now introduce interactions between the four-component electron spinor $\Psi$ and an arbitrary set of dark fields (either the DM itself or a mediator particle). In the ultraviolet (UV), at energies above the electron mass, the general interaction Lagrangian can be written 
\begin{align}
    \label{eq:lint}
    \LUV^{\rm UV}_\text{int} = \bar{\Psi} \, \OUVGeneral \, \Psi \, ,
\end{align}
where $\OUVGeneral$ contains the dark fields and has a Lorentz structure that contracts with $\Psi$. For example, if a dark boson $\phi$ has a Yukawa interaction with electrons, we would write 
\be
\LUV^{\rm UV}_\text{int} = g \, \phi \, \Psibar \Psi ~~,~~ \OUVGeneral = g\, \phi~~,~~~
\ee
where $g$ is a coupling constant. Alternatively, for a four-Fermi interaction with a dark fermion $\chi$, we would write
\be
\LUV^{\rm UV}_\text{int} = \frac{1}{\Lambda^2} [ \chibar \gamma^\mu \chi] \Psibar \gamma_\mu \Psi~~,~~ \OUVGeneral = \frac{1}{\Lambda^2} [ \chibar \gamma^\mu \chi] \gamma_\mu,
\ee
where it is understood that $[\bar \chi \gamma^\mu \chi]$ is a scalar quantity, and therefore $\Psi$ only contracts with the $\gamma_\mu$ outside the brackets.

In this section, our goal is to determine the NR limit of Eq.~\eqref{eq:lint} by applying the \FWspell transformation from Sec. \ref{sec:NRQED} and then express the result in a basis of NR operators that act on the two-component NR electron field $\psi$. Schematically, we identify the mapping between \he and low-energy interactions as
\begin{align}
    \LUV^{\rm UV}_\text{int} = \Psibar \, \OUVGeneral \, \Psi~~~\longrightarrow~~~\LNR_\text{int} = \psi^\dagger \, \ONRGeneral \, \psi \, , 
    \label{eq:schematic_LUV_to_LNR_map}
\end{align}
where $\ONRGeneral$ is the NR equivalent of $\OUVGeneral$ and will be written as a linear combination of NR basis operators. 

\subsection{The NR Interaction Lagrangian}
\label{subsec:fully_general_NR_int}

To derive the NR interaction Lagrangian, we utilize the relation between $\Psi$ and $\psi$ derived in Sec.~\ref{sec:NRQED}. Thus, we apply the unitary FW transformation $U_n$ in Eq.~\eqref{eq:main-trans} on $\Psi$, whose upper and lower components may then be identified as the $\psi$ and the heavy field $\psi_H$
\begin{align}
\label{eq:lintFW}
 \LNR_{\rm int} =   \bar \Psi  \OUVGeneral \Psi ~~\longrightarrow~~ 
& \begin{pmatrix}
    \psi^\dagger & \psi^\dagger_H
\end{pmatrix}
\Big[\hat U_n^\dag \, \gamma^0 \, \OUVGeneral \, \hat U_n\Big] \begin{pmatrix}
    \psi \\ \psi_H
\end{pmatrix}~.
\end{align}
As we found in Sec. \ref{sec:NRQED}, the resulting NR interaction Lagrangian is extracted by projecting out the heavy field $\psi_H$ with the explicit substitution $\Psi \to P_+ \Psi = (\psi~0)^T$ to obtain\footnote{Naively the expression in \Eq{eq:lintFW} is concerning because a generic ${\cal  O}_{\rm UV}$ operator will introduce off-diagonal terms, e.g., $\psi^\dagger(\cdots)\psi_H$, that survive even after the FW transformation has been applied. However, since the FW transformation removes the QED interactions between $\psi$ and $\psi_H$, any interactions introduced from further integrating out the heavy field $\psi_H$ will be quadratic in the DM coupling (and higher order in $1/m_e$). Since such couplings are typically very small these extra terms are therefore suppressed.
These terms may be kept by including $\OUVGeneral$ in the QED Lagrangian and diagonalizing the QED+DM Lagrangian together with the methods presented in Sec.~\ref{sec:NRQED}.} 
\begin{align}
    \LNR_\text{int} = \Psi^\dagger \left( P_+ \, U_n^\dagger \, \gamma^0 \OUVGeneral \,  U_n \, P_+ \right) \Psi = \begin{pmatrix}
        \psi^\dagger & 0
    \end{pmatrix} \left[ U_n^\dagger \, \gamma^0 \OUVGeneral   \, U_n  \right] \begin{pmatrix}
        \psi \\ 0
    \end{pmatrix} = \psi^\dagger \left( \text{Tr} \left[ P_+ \FWU^\dagger_n \, \gamma^0 \, \OUVGeneral \, U_n \right] \right) \psi \, ,
    \label{eq:LNR_int_general}
\end{align}
where the trace is over the $2 \times 2$ blocks of the $4 \times 4$ Dirac matrix structure, and the $P_+$ projection operator selects the top-left $2 \times 2$ block. From the definition in \Eq{eq:schematic_LUV_to_LNR_map}, the NR DM-electron interaction operator $\ONRGeneral$ can then be identified from Eq.~\eqref{eq:LNR_int_general} as
\begin{align}
    \ONRGeneral = \text{Tr} \left[ P_+ \FWU^\dagger_n \, \gamma^0 \, \OUVGeneral \, U_n \right] \, .
    \label{eq:ONR_general_decomposition}
\end{align}
To only keep terms to order $m_e^{-2}$, we use the FW operator from Eq.~\eqref{eq:main-trans} 
\be
\FWU_2 = \exp\left( -i\frac{X_0}{m_e} \right)\exp\left( -i\frac{X_1}{m_e^2} \right) ~~,
\ee
where $\FWX_0$, $\FWX_1$ are defined in \Eq{eq:X0_X1_definitions}. Inserting this expression into Eq.~\eqref{eq:ONR_general_decomposition}, the explicit NR DM-electron interaction operator can be written
\begin{align}
    \ONRGeneral \approx \text{Tr} \left[ P_+\left( \OUVGeneral + \frac{i}{2 m_e}\{\gamma^i D_i, \OUVGeneral \} - \frac{1}{8 m_e^2} \{\gamma^i D_i, \{\gamma^j D_j, \OUVGeneral \}\} - \frac{i e}{ 4m_e^2} \{\gamma^0\gamma^i F_{0i} , \OUVGeneral \} \right) \right] \, .
    \label{eq:ONR_general}
\end{align}
Note that Eq.~\eqref{eq:ONR_general} is general, and does not assume anything about the nature or 
the multiplicity of the dark fields in $\OUVGeneral$, as long as all higher dimension operators in the high-energy theory above $m_e$ are suppressed by energy scales much greater than $m_e$. In the remainder of this section, we will specialize to interactions that are linear in the 
dark sector fields, while retaining full generality in the Lorentz structure in $\OUVGeneral$.

\subsection{A Representative Interaction Lagrangian}
\label{subsec:example_high_to_low_match}

Using the main result of Sec.~\ref{subsec:fully_general_NR_int}, Eq.~\eqref{eq:ONR_general}, any (field-dependent) $\OUVGeneral$ can be related to a corresponding NR operator $\ONRGeneral$ to order $m_e^{-2}$. Our goal now is to decompose the latter as a linear combination of NR basis operators. However, since the physics of the dark sector is currently unknown, there is a vast multiplicity of possible fields and Lorentz structures that could be included in $\OUVGeneral$. Therefore, in this subsection we identify a representative class of UV operators that couple linearly to $\Psi$, and map these onto a basis of NR operators that act on $\psi$.

\vspace{1em}
\begin{center}
    \textit{ \normalsize Representative UV Interaction }
\end{center}
\vspace{1em}

To illustrate the utility of Eq.~\eqref{eq:ONR_general} in developing the NR EFT, consider the UV interaction Lagrangian
\begin{align}
    \LUV^{\rm UV}_\text{int} = y_s \, \phi \, \bar{\Psi} \Psi + i y_p \, \phi \, \bar{\Psi} \gamma^5 \Psi + g_v \, V_\mu \bar{\Psi} \gamma^\mu \Psi + g_a \, V_\mu \bar{\Psi} \gamma^\mu \gamma^5 \Psi + \frac{d_M}{2} \, V_{\mu \nu} \, \bar{\Psi} \sigma^{\mu \nu} \Psi \, ,
    \label{eq:example_LUV_int}
\end{align}
where $y_s, y_p, g_v, g_a$ and $d_M$ are constant coefficients, $\phi$ is a dark scalar field, $V_\mu$ is a dark vector field, and $V_{\mu \nu} = \partial_\mu V_\nu - \partial_\mu V_\nu$ is the dark field strength tensor. The corresponding form of  $\OUVGeneral$ in Eq.~\eqref{eq:lint} can be written 
\begin{align}
    \OUVGeneral = \left( y_s + i y_p \gamma^5 \right) \, \phi + \left( g_v \gamma^\mu + g_a \gamma^\mu \gamma^5 \right) V_\mu + \frac{d_M}{2}  \, \sigma^{\mu \nu} \, V_{\mu \nu} \, ,
    \label{eq:example_OUV}
\end{align}
which contains the most general structure of matrices that can contract with $\Psi$.\footnote{Recall that the 16 matrices $\{1, \gamma^5, \gamma^\mu, \gamma^\mu \gamma^5, \sigma^{\mu \nu} \}$ form a basis for all other $4 \times 4$ Dirac matrices~\cite{Peskin:1995ev}.}

This collection of possible interaction operators can describe a surprisingly versatile range of physics. For example, if only $y_s,~y_p$ are non-zero and $\phi$ is the DM this is the most general dimension-four electron-scalar DM interaction. Similarly if only $g_v,~g_a$ are non-zero and $V_\mu$ is the DM, this is the most general dimension-four electron-vector DM interaction. If only $d_M$ is non-zero, the electron has a ``dark" magnetic dipole moment with respect to $V_\mu$, and it was recently shown that $V_\mu$ can achieve the observed DM abundance through the UV freeze-in mechanisms~\cite{Krnjaic:2022wor}. The operator in Eq.~\eqref{eq:example_OUV} is also useful even when $\phi$ or $V$ are not the DM. For example, if they are mediator particles that couple the electron to a dark fermion, $\chi$, via the Lagrangian $\mathcal{L}^\text{UV}_\text{int} = \bar{\chi} \OUVGeneral \chi + \bar{\Psi} \OUVGeneral \Psi$. Such a Lagrangian would enumerate all possible dimension-four interactions between the dark sector and electron, and more due to the inclusion of the dark magnetic dipole term. The NR limit of $\OUVGeneral$ will be useful in understanding the physics of all these models.

Note that $\phi$ and $V_\mu$ in \Eq{eq:example_OUV} need not be fundamental and can represent operators built from multiple dark fields and/or their derivatives. For example, replacing $V_\mu \rightarrow \partial_\mu a$, where $a$ is an axion-like field, does not change the NR limit calculation; one can simply substitute $V_\mu$ with $\partial_\mu a$ at the end to recover axion-electron phenomenology~\cite{Berlin:2023ubt}. The same reasoning also applies to contact interactions built from products of dark fields. For example, the replacement $V^\mu \rightarrow \bar{\chi} \gamma^\mu \chi$, where $\chi$ is a dark fermion, also does not change the NR limit calculation. Although studying these alternative operators are beyond the scope of this paper, a thorough investigation may be useful for future work. 

\vspace{1em}
\begin{center}
    \textit{ \normalsize A Basis Of NR Electron Operators }
\end{center}
\vspace{1em}

In principle, Eq.~\eqref{eq:ONR_general} alone suffices to define the NR EFT for the UV interaction in \Eq{eq:example_LUV_int}; by directly substituting $\OUVGeneral$ from \Eq{eq:example_OUV} into \Eq{eq:example_LUV_int}, it is straightforward to determine $\ONRGeneral$ as a function of the couplings $y_s, y_p, g_v, g_a$ and $d_M$. 
However, if this NR expansion is not executed carefully, it can yield a complicated and unintuitive collection of terms. 

Therefore we introduce a simple basis of 8 {\it dimensionless} operators $\ONRBasis_\ell$, with $\ell \in \{1,\dots,8\}$, whose linear combinations generate all possible NR operators that will act on $\psi$ to order $m_e^{-2}$. These objects which arise in the NR expansion are built out of $\{ 1, \nabla^i, \bm{\sigma}^i, \Phi \}$, where $\Phi$ is the background potential on the electrons in the target material (for example, the electric potential sourced by the ions in a crystal). In principle we could also include contributions from a background vector potential, which would result in additional terms in this basis set. However, for most targets this is not relevant, so we omit these for simplicity. To order $m_e^{-2}$, this basis of operators acting on $\psi$ can be organized as follows:
\begin{itemize}
    \item {\bf Derivative-Only Operators:}  There are three basis elements which only depend on derivatives that act on $\psi$:
        \be
            &&  \ONRBasis_1 = 1~~,~~\big[   \ONRBasis_2 \big]^i = \frac{\nabla^i}{m_e}~~,~~\big[  \ONRBasis_3 \big]^{ij} = \frac{\nabla^i \nabla^j}{m_e^2}~~,~~ 
              \label{eq:ONR_basis1}
        \ee
    which we have normalized with powers of $m_e$ to make the $\ONRBasis$ dimensionless. Note that $\nabla^i$ has dimension one, which limits the number of terms that can arise at this order.

    \item {\bf Spin-Dependent Operators:} We also have spin-dependent basis elements proportional to Pauli matrices: 
        \be
            \big[  \ONRBasis_4 \big]^i = \bm{\sigma}^i~~,~~
            \big[   \ONRBasis_5 \big]^{ij} = \frac{\bm{\sigma}^i \nabla^j}{m_e}~~,~~\big[   \ONRBasis_6 \big]^{ijk} = \frac{\bm{\sigma}^i \nabla^j \nabla^k}{m_e^2} ~~.~~
              \label{eq:ONR_basis2}
        \ee
    Any term with two Pauli matrices may be reduced to a term with one using the identity $\bm{\sigma}^i \bm{\sigma}^j = \delta^{ij} + i \epsilon^{ijk} \bm{\sigma}^k$.

    \item {\bf Background-Dependent Operators:} We also have two operators that depend on the background electric potential field $\Phi$: 
        \be
            \big[   \ONRBasis_7 \big]^i = \frac{e \, (\nabla^i \Phi)}{m_e^2}~~, 
            ~~\big[   \ONRBasis_8 \big]^{ij} = \frac{e \, \bm{\sigma}^i (\nabla^j \Phi)}{m_e^2} \, ,
            \label{eq:ONR_basis3}
        \ee
    where the brackets around $( \nabla^i \Phi )$ indicate that the enclosed gradient only acts on $\Phi$. Note that $\Phi$ has dimension one, which also limits the terms that can appear at this order; operators containing $(\nabla \nabla \Phi)$ or  $(\nabla \Phi)^2$ terms are higher order. 
\end{itemize}

This set corresponds to all dimension two operators built out of the set $\{ 1, \nabla^i, \bm{\sigma}^i, \Phi \}$ (normalized to be dimensionless with the appropriate factors of $1/m_e$), with the exception of $\ONRBasis = \Phi / m_e$ since $\Phi$ does not appear in $X_0, X_1$ in Eq.~\eqref{eq:X0_X1_definitions}. With the eight dimensionless operators in Eqs.~\eqref{eq:ONR_basis1}-\eqref{eq:ONR_basis3}, any $\ONRGeneral$ can be schematically decomposed as
\begin{align}
    \ONRGeneral = \sum_{\ell=1}^8 \left( \cdots \right)  \ONRBasis_\ell \, ,
\label{eq:ONR_general_decomposition_1}
\end{align}
where we have suppressed the spin and spatial indices on $\ONRBasis_\ell$, and the expression $(\cdots)$ represents any other combination of fields which do not act on $\psi$ or contract with its spinor indices, such that $\psi^\dagger \ONRGeneral \psi = \sum_\ell (\cdots) (\psi^\dagger \ONRBasis_\ell \psi)$.  Note that in our convention, $\ONRGeneral$ has mass dimension 1 to match $\OUVGeneral$, which differs from the dimensionless basis operators $\ONRBasis_\ell$.

\vspace{1em}
\begin{center}
    \textit{ \normalsize Expanding The Field Content In $\ONRGeneral$ }
\end{center}
\vspace{1em}

We now turn to the problem of classifying all the field-dependent terms that can appear inside the $(\cdots)$ brackets of Eq.~\eqref{eq:ONR_general_decomposition_1}. In general, these expressions are combinations of dark sector fields $\{ \phi, V^\mu \}$, the QED photon field $A^\mu$, and derivatives acting on these.\footnote{Note that the photon field $A^\mu$ represents quantized electromagnetic field excitations and is treated as distinct from the background field $\Phi$, which appears in the $\ONRBasis_\sumIndex$ basis elements and corresponds to the background expectation value $\Phi = \langle A^0\rangle$.} However, even with only the UV operators in \Eq{eq:example_OUV}, extracting the equivalent NR operators to order $m_e^{-2}$ in \Eq{eq:ONR_general} yields a large multiplicity of terms, so unlike the $\ONRBasis$ basis operators that act on $\psi$, a simple enumeration of all possible fields and derivatives combinations is unwieldy. 

To manage this we introduce the dimensionless (field-independent) coefficients $\CNR$ to expand any combination of fields and derivatives that may appear in the $(\cdots)$ brackets of Eq.~\eqref{eq:ONR_general_decomposition_1}
\begin{align}
    \ONRGeneral = \sum_{\sumIndex = 1}^8 & \left[ C_{\phi, \sumIndex} \phi + C_{(\nabla \phi), \sumIndex} \frac{(\nabla \phi)}{m_e} + C_{V^0, \sumIndex} V^0 + C_{\Vv, \sumIndex} \Vv + C_{(\nabla \Vv), \sumIndex} \frac{(\nabla \Vv)}{m_e} \right. \nonumber \\
     & \left.\quad +\; C_{\phi A^0, \sumIndex} \frac{\phi A^0}{m_e} + C_{\phi \Av, \sumIndex} \frac{\phi \Av}{m_e} + C_{V^0 \Av, \sumIndex} \frac{V^0 \Av}{m_e} + C_{(\nabla \phi) \Av, \sumIndex}\frac{(\nabla \phi) \Av}{m_e^2} + \cdots \right] \ONRBasis_\sumIndex
     ~,
     \label{eq:ONR_expand_1}
\end{align}
where we have listed an exemplary set of field operators, and decomposed the four-dimensional fields into their $0$ and spatial components explicitly: $A^\mu = (A^0, \bm{A})$, $V^\mu = (V^0, \bm{V})$. Each term in the brackets in Eq.~\eqref{eq:ONR_expand_1} is normalized with the appropriate power of $m_e$ to ensure that the overall mass dimension is one to match $\ONRGeneral$. Each coefficient $\CNR$ in Eq.~\eqref{eq:ONR_expand_1} has two subscripts separated by a comma:
\begin{itemize}
    \item {\bf Left Subscript:} Matches the field content it multiplies in \Eq{eq:ONR_expand_1}. For example, $C_{(\nabla \phi), \cdots}$ multiplies $(\nabla \phi)$.
    \item{\bf Right Subscript:} Matches the subscript of the multiplied basis operator $\ONRBasis_\ell$.
\end{itemize} 
Furthermore note that we have suppressed all Cartesian indices in Eq.~\eqref{eq:ONR_expand_1}. These may be added back in by matching the indices on the field operators and $\ONRBasis_\ell$ with $C$ indices. To illustrate this, we expand Eq.~\eqref{eq:ONR_expand_1} further and keep all indices explicit
\begin{align}
    \ONRGeneral = & \left( C_{\phi, 1} \phi + \left[C_{(\nabla \phi), 1} \right]^{\color{red} i} \frac{(\nabla^{\color{red} i} \phi)}{m_e} + C_{V^0, 1} V^0 + \left[ C_{\Vv, 1} \right]^{\color{red} i} \Vv^{\color{red} i} + \left[ C_{(\nabla \Vv), 1} \right]^{\color{red} ij} \frac{(\nabla^{\color{red} i} \Vv^{\color{red} j})}{m_e} \right) \ONRBasis_1 \nonumber \\
    & + \left( \left[ C_{\phi A^0, 2} \right]^{\color{blue} k} \frac{\phi A^0}{m_e} + \left[ C_{\phi \Av, 2} \right]^{\color{red} i, \color{blue} k} \frac{\phi \Av^{\color{red} i}}{m_e} + \left[ C_{V^0 \Av, 2} \right]^{\color{red} i, \color{blue} k} \frac{V^0 \Av^{\color{red} i}}{m_e} + \left[ C_{(\nabla \phi) \Av, 2} \right]^{\color{red} ij, \color{blue} k}\frac{(\nabla^{\color{red} i} \phi) \Av^{\color{red} j}}{m_e^2}  \right) 
    \big[ \ONRBasis_2 \big]^{\color{blue} k} + \, \cdots \, \, ,
    \label{eq:ONR_expand_2}
\end{align}
where the superscripts on the $C$'s in Eq.~\eqref{eq:ONR_expand_2} follow a similar convention as the subscripts
\begin{itemize}
    \item {\bf Left Superscript:} Matches the indices of the field content ($\color{red} ij$ in \Eq{eq:ONR_expand_2}).
    \item {\bf Right Superscript:} Matches the indices of $\ONRBasis_{\sumIndex}$ ($ \color{blue} k$ in \Eq{eq:ONR_expand_2}).
\end{itemize}
Lastly, note that \textit{all Cartesian indices are raised} in Eq.~\eqref{eq:ONR_expand_2}; the lowered components of a vector $v$ are related by $v_i = \eta_{ij} v^j$, where $\eta^{ij} = -\delta^{ij}$ are the spatial components of the metric tensor with mostly negative signature.

\subsection{Matching The Representative UV-NR Interactions}

\SetTblrInner{rowsep=6pt}
\begin{table}[ht!]
    \begin{tblr}{
        width=\textwidth,
        colspec={ X[-1,c,$$] X[1,c,$$] X[1,c,$$] X[1,c,$$] X[1,c,$$] X[1,c,$$] X[1,c,$$] },
        vline{2} = {1-Z}{1pt,solid},
        hline{2} = {1pt,solid},
        hline{Z} = {1pt,solid},
        hline{4} = {1-Z}{0.5pt, solid},
        cell{5-6}{6-Z} = {light-gray},
        cell{7-Z}{4-Z} = {light-gray}} 
        \SetCell[c=7]{c} \text{\large{ NR EFT Lagrangian Coefficients, $C_{(\text{fields}), (\text{basis operator})}$}} & & & & & & \\
        & \ONRBasis_1 & \big[ \ONRBasis_4 \big]^{i} & \big[ \ONRBasis_2 \big]^i & \big[ \ONRBasis_5 \big]^{ij} & \big[ \ONRBasis_3 \big]^{ij}& \big[ \ONRBasis_8 \big]^{ij} \\ 
        & 1 & \bm{\sigma}^i & \frac{\nabla^i}{m_e} & \frac{\bm{\sigma}^i \nabla^j}{m_e} & \frac{\nabla^i \nabla^j}{m_e^2} & \frac{e \bm{\sigma}^i (\nabla^j \Phi)}{m_e^2} \\
        \phi 
            & y_s & 0 & 0 & 0 & y_s \, \frac{\delta^{ij}}{2} & - y_p \, \frac{\delta^{ij}}{2} \\
        \frac{(\nabla^a \phi)}{m_e} 
            & 0 & -y_p \, \frac{\delta^{ai}}{2} & y_s \, \frac{\delta^{ai}}{2} & y_s \, \left[ \frac{-i \epsilon^{aij}}{4} \right] & & & \\
        \frac{\phi \, \Av^a}{m_e} 
            & 0 & 0 & - i e y_s\, \delta^{ai} & 0 & \\
        \frac{(\nabla^a \nabla^b \phi)}{m_e^2} 
            & y_s \, \frac{\delta^{ab}}{8} & 0 & & & & \\
        \frac{\phi \, ( \partial_t \Av^a )}{m_e^2} 
            & 0 & - e y_p \frac{\delta^{ai}}{2} & & & & \\
        \frac{\phi \, ( \nabla^a A^0 )}{m_e^2} 
            & 0 & - e y_p \frac{\delta^{ai}}{2} & & & & & \\
        \frac{(\nabla^a \phi) \, \Av^b}{m_e^2} 
            & e  y_s \, \left[ \frac{-i \delta^{ab}}{2}\right] &e y_s \, \frac{\epsilon^{abi}}{4} & & & \\
        \frac{\phi \, ( \nabla^a \Av^b )}{m_e^2} 
            & e y_s \, \left[ \frac{-i \delta^{ab}}{2}\right] & e y_s \, \frac{\epsilon^{abi}}{2} & & & \\
        \frac{\phi \Av^a \Av^b}{m_e^2} 
            & - e^2 y_s \, \frac{\delta^{ab}}{2} & 0 & & & & 
    \end{tblr}
    \caption{The $C$ coefficients of the NR interaction Lagrangian derived from the terms with a dark scalar field, $\phi$, in the representative UV Lagrangian (Eq.~\eqref{eq:example_LUV_int} with $y_s, y_p$ non-zero) to order $m_e^{-2}$. Each entry is a $C$ with a left subscript matching the fields to the left, and right subscript matching the basis operator subscript to the top. Superscripts on $C$ match the corresponding field content and basis operator. For example, $\left[ C_{(\nabla \nabla \phi), 1} \right]^{ij} = y_s \delta^{ij}/8$ contributes a term in the NR interaction Lagrangian written abstractly as $\LNR_\text{int} \supset C_{(\nabla \nabla \phi), 1} [ (\nabla \nabla \phi) / m_e^2 ] [\psi^\dagger \ONRBasis_1 \psi ]$, and concretely as $\LNR_\text{int}~\supset~\left[ C_{(\nabla \nabla \phi), 1} \right]^{ij} (\nabla^i \nabla^j \phi) (\psi^\dagger \psi)/m_e^2~=~ y_s (\nabla^2 \phi) \psi^\dagger \psi / 8 m_e^2$. 
    The basis operators and field content are ordered by powers of $1/m_e$; entries farther right contribute terms higher order in electron velocity, and entries farther down contribute terms higher order in incoming/outgoing field four-momenta.
    Grayed entries contribute terms in the NR interaction Lagrangian at order $m_e^{-3}$ and above, and are therefore neglected here;
    columns for $\ONRBasis_6, \ONRBasis_7$ are ignored because all entries are zero for this set of UV interactions.} 
    \label{tab:scalar_coefficients}
\end{table}

\SetTblrInner{rowsep=6pt}
\begin{table}[ht!]
    \begin{tblr}{
        width=\textwidth,
        colspec={ X[-1,c,$$] X[1,c,$$] X[1,c,$$] X[1,c,$$] X[1,c,$$] X[1,c,$$] X[1,c,$$] },
        vline{2} = {1-Z}{1pt,solid},
        hline{2} = {1pt,solid},
        hline{4} = {0.5pt,solid},
        hline{Z} = {1pt,solid},
        cell{6-9}{6-Z} = {light-gray},
        cell{10-Z}{4-Z} = {light-gray}
        } 
            \SetCell[c=7]{c} \text{\large NR EFT Lagrangian Coefficients, $C_{(\text{fields}), (\text{basis operator}) }$} & & & & & & \\
            & \ONRBasis_1 & \big[ \ONRBasis_4 \big]^i & \big[ \ONRBasis_2 \big]^{i} & \big[ \ONRBasis_5 \big]^{ij} & \big[ \ONRBasis_6 \big]^{ijk} & \big[ \ONRBasis_8 \big]^{ij} \\ 
            & 1 & \bm{\sigma}^i & \frac{\nabla^i}{m_e} & \frac{\bm{\sigma}^i \nabla^j}{m_e} & \frac{\bm{\sigma}^i \nabla^j \nabla^k}{m_e^2} & \frac{e \bm{\sigma}^i (\nabla^j \Phi)}{m_e^2} \\
            V^0 & 
                g_v & 0 & 0 & - i g_a \delta^{ij} & 0 & 0 \\ 
            \Vv^a &
                0 & - g_a \delta^{ai} & i g_v \delta^{ai} & 0 & \frac{g_a}{2} H_1^{aijk} & - \frac{g_v}{2} \epsilon^{aij} \\ 
            \frac{\nabla^a V^0}{m_e} & 
                0 & - \frac{i g_a}{2} \delta^{ai} & 0 & \frac{i g_v}{4} \epsilon^{aij} & & \\
            \frac{\nabla^a \Vv^b}{m_e} & 
                \frac{i g_v}{2} \delta^{ab} & - \frac{g_v}{2} \epsilon^{abi} & \frac{i g_a}{4} \epsilon^{abi} & \frac{g_a}{4} H_2^{abij} & & \\
            \frac{V^0 \Av^a}{m_e} & 
                0 & - e g_a \delta^{ai} & 0 & 0 & & & \\
            \frac{\Vv^a \Av^b}{m_e} & 
                e g_v \delta^{ab} & 0 & 0 & - \frac{i e g_a}{2} H_2^{abji} & & & \\
            \frac{\nabla^a \nabla^b V^0}{m_e^2} &
                \frac{g_v}{8} \delta^{ab} & 0 & & & & \\ 
            \frac{\nabla^a \nabla^b \Vv^c}{m_e^2} &
                0 & - \frac{g_a}{8} \delta^{ab} \delta^{ci} & & & & \\ 
            \frac{(\nabla^a V^0) \Av^b}{m_e^2} &
                0 & - \frac{e g_v}{4} \epsilon^{abi} & & & & \\ 
            \frac{(\nabla^a \Vv^b) \Av^c}{m_e^2} &
                \frac{e g_a}{4} \epsilon^{abc} & - \frac{i e g_a}{4} H_2^{abic} & & & & \\ 
            \frac{\Vv^a (\partial_t \Av^b) }{m_e^2} &
                 0 & \frac{e g_v}{2} \epsilon^{abi} & & & & \\ 
             \frac{\Vv^a (\nabla^b \Av^c) }{m_e^2} &
                 - \frac{e g_a}{2} \epsilon^{abc} & - \frac{i e g_a}{4} H_2^{abic} & & & & \\ 
            \frac{\Vv^a (\nabla^b A^0) }{m_e^2} &
                  0 & \frac{e g_v}{2} \epsilon^{abi} & & & & \\ 
            \frac{\Vv^a \Av^b \Av^c }{m_e^2} &
                 0 & -\frac{e^2 g_a}{2} H_1^{abic} & & & & \\ 
    \end{tblr}
    \caption{
    The $C$ coefficients of the NR interaction Lagrangian derived from the terms with a dark vector field, $V$, in the representative UV Lagrangian (Eq.~\eqref{eq:example_LUV_int} with $g_v, g_a$ non-zero) to order $m_e^{-2}$. Table entries ($C$'s) follow the same convention as in Table~\ref{tab:scalar_coefficients}, and is discussed in detail in Sec.~\ref{sec:NR_DM_electron_interaction}.
    Grayed entries contribute terms in the NR interaction Lagrangian at order $m_e^{-3}$ and above, and are therefore neglected here. 
    Columns for $\ONRBasis_3, \ONRBasis_7$ are ignored because all entries are zero. Furthermore, to help with presentation we define some four index tensors, $H_1^{ijkl} \equiv \delta^{ik} \delta^{jl} - \delta^{ij} \delta^{kl}$, $H_2^{ijkl} \equiv \delta^{ij} \delta^{kl} + \delta^{ik} \delta^{jl} - 2 \delta^{il} \delta^{jk}$.
    \vspace{1em}
    }
    \label{tab:vector_coefficients}
\end{table}

With this notation our goal of taking NR limit crystallizes: after substituting the general $\OUVGeneral$ in Eq.~\eqref{eq:example_OUV} into Eq.~\eqref{eq:ONR_general}, find the $\CNR$'s as functions of $y_s, y_p, g_v$, and $g_a$; we present analogous expressions for $d_M$ in App.~\ref{app:summary_tables}. This defines the NR EFT of DM-electron interactions to order $m_e^{-2}$ determined by the \he Lagrangian in Eq.~\eqref{eq:example_LUV_int}. This is a straightforward, albeit tedious, task which is greatly simplified with the help of \textsf{FeynCalc}~\cite{Mertig:1990an,Shtabovenko:2016sxi,Shtabovenko:2020gxv,Brambilla:2020fla,Shtabovenko:2023idz}. The results for the scalar coupled models ($y_s, y_p$ non-zero) and the vector coupled models ($g_v, g_a$ non-zero) are shown in Tables~\ref{tab:scalar_coefficients} and~\ref{tab:vector_coefficients}, respectively. 

To further elucidate the interactions that arise from the $C$'s in Tables~\ref{tab:scalar_coefficients} and~\ref{tab:vector_coefficients} we can separate out the scalar and vector contributions to $\ONRGeneral$ as
\be
    \ONRGeneral = \ONRGeneral^\phi + \ONRGeneral^{V^0}+ \ONRGeneral^{\bm V}.
\ee
All non-zero entries in the tables can be written as terms in either $\ONRGeneral^\phi, \ONRGeneral^{V^0}$ or $\ONRGeneral^{\bm V}$. For the scalar coefficients, this decomposition yields
\be
    \ONRGeneral^\phi &=&
    \sum_{\ell = 1}^8 \biggl[ 
    C_{\phi, \ell} \, \phi + 
    C_{(\nabla \phi),\ell} \frac{( \nabla \phi )}{m_e} + 
    C_{\phi \bm{A}, \ell} \, \frac{\, \phi \bm{A}}{m_e}  + 
    C_{(\nabla\nabla \phi), \ell} \, \frac{(\nabla\nabla \phi)}{m_e^2} + 
    C_{\phi (\partial_t \Av), \ell} \, \frac{\, \phi (\partial_t \Av)}{m^2_e}
    \nonumber \\ 
    && \quad ~ + 
    C_{\phi (\nabla A^0), \ell} \, \frac{\phi (\nabla A^0)}{m_e^2} + 
    C_{( \nabla \phi ) \Av, \ell} \, \frac{ (\nabla \phi) \Av}{m_e^2} + 
    C_{ \phi (\nabla \bm A), \ell} \, \frac{\phi (\nabla \bm{A})}{m_e^2} +
     C_{\phi  {\bm A \bm A}, \ell} \, \frac{\phi  \bm{A}  \bm{A} }{m_e^2}  
    \,  \biggr]  \ONRBasis_\ell ~,
        \label{eq:ONR_general_decomposition_2}
\ee
The corresponding expression for vector time component $V^0$ is 
 \be
    \ONRGeneral^{V^0} &=&
    \sum_{\ell = 1}^8 \biggl[ 
    C_{V^0, \ell} \, V^0 + 
    C_{(\nabla V^0),\ell} \frac{( \nabla V^0 )}{m_e} +
     C_{V^0 \bm A,\ell} \frac{V^0 \bm A}{m_e} + 
     C_{(\nabla \nabla V^0),\ell} \frac{( \nabla \nabla V^0 )}{m_e^2} +
     C_{(\nabla V^0) \Av,\ell} \frac{( \nabla V^0 )\Av }{m_e^2}
    \,  \biggr]  \ONRBasis_\ell ~,
        \label{eq:ONR_general_decomposition_3}
\ee
and finally, the analogous field expansion for the vector spatial components $\Vv$ is
\be
    \ONRGeneral^{\bm V} &=&
    \sum_{\ell = 1}^8 \biggl[ 
    C_{\bm V, \ell} \, \bm V + 
    C_{(\nabla \bm V),\ell} \frac{( \nabla \bm V )}{m_e} +
    C_{\bm V \bm A,\ell} \frac{\bm V \bm A}{m_e} + 
    C_{(\nabla \nabla \bm V) ,\ell} \frac{ (\nabla\nabla \bm V) }{m^2_e} + 
    C_{(\nabla  \bm V) \bm A,\ell} \frac{(\nabla \bm V) \bm A}{m^2_e}  \nonumber \\
    &&  \quad ~ 
    + C_{\bm V (\partial_t \bm A),\ell} \frac{ \bm V (\partial_t \bm A)}{m^2_e}  
    + C_{\bm V (\nabla  \bm A),\ell} \frac{ \bm V (\nabla \bm A)}{m^2_e}    
    + C_{ \bm V \bm A \bm A ,\ell} \frac{\bm V \bm A \bm A}{m^2_e}     \,  \biggr]  \ONRBasis_\ell ~~.
    \label{eq:ONR_general_decomposition_4}
\ee
In addition to presenting the $\CNR$ coefficients, we also summarize the results in App.~\ref{app:summary_tables} for the cases where only one of the $y_s, y_p, g_v, g_a, d_M$ coefficients is non-zero. For each table in App.~\ref{app:summary_tables} the first row is the assumed \he interaction Lagrangian and the second row is the corresponding NR EFT interaction Lagrangian. The later sections of the tables give the Feynman rules for different vertices that arise in NR EFT, and their derivation is discussed in detail next in Sec.~\ref{sec:feynman_rules}.


\section{Feynman Rules In The NR EFT}
\label{sec:feynman_rules}

To understand the phenomenology of a given DM interaction, one typically uses Feynman diagrams to compute matrix elements associated with the observable of interest. This procedure is usually straightforward since the Feynman rules for usual relativistic QFTs are well known textbook material~\cite{Peskin:1995ev,Schwartz}. However, here our NR EFT Lagrangian (the theory defined by the NR operators from Eqs.~\eqref{eq:ONR_general_decomposition_2}~-~\eqref{eq:ONR_general_decomposition_4}) is not a familiar relativistic QFT, and therefore the corresponding Feynman rules do not apply in the usual way. 

Usually in relativistic QFT, one perturbs around the free theory, which is defined as the limit where interactions and background fields are absent. This means that the building blocks for relativistic Feynman rules (incoming/outgoing states, propagators, and vertex insertions) assume relatively simple forms since the states involved are mainly plane waves. For NR electrons in generic direct detection targets, these simple elements are inappropriate for two key reasons: 
\begin{itemize}
    \item Unlike in the free theory, here the electrons in atoms, molecules or crystals are embedded in background potentials, which alter their wavefunctions and energy levels. 
    \item Electrons occupy filled states within the target, so the background here is no longer the pure vacuum. This situation bears conceptual resemblance to 
    QFT at finite temperature, where the background is thermally populated~\cite{Bellac,Hardy:2016kme}, so the primitive elements that define the Feynman rules differ from their familiar vacuum expressions.
\end{itemize}
However, despite these differences, there are also many recognizable similarities, as the central quantity of interest is a transition matrix element,
which is systematically calculated using intuitive Feynman diagrams.

The purpose of this section is to derive the Feynman rules for an NR EFT Lagrangian with electrons in an arbitrary environment at zero temperature. Similar discussions can also be found in condensed matter textbooks~\cite{Mahan}, although these usually focus on the NR QED Lagrangian without including the effects of new physics. Furthermore, as emphasized in Sec.~\ref{sec:NR_DM_electron_interaction}, because the NR EFT prescription only modifies the electron degrees of freedom, no other fields in the \he Lagrangian need to change. Thus, in this section we develop new Feynman rules for the NR electrons, without affecting the familiar QFT Feynman rules that apply to dark fields, which can even be relativistic in our treatment.

This section is organized as follows: in Sec.~\ref{subsec:nr_matrix_element_def} we define the transition matrix element. In Sec~\ref{subsec:nr_electron_field_quantization} we expand the electron field in terms of the ``in-medium" eigenstates. In Secs.~\ref{subsec:3_pt_vertex},~\ref{subsec:4_pt_vertex} we use these definitions to derive the three and four point Feynman rules, respectively. Next in Sec.~\ref{subsec:nr_electron_propagator} we derive the NR electron propagator needed to connect more complicated diagrams. In Sec.~\ref{subsec:loop_diagram} we use the three point, four point, and propagator Feynman rules to derive loop diagrams, and provide new Feynman rules for those diagrams as well. The Feynman rules will be introduced along the way in each subsection, and bulleted. In Sec.~\ref{subsec:feynman_rules_summary} we summarize all the new Feynman rules introduced. 

\subsection{NR Matrix Element Definition}
\label{subsec:nr_matrix_element_def}

We begin by defining the matrix element $\mathcal{M}$ which we will build using Feynman rules. In familiar relativistic QFT this quantity is defined as
\begin{align}
    \langle \mathcal{F} | \, i \mathcal{T}_{\rm UV} \, | \mathcal{I} \rangle = (2 \pi)^4 \, \delta^{4}\left( {\textstyle \sum} p
\right) (i \mathcal{M}_\text{UV})~~~,~~~1 + i \mathcal{T}_{\rm UV} \equiv T\left\{   \exp \left(  i \int d^4 x \, \LUV_\text{int} \right) \right\} \, ,
    \label{eq:matrix_element_QFT}
\end{align}
where $| \mathcal{I} \rangle$ and $| \mathcal{F} \rangle$ are the initial and final states, respectively, $\cal M_{\rm UV}$ is the relativistic matrix element, $\mathcal{T}_{\rm UV}$ is the transfer matrix, $T\{\}$ is a time-ordered product and $\LUV_\text{int}$ is interaction Lagrangian. The sum over all external
state four-momenta ${\textstyle \sum} p$ is shorthand for $\sum p \equiv \sum_{\rm in} p_{\rm in} - \sum_{\rm out} p_{\rm out}$, which enforces energy-momentum conservation for the $\cal I \to F$ transition. 

In the presence of a background field, or if boundary conditions apply to the system, spatial momentum is not necessarily conserved since continuous translation symmetry is absent. Therefore, the factor of $\delta^{3}(\sum \bm{p})$ does not necessarily appear in NR transition amplitudes, so we define
\begin{align}
    \left\langle \mathcal{F} | \, i \mathcal{T}_\text{NR} \, | \mathcal{I} \right\rangle \equiv 2 \pi \, 
     \delta \! \left({\textstyle \sum E} \right) 
    \, (i \mathcal{M}_{\rm NR})~~~,~~~1 + i \mathcal{T}_\text{NR}= T\left\{ \exp\left(i \int d^4 x \, \LNR_\text{int} \right) \right\} \, , 
    \label{eq:matrix_element}
\end{align}
where $\delta\left({\textstyle \sum E} \right)$ is an energy conserving delta function, the transfer matrix is written analogously with its UV counterpart, and $\LNR_\text{int}$ is the interaction part of an NR EFT Lagrangian. Note that due to the definition in Eq.~\eqref{eq:matrix_element}, $\mathcal{M}_\text{NR}$ and $\mathcal{M}_\text{UV}$ have different mass dimensions.

The absence of momentum conservation is important because it changes how four-momentum flows through a diagram. In relativistic QFT four-momentum is conserved at each vertex, so the delta function can be removed when the remaining undetermined four-momenta are integrated over. 
By contrast, in the NR theory developed here, four-momentum conservation is reduced to just energy conservation, so the corresponding Feynman rules for calculating the NR matrix element $\mathcal{M}_\text{NR}$ become
\begin{itemize}
    \item Build $ i \mathcal{M}_\text{NR}$ out of NR vertex factors and propagators (as discussed below) 
    \item Conserve {\it energy} at each vertex
    \item For each undetermined energy $E$, insert factors of $\displaystyle \left( \int \frac{dE}{ 2\pi} \right)$ and perform the corresponding integral
    \item Multiply the resulting expression by $-i$
\end{itemize}
These rules consistently remove all remaining energy conserving delta functions that appear at intermediate steps as we will see below.

\subsection{NR Electron Field Quantization}
\label{subsec:nr_electron_field_quantization}

Before computing the matrix elements in Eq.~\eqref{eq:matrix_element} we must first quantize the electron field, $\psi$. In relativistic QFT, this is done by expanding the field in a basis of solutions to the Dirac equation and imposing anti-commutation relation on the raising and lowering operators that create single-particle states~\cite{Peskin:1995ev,Schwartz}. Here in the NR EFT, the situation is different since the NR electron field $\psi$ no longer satisfies the Dirac equation.
Using Eq.~\eqref{eq:NR_QED_lagrangian_2} to extract the equation of motion for $\psi$, to leading order in $1/m_e$, we recover the Schr\"odinger equation
\begin{align}
    i \partial_t \psi \approx \left( \frac{{\bm k}^2}{2 m_e} + e  \Phi \right) \psi  \, ,
    \label{eq:electron_EOM_leading_order}
\end{align}
where $\bm k$ is the electron momentum. In general, solving Eq.~\eqref{eq:electron_EOM_leading_order} in a many-electron system is a challenging task, primarily due to electron-electron interactions. However in most of the targets relevant for DM direct detection, independent electron, or single-particle, approximations~\cite{Girvin_2019} can be made to simplify the system (see, for example, Refs.~\cite{Essig:2015cda,Blanco:2019lrf,Griffin:2021znd}), such that the electron field can then be expanded in a basis of single-electron eigenstates as~\cite{weinberg}
\begin{align}
    \psi({\bm x}, t) = \sum_J \, e^{- i E_J t} \, \psi_J({\bm x}) \, b_J \, ,
    \label{eq:electron_field_quantization}
\end{align}
where $\psi(\vec{x}, t)$ satisfies Eq.~\eqref{eq:electron_EOM_leading_order}, $\psi_J(\vec{x})$ satisfies the time-independent version of Eq.~\eqref{eq:electron_EOM_leading_order}, and the 
$b_J$ are single-electron annihilation operators. Computing these $\psi_J(\vec{x})$ is a non-trivial task, and a variety of analytic and numeric approaches, e.g., in Refs.~\cite{Graham:2012su,Lee:2015qva,Essig:2015cda,Blanco:2019lrf,Griffin:2021znd},  have been used to model them. We emphasize that the goal here is not to discuss specific realizations of the $\psi_J(\vec{x})$, but rather provide results which can be applied for any $\psi_J(\vec{x})$, as long as the electron field is well approximated by Eq.~\eqref{eq:electron_field_quantization} and the $\psi_J(\vec{x})$ wavefunctions are known.

To satisfy the equal time anti-commutation relations, we impose canonical anti-commutation relations for $b_J$ and require the states form a complete basis
\begin{align}
    \{ \psi({\bm x}, 0), \psi^\dagger({\bm y}, 0) \} = \delta^{3}({\bm x} - {\bm y})~~~\Longleftrightarrow~~~\left\{ b_J, b_{K}^\dagger \right\} = \delta_{J K}~~~,~~~\sum_J \, \psi_J({\bm x}) \, \psi_J^\dagger({\bm y}) = \delta^3({\bm x} - {\bm y}) \, ,
    \label{eq:canonical_commutation_relations_b}
\end{align}
and these conditions demand that the wavefunctions form an orthonormal basis
\begin{align}
    \int d^3{\bm x} \, \psi^\dagger_J({\bm x}) \, \psi_K({\bm x}) = \delta_{IJ}
    ,
    \label{eq:completeness_and_orthonormalization}
\end{align}
where the single-electron states are $| J \rangle = b_J^\dagger \, | 0 \rangle$ and $| 0 \rangle$ is the vacuum state of the target. 

While these expressions are similar to their relativistic analogues, there are two key differences: 1) the appearance of a state sum over $J$ instead of an integral over spatial momenta in \Eq{eq:electron_field_quantization}, and 2) the Kronecker delta symbol $\delta_{JK}$ instead of a Dirac delta function in \Eq{eq:completeness_and_orthonormalization}. Although free-electron states can be indexed with their spatial momenta, this is not possible for general electronic states in direct detection targets (e.g., bound states with discrete energy levels). In general, NR electronic states must be indexed with some combination of discrete and continuous variables, and our formalism is sufficiently flexible to accommodate both cases: if a given state is continuously indexed (as in the free-electron case), then we replace the discrete elements with their continuous counterparts
\begin{align}
    \delta_{JK} \rightarrow 2 \pi \, \delta ( J - K )~~~,~~~\sum_J \rightarrow \int \frac{d J }{2 \pi} \, .
    \label{eq:discrete_to_continuous}
\end{align}
For example, if $J = \bm{k}$, then all the indices are continuous and this replacement yields 
\begin{align}
    \delta_{\bm{k}\bm{k}'} \rightarrow (2 \pi)^3 \delta^3 ( \bm{k} - \bm{k}' )~~~,~~~\sum_{\bm{k}} \rightarrow \int \frac{d^3 \bm{k} }{(2 \pi)^3} \, ~ ,
\label{eq:discrete_to_continuous_example}
\end{align}
which recovers the familiar relativistic formalism. However, note that with these conventions, free-electron states are normalized according to 
\be
\label{eq:electron-state-normalization}
\langle {\bm k} | {\bm k} \rangle = \vol~,
\ee
which differs from the commonly used $\langle {\bm k} | {\bm k} \rangle_{\rm rel} = 2 E_{\vec{k}} \vol$ relativistic convention \cite{Schwartz},
but has the virtue of simplifying the NR electron Feynman rules that we derive below.

\subsection{Three-Point Vertices}
\label{subsec:3_pt_vertex}

With the NR electron states defined, we can now begin to derive vertex Feynman rules for a general three-point interaction between NR electrons and linearly coupled dark fields. 

\vspace{1em}
\begin{center}
    \textit{ \normalsize Warmup }
\end{center}
\vspace{1em}

We begin with the simplest three-point vertex example: a Yukawa coupling in the NR Lagrangian
\begin{align}
    \LNR_\text{int} = g\, \phi \,\psi^\dagger \psi \, ,
    \label{eq:example_l_int_3_point}
\end{align}
where $\phi$ is a scalar field and $g$ is a small coupling constant. To derive the Feynman rule for $\mathcal{M}_{\rm NR}$, we use Eq.~\eqref{eq:matrix_element} and expand $\mathcal{T}_\text{NR}$ to leading order in $g$, which yields
\be
\label{eq:matrix_element_g}
  \left\langle \mathcal{F} | \, i \mathcal{T}_\text{NR} \, | \mathcal{I} \right\rangle = 2 \pi \delta \!\left({\textstyle \sum E} \right)  ( i \mathcal{M}_{\rm NR} ) 
    \approx \langle {\cal F} | \left( i g \! \int d^4 x \, \phi \,\psi^\dagger \psi \right) |\cal I\rangle,
\ee
where the initial state consists of a {\it free} (and possibly relativistic) $\phi$ particle in an eigenstate of momentum $\bm{q}$, and an electron in some arbitrary NR state $|J \rangle$; the 
final state consists only of an electron in an NR state $| K \rangle$. We can write the corresponding $| \mathcal{I} \rangle$ and $| \mathcal{F} \rangle$ as the product states
\be
    |\mathcal{I} \rangle = | \bm{q} \rangle \otimes | J \rangle~~,~~
    | \mathcal{F} \rangle = | 0 \rangle \otimes | K \rangle,
\ee
so the matrix element from Eq.~\eqref{eq:matrix_element_g} satisfies
\begin{align}
    2 \pi \delta \!\left({\textstyle \sum E} \right)  ( i \mathcal{M}_{\rm NR} ) & = i g \int dt \int d^3 {\bm x} \, \langle 0 | \, \phi(x) \, | \bm{q} \rangle \, \langle K | \, \psi^\dagger(x) \, \psi(x) \, | J \rangle \, ,
    \label{eq:3_pt_matrix_element_1}
\end{align}
and $\phi$ is quantized according to the usual relativistic formalism~\cite{Peskin:1995ev,Schwartz}
\begin{align}
    |\bm{p} \rangle  = \sqrt{2 E_{\bm{p}}} \, a^\dagger_{\bm{p}} | 0 \rangle~~~,~~~
    \phi({\bm x}, t)  = \int \frac{d^3\bm{p}}{(2 \pi)^3} \frac{1}{\sqrt{2 E_{\bm{p}}}} e^{i \bm{p} \cdot {\bm x}} \left( a_{\bm{p}} \, e^{- i E_{\bm{p}} t} + a_{-\bm{p}}^\dagger \, e^{i E_{\bm{p}} t} \right) \, ,
    \label{eq:phi_definition}
\end{align}
where $a^\dagger_{\bm p}$ and $a_{\bm p}$ are respectively the creation and annihilation operators for $\phi$ states, and the normalization convention satisfies
\be
\langle \bm p | \bm p \rangle = 2 E_{\vec{p}} \vol,
\ee
which {\it differs} from the corresponding convention for the electron states in \Eq{eq:electron-state-normalization}.
Inserting \Eq{eq:phi_definition} into Eq.~\eqref{eq:3_pt_matrix_element_1} and expanding $\psi$ in a set of basis states from \Eq{eq:electron_field_quantization} transforms the time integral into an energy conserving delta function. As noted in Sec. \ref{sec:feynman_rules}, the delta function cancels out of the final expression for $\cal M_{\rm NR}$, so we obtain
\begin{align}
\label{eq:matrix_element_NR_Feynman}
    \mathcal{M}_{\rm NR} = g \int d^3 {\bm x} \, e^{i \bm{q} \cdot {\bm x}} \, \psi_K^\dagger({\bm x}) \psi_J({\bm x}) \, .
\end{align}
Thus the Feynman rules for calculating $\cal M_{\rm NR}$ for a Yukawa interaction are: at each vertex insert a factor of 
\begin{align}
   i g \int d^3 {\bm x} \, e^{i \bm{q} \cdot {\bm x}} \psi^\dagger_K({\bm x}) \psi_J({\bm x})  \, ~~,~~~
    \label{eq:example_feyn_rule}
\end{align}
and then multiply the final expression by $-i$, as noted in Sec.~\ref{subsec:nr_matrix_element_def}. This procedure recovers the desired NR matrix element in \Eq{eq:matrix_element_NR_Feynman}
and justifies the steps outlined in Sec. \ref{subsec:nr_matrix_element_def}.

Note that when $\psi_{J}, \psi_{K}$ are free-electron states with $J = \bm{p}$ and $ K = \bm{p}'$, \Eq{eq:example_feyn_rule} can be reduced using the integral representation of the delta function
\be
 (2 \pi)^3\delta^3(\bm{p} + \bm{q} - \bm{p}') \equiv \int d^3 \bm x \, e^{-i \cdot (\bm{p} + \bm{q} - \bm{p}') \cdot \bm x},
 \label{eq:vol_cancel}
\ee
which yields $i g \, (2 \pi)^3 \delta^3(\bm{p} + \bm{q} - \bm{p}')$. As expected from the usual relativistic calculation a momentum conserving delta function appears.

\vspace{1em}
\begin{center}
    \textit{ \normalsize Three-Point Scalar Vertex }
\end{center}
\vspace{1em}

We now consider a more general three-point interaction in the NR EFT Lagrangian using the notation introduced in Sec.~\ref{sec:NR_DM_electron_interaction}
\begin{align}
    \LNR_\text{int} = \sum_{\ell=1}^8 \left[ \CNR_{\phi, \ell} \phi + C_{\nabla \phi, \ell} \frac{(\nabla \phi)}{m_e} +  \cdots \right] \left[ \psi^\dagger \, \ONRBasis_\ell \, \psi \right] \, ,
    \label{eq:general_3_pt_L_int}
\end{align}
where the terms in the left square bracket contain all field content linear in $\phi$ (or its derivatives), and all dimensionless coefficients $C$ contain only a single $\phi$ (including $(\nabla \phi)$ or $(\nabla \nabla \phi)$, etc.) in their left subscript. Following the same steps as in the Yukawa example above, after removing the energy conserving delta function, the NR matrix element becomes
\begin{align}
    i \mathcal{M}_{\rm NR} & =  i \sum_{\ell=1}^8 \left[ \CNR_{\phi, \ell} + C_{\nabla \phi, \ell} \left( \frac{i \bm{q}}{m_e} \right) +  \cdots \right] \left[ \int d^3 \bm{x} \, e^{i \bm{q} \cdot \bm{x}} \, \psi^\dagger_K(\bm{x}) \, \ONRBasis_\ell \, \psi_J(\bm{x}) \right]  \, ,
    \label{eq:M_3_general}
\end{align}
where $\nabla^i = - \partial^i \rightarrow i \bm{q}^i$ in momentum space. This organizational scheme generalizes straightforwardly to an NR EFT interaction containing terms with more derivatives acting on $\phi$: replace each $\nabla$ with $i \bm{q}$.

Since three-point vertices appear frequently in standard calculations, for future convenience we define the dimensionless combination of NR Lagrangian coefficients $\CNR$ and momenta $\qVec$ to be 
\begin{align}
    \fFR_{\phi, \ell}(\bm{q}) & \equiv \CNR_{\phi, \ell} + C_{\nabla \phi, \ell} \left( \frac{i \bm{q}}{m_e} \right) + \, \cdots  \, ,
    \label{eq:3_pt_Feynman_coefficient}
\end{align}
which collect all $C$ terms with only one $\phi$ subscript (including derivatives acting on $\phi$) and all higher order terms represented as $(\cdots) $ are normalized with appropriate powers of $m_e$ to maintain consistent dimensionality.
Note that this coefficient $\fFR$ contains all the \textit{model-dependent} information about the DM-electron interaction and does not depend on 
the details of the target. Furthermore we define the basis integral in Eq.~\eqref{eq:M_3_general} as
\begin{align}
    \MFR_{JK, \ell}(\bm{q}) & \equiv \int d^3 \bm{x} \, e^{i \bm{q} \cdot \bm{x}} \, \psi^\dagger_K(\bm{x}) \, \ONRBasis_\ell \, \psi_J(\bm{x}) \, ,
    \label{eq:M_hat_definition}
\end{align}
which contains all the \textit{target-dependent} information about the DM-electron interaction. Each NR basis operator $\ONRBasis_\ell$ has a corresponding amplitude, $\MFR_{ { J K}, \ell}$, once the external electron states are specified. Thus, the general three-point matrix element is
\begin{align}
  i \mathcal{M}_{\rm NR} =  i \sum_{\ell = 1}^8 \fFR_{\phi, \ell}(\qVec) \, \MFR_{JK, \ell}(\qVec) \, ,
\end{align}
and in terms of $\fFR$ and $\MFR$, the general three-point scalar vertex Feynman rule becomes
\begin{itemize}
    \item \textbf{Three-Point Scalar Vertex}: At each three-point vertex between an incoming particle $\phi$ with momentum $\qVec$ and electronic states $J, K$ insert
    \vspace{1em}
        \begin{align*}
            \begin{gathered}
                \begin{fmffile}{3pt_scalar}
    \begin{fmfgraph*}(100,75)
        \fmfleft{i} \fmfright{f1,f2}
        \fmf{dashes,label=$\mArrow{0.4cm}{q}$,label.side=left}{i,v}
        \fmf{fermion}{v,f2}
        \fmf{fermion}{f1,v}
        \fmfv{decor.shape=circle,decor.filled=hatched,decor.size=0.125w}{v}
        \fmflabel{$\phi$}{i}
        \fmflabel{$J$}{f1}
        \fmflabel{$K$}{f2}
    \end{fmfgraph*}
\end{fmffile}
            \end{gathered} \quad\quad\quad \Longrightarrow \quad\quad\quad i \sum^8_{\ell = 1} \fFR_{\phi, \ell}(\qVec) \; \MFR_{JK, \ell}(\qVec) \, ,      
    \end{align*}
    \vspace{0.5em}
\end{itemize}

\noindent where the hashed circle represents the sum of all three-point interactions in the NR EFT and arrows represent the flow of momentum through the diagram. 

\vspace{1em}
\begin{center}
    \textit{ \normalsize Three-Point Vector Vertex }
\end{center}
\vspace{1em}

A vector particle may also appear on the external line of a three-point vertex. Consider the NR interaction Lagrangian
\begin{align}
    \LNR_\text{int} = \sum_{ \sumIndex } \left[ C_{ V^0, \sumIndex } \, V^0 + C_{(\nabla V^0), \sumIndex} \frac{(\nabla V^0)}{m_e} + C_{ \Vv, \sumIndex } \,  \Vv + C_{(\nabla \Vv), \sumIndex} \frac{(\nabla \Vv)}{m_e} +  \, \dots \right] \left[ \psi^\dagger \, \ONRBasis_\sumIndex \, \psi \right] \, ,
\end{align}
which arises in the NR limit from the interaction in Sec. \ref{sec:NR_DM_electron_interaction},
where $V^\mu = (V^0, \Vv)$ is a real DM particle quantized according the relativistic QFT prescription
\begin{align}
    | \vec{q}, \lambda \rangle = \sqrt{2 E_{\vec{p}}} \, a_{\vec{p},\lambda}^\dagger | 0 \rangle~~~,~~~V^\mu(x) = \sum_\lambda \int \frac{d^3 \vec{p} }{(2 \pi)^3} 
    \frac{\epsilon_\lambda^\mu}{ \sqrt{ 2 E_{\vec{p}} } } e^{i \vec{p} \cdot \vec{x}} \left( a_{\vec{p}, \lambda} e^{- i E_{\vec{p}} t} + a_{-\vec{p},\lambda}^\dagger e^{i E_{\vec{p}} t} \right) \, ,
    \label{eq:vector_field_definition}
\end{align}
where $\epsilon_\lambda^\mu$ is a polarization four-vector, the sum is over $\lambda$ polarization states, and the state normalization convention is 
\be
\langle \bm p, \lambda  |\bm p, \lambda \rangle = 2 E_p \vol,
\ee 
which, again, differs from the NR electron state normalization from \Eq{eq:electron-state-normalization}.
Following the same steps as in the general three-point scalar vertex calculation, and adding a polarization index $\lambda$ to the initial state, the NR matrix element is 
\begin{align}
    i \mathcal{M}_\text{NR} = i \, \sum_{\sumIndex} \left\{
    \epsilon_\lambda^0 \left( C_{ V^0, \sumIndex } + \left[ C_{(\nabla V^0), \sumIndex} \right]^{i} \left( \frac{i \bm{q}^i}{m_e} \right) + \, \dots \right) + 
    \epsilon_\lambda^i \left( \left[ C_{ \Vv, \sumIndex } \right]^i + \left[ C_{(\nabla \Vv), \sumIndex} \right]^{ji} \left( \frac{i \bm{q}^j}{m_e} \right) + \, \dots \right) \right\}\MFR_{JK,\ell} \, ,
    \label{eq:MNR_vector}
\end{align}
where we have kept the vector field spatial indices explicit. 

The key difference here relative to the scalar case concerns the Lorentz indices: in relativistic QFT, a matrix element with an external vector line has its Lorentz index contracted with a polarization vector via $\mathcal{M} \sim \epsilon_\mu \mathcal{M}^\mu$, so the Cartesian index is lowered relative to Eq.~\eqref{eq:MNR_vector}. For consistency with familiar relativistic QFT conventions, we define the Feynman rule coefficients of the spatial vector components with a minus sign
to yield
\begin{align}
\label{eq:f_vector}
    [ f_{V, \sumIndex} ]^0 & \equiv C_{V^0, \sumIndex} + C_{(\nabla V^0), \sumIndex}^{i} \left( \frac{i \bm{q}^i}{m_e} \right) + \cdots \, ~~~,~~~\left[ f_{V, \sumIndex} \right]^i \equiv - \left[ C_{\Vv, \sumIndex}^{i} + C_{(\nabla \Vv), \sumIndex}^{ji} \left( \frac{i \bm{q}^j}{m_e} \right) + \cdots \, \right] \, ,
\end{align}
which respectively collect all $C$ coefficients with one $V^0$ and one $\bm V$ (plus their derivatives) in their subscripts,  which 
can be explicitly related to the corresponding expressions in 
Table \ref{tab:vector_coefficients}. The expressions in \Eq{eq:f_vector} may also be combined into a four-vector Feynman rule coefficient
\be
\left[ f_{V, \sumIndex} \right]^\mu \equiv \left( \, [f_{V, \ell}]^0 \, , \, \left[ f_{V, \ell} \right]^i \, \right)~,
\ee
so the three-point NR matrix element with an external vector becomes
\begin{align}
    \mathcal{M}_\text{NR} = \epsilon^\lambda_\mu \, \mathcal{M}_\text{NR}^\mu~~~,~~~i \mathcal{M}_\text{NR}^\mu = i \sum_{\ell = 1}^{8} \left[f_{V, \sumIndex} \right]^\mu \MFR_{JK, \sumIndex} \, ,
\end{align}
which is the same Feynman rule as in the previous section, but with an added Lorentz index to the Feynman coefficients, which matches that of the incoming vector DM particle
\begin{itemize}
    \item \textbf{Three-Point Vector Vertex}: At each three-point vertex between an incoming vector particle $V$ with momentum $\qVec$ and electronic states $J, K$ insert
    \vspace{1em}
    \begin{align*}
        \begin{gathered}
             \begin{fmffile}{3pt_vector}
    \begin{fmfgraph*}(100,75)
        \fmfleft{i} \fmfright{f1,f2}
        \fmf{photon,label=$\mArrow{0.4cm}{q}$,label.side=left}{i,v}
        \fmf{fermion}{v,f2}
        \fmf{fermion}{f1,v}
        \fmfv{decor.shape=circle,decor.filled=hatched,decor.size=0.125w}{v}
        \fmflabel{$V^\mu$}{i}
        \fmflabel{$J$}{f1}
        \fmflabel{$K$}{f2}
    \end{fmfgraph*}
\end{fmffile}
        \end{gathered} \quad\quad\quad \Longrightarrow \quad\quad\quad i \sum^8_{\ell = 1} \left[ \fFR_{V, \ell}(\qVec) \right]^{\mu} \MFR_{JK, \ell}(\qVec)  
    \end{align*}
    \vspace{0.5em}
\end{itemize}

\subsection{Four-Point Vertices}
\label{subsec:4_pt_vertex}

In general there are many possible four-point interactions (involving two fields in $\ONRGeneral$) that can arise in the
NR EFT developed in Sec.~\ref{sec:NR_DM_electron_interaction}. However, given the interactions in our UV example Lagrangian presented in Sec.~\ref{subsec:example_high_to_low_match},
all dark fields are linear in $\mathcal{O}_\text{UV}$ in the UV, so there are only two four-point interactions that derive in the NR from this UV example to order $m_e^{-2}$: $\psi^\dagger \hat{\mathcal{O}} \psi$ coupled to a photon plus one dark scalar or one dark vector.

\vspace{1em}
\begin{center}
    \textit{ \normalsize Four-Point Scalar-Vector Vertex }
\end{center}
\vspace{1em}

As in Sec.~\ref{subsec:3_pt_vertex}, we consider a general four-point interaction in the NR EFT Lagrangian, where we collect all operators that contain one $\phi$ and one photon $A$ (including both $A^0$ and $\Av$) plus derivatives acting on these fields. Using the notation of Sec.~\ref{sec:NR_DM_electron_interaction}, we can write these terms as 
\begin{align}
    \LNR_\text{int} = \sum_{\sumIndex} & \left[ \CNR_{\phi \Av, \ell} \frac{\phi \Av}{m_e} + C_{(\nabla \phi) \Av, \ell} \frac{(\nabla \phi) \Av}{m_e^2} + C_{\phi (\nabla \Av), \ell} \frac{\phi (\nabla \Av)}{m_e^2} \right.\nonumber \\
    & \left. \quad + \; \CNR_{\phi A^0, \ell} \frac{\phi A^0}{m_e} + C_{(\nabla \phi) A^0, \ell} \frac{(\nabla \phi) A^0}{m_e^2} + C_{\phi (\nabla A^0), \ell} \frac{\phi (\nabla A^0)}{m_e^2} + \cdots \right] \left[ \psi^\dagger \ONRBasis_\ell \psi \right] \, ,
    \label{eq:general_4_pt_L_int}
\end{align}
which arise in the NR limit from the interaction in Sec.~\ref{sec:NR_DM_electron_interaction}. We now repeat the steps from Sec.~\ref{subsec:3_pt_vertex}, taking the incoming $\phi$ particle to have momentum $\vec{p}_1$, and outgoing photon (quantized according to Eq.~\eqref{eq:vector_field_definition}) to have momentum $\vec{p}_2$ and polarization $\lambda$. Keeping the field indices explicit, the four-point NR amplitude is
\begin{align}
    i \mathcal{M}_{\rm NR} = i  \sum_{\sumIndex} & \left\{ \epsilon_\lambda^j \left( \left[ \CNR_{\phi \Av, \ell} \right]^j \frac{1}{m_e} + \left[ C_{(\nabla \phi) \Av, \ell} \right]^{ij} \frac{(i \bm{p}_1^i)}{m_e^2} + \left[ C_{\phi (\nabla \Av), \ell} \right]^{ij} \frac{(-i\bm{p}_2^i)}{m_e^2} + \, \cdots \right) \right. \nonumber \\ 
    & \left. \quad + \; \epsilon_\lambda^0 \left( \CNR_{\phi A^0, \ell} \frac{1}{m_e} + \left[( C_{(\nabla \phi) A^0, \ell} \right]^{i} \frac{(i \bm{p}_1^i)}{m_e^2} + \left[ C_{\phi (\nabla A^0), \ell} \right]^{i} \frac{(-i\bm{p}_2^i)}{m_e^2} + \, \cdots \right) \right\} \MFR_{JK, \ell}(\bm{p}_1 - \bm{p}_2) \, ,
    \label{eq:general_4_pt_expanded}
\end{align}
where the terms with $\vec{p}_1$ and $\vec{p}_2$ have different signs because $A$ is an outgoing particle. 

As in the previous derivation for the three-point vertex, each term in the general four-point vertex from Eq.~\eqref{eq:general_4_pt_expanded} factorizes into a model-dependent function (the term inside the brackets) and a target-dependent expression, $\MFR_{JK, \ell}$. 
Thus, following the notation convention from Sec.~\ref{subsec:3_pt_vertex}, we group the model-dependent pieces into an overall coefficient
\begin{align}
    \left[ \fFR_{\phi A, \ell}(\bm{p}_1, \bm{p}_2) \right]^0 & \equiv~  \CNR_{\phi A^0, \ell} \frac{1}{m_e} + C_{(\nabla \phi) A^0, \ell} \frac{(i \bm{p}_1)}{m_e^2} + C_{\phi (\nabla A^0), \ell} \frac{(-i\bm{p}_2)}{m_e^2} + \, \cdots  \nonumber \\
    \left[ \fFR_{\phi A, \ell}(\bm{p}_1, \bm{p}_2) \right]^i & \equiv -\left[ [\CNR_{\phi \Av, \ell}]^i \frac{1}{m_e} + [C_{(\nabla \phi) \Av, \ell}]^{ji} \frac{(i \bm{p}_1^j)}{m_e^2} + [C_{\phi (\nabla \Av), \ell}]^{ji} \frac{(-i\bm{p}_2^j)}{m_e^2} + \, \cdots \right] \, ,
\end{align}
and the four-point NR matrix element can be written 
\begin{align}
    \mathcal{M}_{\rm NR} = \epsilon^\lambda_\mu \, \mathcal{M}_\text{NR}^\mu~~~,~~~ i \mathcal{M}_\text{NR}^\mu = i \sum_{\sumIndex} \left[ \fFR_{\phi A, \ell}(\bm{p}_1, \bm{p}_2) \right]^\mu \, \MFR_{JK, \ell}(\bm{p}_1 - \bm{p}_2) \, ,
    \label{eq:4pt_feynman_rule}
\end{align}
from which we read off the Feynman rule:
\begin{itemize}
    \item \textbf{Four-Point Scalar-Vector Vertex}: At each four-point vertex between an incoming scalar particle $\phi$ with momentum $\bm{p}_1$, an outgoing vector particle $A$ with momentum $\bm{p}_2$, and electronic states $J, K$, insert
    \vspace{2em}
    \begin{align*}
        \begin{gathered}
            \begin{fmffile}{4pt_scalar_vector}
    \begin{fmfgraph*}(100,75)
        \fmfleft{i1,i2} \fmfright{f1,f2}
        \fmf{dashes,label=$\rotatebox{-45}{ $\xrightarrow{\makebox[0.4cm]{}}$ }$,label.dist=-0.08w}{i2,v}
        \fmf{fermion}{i1,v}
        \fmf{fermion}{v,f1}
        \fmf{photon,label=$\rotatebox{45}{ $\xrightarrow{\makebox[0.4cm]{}}$ }$,label.dist=-0.08w,label.side=right}{v,f2}
        \fmflabel{$\phi$}{i2}
        \fmflabel{$A^\mu$}{f2}
        \fmflabel{$J$}{i1}
        \fmflabel{$K$}{f1}
        \fmfv{decor.shape=circle,decor.filled=hatched,decor.size=0.125w}{v}
        \fmffreeze
        \fmfforce{0.5*vloc(__v)+0.5*vloc(__i2) shifted (-0.75mm,-4mm)}{l}
        \fmflabel{$\vec{p}_1$}{l}
        \fmfforce{0.5*vloc(__v)+0.5*vloc(__f2) shifted (+1mm,-4mm)}{l2}
        \fmflabel{$\vec{p}_2$}{l2}
    \end{fmfgraph*}
\end{fmffile}
        \end{gathered} \quad\quad\quad \Longrightarrow \quad\quad\quad i \sum_{\ell=1}^8 \left[ f_{\phi A, \ell}(\bm{p}_1, \bm{p}_2) \right]^{\mu} \, \MFR_{JK, \ell}(\bm{p}_1 - \bm{p}_2) \, 
    \end{align*}
    \vspace{0.5em}
\end{itemize}

\vspace{1em}
\begin{center}
    \textit{ \normalsize Four-Point Vector-Vector Vertex }
\end{center}
\vspace{1em}

The final four-point vertex we consider is between an incoming vector, $V^\mu = (V^0, \Vv)$, and outgoing photon, $A^\mu = (A^0, \Av)$. The general interaction Lagrangian which generates a four-point interaction between these fields can be written abstractly using the notation from Sec.~\ref{sec:NR_DM_electron_interaction} as
\begin{align}
    \LNR_\text{int} = \sum_{\ell} & \left[ \CNR_{V^0 A^0, \ell} \frac{V^0 A^0}{m_e} + C_{(\nabla V^0) A^0, \ell} \frac{(\nabla V^0) A^0}{m_e^2} + C_{V^0 (\nabla A^0), \ell} \frac{V^0 (\nabla A^0)}{m_e^2} \right.\nonumber \\
    & \qquad + \; \CNR_{\Vv A^0, \ell} \frac{\Vv A^0}{m_e} + C_{(\nabla \Vv) A^0, \ell} \frac{(\nabla \Vv) A^0}{m_e^2} + C_{\Vv (\nabla A^0), \ell} \frac{\Vv (\nabla A^0)}{m_e^2} \nonumber\\ 
    & \qquad \qquad  + \; \CNR_{V^0 \Av, \ell} \frac{V^0 \Av}{m_e} + C_{(\nabla V^0) \Av, \ell} \frac{(\nabla V^0) \Av}{m_e^2} + C_{V^0 (\nabla \Av), \ell} \frac{V^0 (\nabla \Av)}{m_e^2} \nonumber\\ 
    & \qquad \qquad \qquad \left. + \; \CNR_{\Vv \Av, \ell} \frac{\Vv \Av}{m_e} + C_{(\nabla \Vv) \Av, \ell} \frac{(\nabla \Vv) \Av}{m_e^2} + C_{\Vv (\nabla \Av), \ell} \frac{\Vv (\nabla \Av)}{m_e^2} + \;\cdots \right] \left[ \psi^\dagger \ONRBasis_\ell \psi \right] ~.
\end{align}
We now repeat the steps from the four-point scalar-photon matrix element calculation. The incoming $V$ particle has momentum $\vec{p}_1$ and polarization $\lambda$, and the outgoing photon has momentum $\vec{p}_2$ and polarization $\lambda'$. The four-point NR amplitude is then
\be
    i \mathcal{M}_\text{NR} &=& i \sum_{\ell}  \biggl[ \epsilon_{V, \lambda}^0 \epsilon_{A, \lambda'}^0 \left( \CNR_{V^0 A^0, \ell} \frac{1}{m_e} + \left[ C_{(\nabla V^0) A^0, \ell} \right]^i \frac{(i \vec{p}_1^i)}{m_e^2} + \left[ C_{V^0 (\nabla A^0), \ell} \right]^i \frac{ (-i\vec{p}_2^i)}{m_e^2} \right) \nonumber \\
    && ~~~~ + \epsilon_{V, \lambda}^i \epsilon_{A, \lambda'}^0 \left( \left[ \CNR_{\Vv A^0, \ell} \right]^{i} \frac{1}{m_e} + \left[ C_{(\nabla \Vv) A^0, \ell} \right]^{ji} \frac{(i \vec{p}_1^j)}{m_e^2} + \left[ C_{\Vv (\nabla A^0), \ell} \right]^{ij} \frac{(-i\vec{p}_2^j)}{m_e^2} \right) \nonumber\\ 
    &&  ~~~~+ \; \epsilon_{V, \lambda}^0 \epsilon_{A, \lambda'}^i \left( \left[ \CNR_{V^0 \Av, \ell} \right]^i \frac{1}{m_e} + \left[ C_{(\nabla V^0) \Av, \ell} \right]^{ji} \frac{(i \vec{p}_1^j)}{m_e^2} + \left[ C_{V^0 (\nabla \Av), \ell} \right]^{ji} \frac{(-i\vec{p}_2^j)}{m_e^2} \right) \nonumber\\ 
    &&  ~~~~+ \; \epsilon_{V, \lambda}^i \epsilon_{A, \lambda'}^j \left( \left[ \CNR_{\Vv \Av, \ell} \right]^{ij} \frac{1}{m_e} + \left[ C_{(\nabla \Vv) \Av, \ell} \right]^{kij} \frac{(i \vec{p}_1^k)}{m_e^2} + \left[ C_{\Vv (\nabla \Av), \ell} \right]^{ikj} \frac{(-i \vec{p}_2^k)}{m_e^2} \right) + \;\cdots \biggr] \MFR_{JK, \ell}(\vec{p}_1 - \vec{p}_2),~~~~~~
\ee
where $\epsilon_{A, \lambda}^\mu, \epsilon_{V, \lambda'}^\mu$ are the photon and $V$ polarization vectors, respectively.

As in the previous sections this matrix element is greatly simplified by defining the Feynman rule coefficients $[f_{VA}]^{\mu \nu}$ for 
which
\begin{align}
    [f_{V A, \ell}]^{00} & =  \,~~\left(  [\CNR_{V^0 A^0, \ell} ]\frac{1}{m_e} + \left[ C_{(\nabla V^0) A^0, \ell} \right]^i \frac{(i \vec{p}_1^i)}{m_e^2} + \left[ C_{V^0 (\nabla A^0), \ell} \right]^i \frac{(-i\vec{p}_2^i)}{m_e^2}\right) \\ 
    [f_{VA, \ell}]^{i0} & = -\left( \left[ \CNR_{\Vv A^0, \ell} \right]^{i} \frac{1}{m_e} + \left[ C_{(\nabla \Vv) A^0, \ell} \right]^{ji} \frac{(i \vec{p}_1^j)}{m_e^2} + \left[ C_{\Vv (\nabla A^0), \ell} \right]^{ij} \frac{(-i\vec{p}_2^j)}{m_e^2} \right)\\ 
    [f_{VA, \ell}]^{0 i} & = -\left( \left[ \CNR_{V^0 \Av, \ell} \right]^i \frac{1}{m_e} + \left[ C_{(\nabla V^0) \Av, \ell} \right]^{ji} \frac{(i \vec{p}_1^j)}{m_e^2} + \left[ C_{V^0 (\nabla \Av), \ell} \right]^{ji} \frac{(-i\vec{p}_2^j)}{m_e^2} \right)\\ 
    [f_{VA, \ell}]^{ij} & = \,~~\left( \left[ \CNR_{\Vv \Av, \ell} \right]^{ij} \frac{1}{m_e} + \left[ C_{(\nabla \Vv) \Av, \ell} \right]^{kij} \frac{(i \vec{p}_1^k)}{m_e^2} + \left[ C_{\Vv (\nabla \Av), \ell} \right]^{ikj} \frac{(-i \vec{p}_2^k)}{m_e^2}  \right)
\end{align}
With this notation the vector-vector four-point matrix element can be written succinctly as, 
\begin{align}
    \mathcal{M}_{\rm NR} = \epsilon^\lambda_{V, \mu} \epsilon^{\lambda'}_{A, \nu} \, \mathcal{M}_\text{NR}^{\mu \nu}~~~,~~~ i \mathcal{M}_\text{NR}^{\mu \nu} = i \sum_{\sumIndex} \left[ \fFR_{V A, \ell}(\bm{p}_1, \bm{p}_2) \right]^{\mu \nu} \, \MFR_{JK, \ell}(\bm{p}_1 - \bm{p}_2) \, ,
    \label{eq:4pt_feynman_rule_2}
\end{align}
from which we read off the Feynman rule:
\begin{itemize}
    \item \textbf{Four-Point Vector-Vector Vertex}: At each four-point vertex between an incoming vector particle $V$ with momentum $\bm{p}_1$, an outgoing photon $A$ with momentum $\bm{p}_2$, and electronic states $I, J$, insert
    \vspace{2em}
    \begin{align*}
        \begin{gathered}
            \begin{fmffile}{4pt_vector_vector}
    \begin{fmfgraph*}(100,75)
        \fmfleft{i1,i2} \fmfright{f1,f2}
        \fmf{photon,label=$\rotatebox{-45}{ $\xrightarrow{\makebox[0.4cm]{}}$ }$,label.dist=-0.08w}{i2,v}
        \fmf{fermion}{i1,v}
        \fmf{fermion}{v,f1}
        \fmf{photon,label=$\rotatebox{45}{ $\xrightarrow{\makebox[0.4cm]{}}$ }$,label.dist=-0.08w,label.side=right}{v,f2}
        \fmflabel{$V^\mu$}{i2}
        \fmflabel{$A^\nu$}{f2}
        \fmflabel{$J$}{i1}
        \fmflabel{$K$}{f1}
        \fmfv{decor.shape=circle,decor.filled=hatched,decor.size=0.125w}{v}
        \fmffreeze
        \fmfforce{0.5*vloc(__v)+0.5*vloc(__i2) shifted (-0.75mm,-4mm)}{l}
        \fmflabel{$\vec{p}_1$}{l}
        \fmfforce{0.5*vloc(__v)+0.5*vloc(__f2) shifted (+1mm,-4mm)}{l2}
        \fmflabel{$\vec{p}_2$}{l2}
    \end{fmfgraph*}
\end{fmffile}
        \end{gathered} \quad\quad\quad \Longrightarrow \quad\quad\quad i \sum_{\ell=1}^8 \left[ f_{V A, \ell}(\bm{p}_1, \bm{p}_2) \right]^{\mu \nu} \, \MFR_{JK, \ell}(\bm{p}_1 - \bm{p}_2) \, 
    \end{align*}
    \vspace{0.5em}
\end{itemize}

\subsection{NR Electron Propagator}
\label{subsec:nr_electron_propagator}

Since higher order diagrams involve internal, off-shell electron lines, we need to determine the Dirac propagator in the NR limit. Since the  relativistic propagator can be written 
\begin{align}
    D_\text{QFT}(p) = \int d^4 x \, e^{i p \cdot x} \, \langle 0 | T\left\{ \Psi(x) \bar{\Psi}(0) \right\} | 0 \rangle = \frac{i (\slashed{p} + m_e)}{p^2 - m_e^2 + i \epsilon} \, ,
    \label{eq:electron_propagator_QFT}
\end{align}
naively it would seem that the corresponding quantity in the NR EFT would be the Fourier transform of $\langle T\{ \psi({\bm x}, t) \psi^\dagger({\bm y}, t') \} \rangle$. However, defining the NR propagator in this way would {\it double count}  the electronic wavefunctions $\psi_{J}(\bm x)$, which are already included at the vertices of a given diagram inside $\MFR_{JK, \ell}$ (see, for example, \Eq{eq:example_feyn_rule}). 

Therefore, instead of taking the NR limit of $D_\text{QFT}$  in \Eq{eq:electron_propagator_QFT}, we first define the {\it time-dependent} creation/annihilation operators $b_J^\dagger(t)$ and $b_J(t)$ by absorbing the time-dependent phase factors in \Eq{eq:electron_field_quantization}. The electron field is then
\begin{align}
    \psi({\bm x}, t) & = \sum_{J} \psi_{J}({\bm x}) \, b_J(t)~~~,~~~ b_J(t) \equiv e^{- i E_J t} \, b_J \, , 
\end{align}
and we can use $b_J(t)$ to define the NR propagator in both time and energy
\begin{align}
    D_{JK}(t) \equiv  \langle \, T\{ \, b_J(t) \, b_K^\dagger(0) \, \} \,  \rangle ~~,~~
    D_{JK}(\omega) \equiv \int dt \, e^{i \omega t} \, D_{JK}(t) ~,~\label{eq:NR_electron_propagator_definition}
\end{align}
which connects the time evolution between the $J^\text{th}$ and $K^\text{th}$ eigenstates without the spatial wavefunctions. The propagator in Eq.~\eqref{eq:NR_electron_propagator_definition} appears frequently in condensed matter field theory and should be regarded as a constituent part of the full propagator, $\langle  T\{ \psi({\bm x}, t) \psi^\dagger({\bm y}, t') \} \rangle$, which can be derived in terms of $D_{JK}$ -- for a full discussion of this relationship, see Ch. 2 in Ref.~\cite{Mahan}.

Another subtle difference between $D_{\rm QFT}$ and $D_{JK}$ is that the ground state of the former is the zero-particle vacuum, whereas the latter ground 
state contains electrons at finite density. 
Thus, the expectation value in \Eq{eq:NR_electron_propagator_definition} must be evaluated with respect to the ground state $|\Omega\rangle$, defined according to 
\be
| \Omega \rangle \equiv \prod_{L} b_L^\dagger | 0 \rangle~,
\label{eq:vacuum-nonzero}
\ee
where $|0\rangle$ is the zero-particle vacuum and the product of $b_L^\dagger$ operators defines the target ground state.\footnote{This distinction is not important for the three and four-point vertex Feynman rules discussed in Secs.~\ref{subsec:3_pt_vertex}, and~~\ref{subsec:4_pt_vertex}, respectively. One can show that the transition probability for $| J \rangle \rightarrow | K \rangle$ is the same as taking the initial state be the in-medium vacuum, $| \Omega \rangle$, and the final state to be the in-medium vacuum with $J$ removed, and $K$ added, i.e., $| \Omega \rangle \rightarrow b_K^\dagger b_J | \Omega \rangle$.} We define the Fermi surface to be the zero energy point, $E = 0$, and therefore
all electronic states in the product in Eq.~\eqref{eq:vacuum-nonzero} have negative energy. Furthermore, the existence of this ground state electron population in \Eq{eq:vacuum-nonzero} affects how creation and annihilation operators act on the ground state
\begin{align}
    b_J^\dagger b_J | \Omega \rangle = \theta(- E_J) |\Omega \rangle\, ,
\end{align}
where $\theta$ is the Heaviside theta function. Note that $b_J$ only annihilates the vacuum if $E_J$ is positive. 

Incorporating these NR subtleties, we can directly evaluate the integral in Eq.~\eqref{eq:NR_electron_propagator_definition} to obtain 
\be
    D_{JK}(\omega)  &=& \int dt \, e^{i (\omega - E_J) t} \, \left[ \theta(t) \,  \langle  \Omega | \, b_J b_K^\dagger \, | \Omega \rangle - \theta(-t) \, \langle \Omega | \, b_K^\dagger b_J \, | \Omega \rangle \right] 
\nonumber \\ 
     &=&\delta_{JK} \int dt \, e^{i ( \omega - E_J ) t } \left[ \theta(t) \theta(E_J) - \theta(-t) \theta(-E_J) \right] ,
     \label{eq:NR_electron_propagator1}
\ee
and we introduce an $i\epsilon$ prescription to render the integrals finite
\be
   D_{JK}(\omega)   = i \delta_{JK} \left[ \frac{\theta(E_J)}{\omega - E_J + i \epsilon} + \frac{\theta(-E_J)}{\omega - E_J - i \epsilon} \right] 
     = \frac{i \delta_{JK}}{\omega - E_J + i \, \epsilon_J}~~~,~~~\epsilon_J \equiv \epsilon \, \text{sign}(E_J) \, ,
    \label{eq:NR_electron_propagator2}
\ee
so the Feynman rule for propagating internal electron lines can be written 
\begin{itemize}
    \item \textbf{NR Electron Propagator}: For every internal electron line labelled by $J$, insert the propagator from \Eq{eq:NR_electron_propagator1}  
    \begin{align*}
        \begin{gathered}
            \vspace{-0.6em}
                \begin{fmffile}{e_prop}
    \begin{fmfgraph*}(100,75)
        \fmfleft{i} \fmfright{f}
        \fmf{fermion,label=$J$,label.dist=-0.15w}{i,f}
    \end{fmfgraph*}
\end{fmffile}
        \end{gathered} \quad\quad\quad \Longrightarrow \quad\quad\quad \frac{i}{\omega - E_{J} + i \epsilon_J}~~~,~~~\epsilon_J = \epsilon \, \text{sign}\left( E_J \right) \, .
    \end{align*}
\end{itemize}

\subsection{Loop Diagrams}
\label{subsec:loop_diagram}

With the vertex functions and electron propagators defined above, we now develop the formalism for calculating loop diagrams. Following the conventions of earlier subsections, we begin with a ``warmup" example involving only one NR interaction and then generalize this to an arbitrary set of NR operators.

\vspace{1em}
\begin{center}
    \textit{ \normalsize Warmup }
\end{center}
\vspace{1em}

We now revisit the simple NR interaction Lagrangian from \Eq{eq:example_l_int_3_point} with the Yukawa coupling
\begin{align}
    \LNR_\text{int} = g \, \phi \, \psi^\dagger \psi \, ,
\end{align}
and consider an initial state with an incoming $\phi$ of momentum $\bm{p}$, where $| \mathcal{I} \rangle = | \bm{p} \rangle$, and a final state with outgoing $\phi$ of $\bm{p}'$, where $| \mathcal{F} \rangle = | \bm{p}' \rangle$. The simplest self-energy diagram for this interaction is
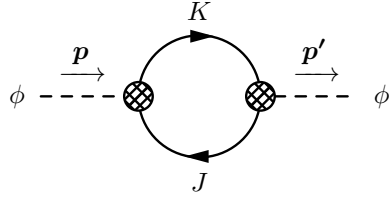
\begin{align*}
    \begin{fmffile}{loop_warmup}
    \begin{fmfgraph*}(120,75)
        \fmfleft{i} \fmfright{f}
        \fmf{dashes,label=$\mArrow{0.4cm}{p}$,label.side=left}{i,m1}
        \fmf{plain,left,tension=.4,label=$K$}{m1,m2}
        \fmf{plain,left,tension=.4,label=$J$}{m2,m1}
        \fmf{dashes,label=$\mArrow{0.4cm}{p'}$,label.side=left}{m2,f}
        \fmfv{decor.shape=circle,decor.filled=hatched,decor.size=0.09w}{m1}
        \fmfv{decor.shape=circle,decor.filled=hatched,decor.size=0.09w}{m2}
        \fmflabel{$\phi$}{i}
        \fmflabel{$\phi$}{f}
        \fmffreeze
        \fmfset{arrow_len}{4mm}
        \fmfforce{0.5*vloc(__m1)+0.5*vloc(__m2) shifted (-2mm,8mm)}{al}
        \fmfforce{0.5*vloc(__m1)+0.5*vloc(__m2) shifted (2mm,8mm)}{ar}
        \fmf{phantom_arrow}{al,ar}
        \fmfforce{0.5*vloc(__m1)+0.5*vloc(__m2) shifted (-2mm,-8mm)}{al1}
        \fmfforce{0.5*vloc(__m1)+0.5*vloc(__m2) shifted (2mm,-8mm)}{ar1}
        \fmf{phantom_arrow}{ar1,al1}
    \end{fmfgraph*}
\end{fmffile}
\end{align*}
\noindent which arises from  expanding $\mathcal{T}_\text{NR}$ to second order in the coupling $g$ and inserting the result into Eq.~\eqref{eq:matrix_element}. The corresponding matrix element satisfies 
\begin{align}
    2 \pi \delta\left( {\textstyle \sum} E \right) i \mathcal{M}_\text{NR} = \frac{(i g)^2}{2!} \int d^4 x d^4 x' \langle \bm{p}' | T\{ \phi(x) \psi^\dagger(x) \psi(x) \, \phi(x') \psi^\dagger(x') \psi(x') \} | \bm{p} \rangle \, ,
    \label{eq:loop_1}
\end{align}
so using Wick's theorem and inserting an overall factor of $-1$ for a closed electron loop (as in relativistic QFT~\cite{Peskin:1995ev,Schwartz}), the right hand side (RHS) of  Eq.~\eqref{eq:loop_1} becomes
\be
    \text{RHS} = g^2 \int d^3 {\bm x} \, d^3 {\bm x}' \, e^{i \bm{p} \cdot {\bm x}- i \bm{p}' {\bm x}'} \sum_{JKLM} \psi_{J}^\dagger({\bm x}) \psi_{K}({\bm x}) \psi^\dagger_{L}({\bm x}') \psi_{M}({\bm x}') \int dt dt' \, e^{i ( E_{\bm{p}} t - E_{\bm{p}'} t' ) } D_{KL}(t - t') D_{MJ}(t' - t) ,~~~
\ee 
where we have used the relativistic form of $\phi$ from \Eq{eq:phi_definition} and the definition of the propagator $D_{JK}(t)$ from \Eq{eq:NR_electron_propagator1}. Using the energy representation for $D_{JK}(\omega)$, this becomes
\begin{align}
    \text{RHS} = - g^2 \, 2 \pi \delta\left( { \textstyle \sum E} \right) \sum_{JK} \left[ i \MFR_{JK, 1}(\bm{p}) \right] \left[ i \MFR_{KJ, 1}(-\bm{p}') \right] \int \frac{dE}{2 \pi} \left( \frac{i}{E - E_J + i \epsilon_J} \right) \left( \frac{i}{\omega + E - E_K + i \epsilon_K} \right) \, ,\label{eq:T_loop_pre}
\end{align}
where we have used the integral representation of the Dirac delta function to trade the time integrals for an energy conserving delta function and an integration over the undetermined energy $E$.\footnote{Note that to maintain consistency with the overall EFT prescription the energy of the electronic states involved in Eq.~\eqref{eq:T_loop_pre} must be much less than the electron mass.}

Putting it all together,  we can now write the loop amplitude from Eq.~\eqref{eq:T_loop_pre} in terms of the explicit Feynman rules given in Sec. \ref{subsec:nr_matrix_element_def}: 
\be
    \mathcal{M}_{\rm NR} = \left( -i \right) (-1) \left( \sum_{JK} \int \frac{dE}{2\pi} \right) \left[ i g \MFR_{JK, 1}(\bm{p}) \right] \left[ i g \MFR_{KJ, 1}(-\bm{p}')  \right] \left( \frac{i}{E - E_J + i \epsilon_J} \right) \left( \frac{i}{\omega + E - E_K + i \epsilon_K} \right) ,~~~~~
    \label{eq:m_loop_decompose}
\ee
where the first factor comes from the rule to multiply by $-i$ (Sec.~\ref{subsec:nr_matrix_element_def}), the second factor of $(-1)$ is from a closed electron loop, the integral over $E$ is over the undetermined loop energy (Sec.~\ref{subsec:nr_matrix_element_def}), the fourth and fifth terms are the three-point Feynman rules (Sec.~\ref{subsec:3_pt_vertex}), and the last two factors are from the NR electron propagators (Sec.~\ref{subsec:nr_electron_propagator}). 

Note that the only new aspect of Eq.~\eqref{eq:m_loop_decompose}, which is not determined by the previously stated Feynman rules, is the sum over electron states. Therefore we explicitly add the additional rule: 
\begin{itemize}
    \item{\bf Loop Feynman Rule:} Sum over all internal electron lines labelled by $J,K,\dots$ by inserting $\displaystyle\sum_{~~JK \dots}$.
\end{itemize}
This prescription replaces the three-momentum integrals which are familiar from relativistic loop calculations. 
This correspondence is restored in the free-electron limit, in which the states are indexed by momentum $J \rightarrow \bm{p}$, and the sum over the electronic states becomes an integral over spatial momentum $\bm{p}$ with the replacement $\sum_J \rightarrow \int d^3 \bm{p} / (2 \pi)^3$, as discussed in Sec.~\ref{subsec:nr_electron_field_quantization}. Combining this three-dimensional integral with the integral over the undetermined energy $E$ in \Eq{eq:m_loop_decompose} yields the replacement
\begin{align}
    \sum_J \int \frac{dE}{2 \pi} \rightarrow \int \frac{d^4 p}{(2 \pi)^4} \, ,
\end{align}
thereby recovering the familiar four-dimensional phase space integral over an undetermined loop four-momentum from relativistic QFT. 

\vspace{1em}
\begin{center}
    \textit{ \normalsize Scalar Loop Diagram (Type I) }
\end{center}
\vspace{1em}

We now generalize the scalar field loop calculation to the NR EFT interaction Lagrangian given in \Eq{eq:general_3_pt_L_int}. The leading self-energy Feynman diagram arising from these interactions is
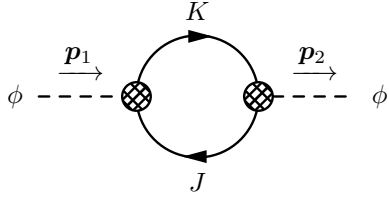
\begin{align*}
    \begin{fmffile}{loop_warmup_2}
    \begin{fmfgraph*}(120,75)
        \fmfleft{i} \fmfright{f}
        \fmf{dashes,label=$\mArrowS{0.4cm}{p}{1}$,label.side=left}{i,m1}
        \fmf{plain,left,tension=.4,label=$K$}{m1,m2}
        \fmf{plain,left,tension=.4,label=$J$}{m2,m1}
        \fmf{dashes,label=$\mArrowS{0.4cm}{p}{2}$,label.side=left}{m2,f}
        \fmfv{decor.shape=circle,decor.filled=hatched,decor.size=0.09w}{m1}
        \fmfv{decor.shape=circle,decor.filled=hatched,decor.size=0.09w}{m2}
        \fmflabel{$\phi$}{i}
        \fmflabel{$\phi$}{f}
        \fmffreeze
        \fmfset{arrow_len}{4mm}
        \fmfforce{0.5*vloc(__m1)+0.5*vloc(__m2) shifted (-2mm,8mm)}{al}
        \fmfforce{0.5*vloc(__m1)+0.5*vloc(__m2) shifted (2mm,8mm)}{ar}
        \fmf{phantom_arrow}{al,ar}
        \fmfforce{0.5*vloc(__m1)+0.5*vloc(__m2) shifted (-2mm,-8mm)}{al1}
        \fmfforce{0.5*vloc(__m1)+0.5*vloc(__m2) shifted (2mm,-8mm)}{ar1}
        \fmf{phantom_arrow}{ar1,al1}
    \end{fmfgraph*}
\end{fmffile}
\end{align*}

\noindent and the corresponding matrix element can be written
\be
    i \mathcal{M}_\text{NR} = -\sum_{JK} \int \frac{dE}{2 \pi} 
      \left[ i \sum_\ell \fFR_{\phi, \ell}(\bm{p}_1) \MFR_{JK, \ell}(\bm{p}_1) \right] \!\! \left[i \sum_{m} \fFR_{\phi, m}(-\bm{p}_2) \MFR_{KJ, m}(-\bm{p}_2) \right]
       \!\left[ \frac{i}{E - E_J + i \epsilon_J} \right]\! \left[ \frac{i}{\omega + E - E_K + i \epsilon_K} \right] \! , ~~~~~~~
    \label{eq:se_1}
\ee
which is analogous to \Eq{eq:m_loop_decompose}, but features sums over the NR basis elements indexed with $\ell, m$ to include additional operators beyond $\ONRBasis_1$ which defined the NR Yukawa coupling in the warmup example above. 
After performing the $E$ integral with the identity~\cite{Mitridate:2021ctr}
\begin{align}
    \int \frac{dE}{2 \pi} \left( \frac{1}{E - E_J + i \epsilon_J} \right) \left( \frac{1}{\omega + E - E_K + i \epsilon_K} \right) = i \left[ \frac{\theta(-E_J)\theta(E_K) - \theta(E_J)\theta(-E_K)}{\omega - E_K + E_J + i \epsilon \, \text{sign}(E_K - E_J)}
    \right]~,
\end{align}
Eq.~\eqref{eq:se_1} simplifies to become
\begin{align}
    i \mathcal{M}_\text{NR} = -i \sum_{\ell m} \fFR_{\phi,\ell}(\bm{p}_1) \fFR_{\phi,m}(-\bm{p}_2) \sum_{IF} \left[ \frac{ \MFR_{IF, \ell}(\bm{p}_1) \, \MFR_{FI, m}(-\bm{p}_2) }{\omega - E_F + E_I + i \epsilon} - \frac{ \MFR_{IF, m}(-\bm{p}_2) \, \MFR_{FI, \ell}(\bm{p}_1) }{\omega + E_F - E_I - i \epsilon} \right],
    \label{eq:se_2}
\end{align}
where $I$ and $F$ only sum over the filled ($E_I < 0$), and unfilled ($E_F > 0$) states, respectively.\footnote{Note that $I$ indexes the negative energy states of the \textit{entire} target, and $\sum_I$ sums over all filled states in the entire target. For example, consider a target of $N$ hydrogen atoms, labelled by $j$, each with an electron in the $n = 2, \ell = 1, m = 0$ state. Then $I = \{ j, n, \ell, m \}$, and $\sum_I = \sum_{j} \sum_{n \ell m} \delta_{n 2} \, \delta_{\ell 1} \, \delta_{m 0} = N \sum_{n \ell m} \delta_{n 2} \, \delta_{\ell 1} \, \delta_{m 0}$.} As with the three-point and four-point vertex Feynman rules derived in Secs.~\ref{subsec:3_pt_vertex},~\ref{subsec:4_pt_vertex}, respectively, the self-energy factorizes in to a model-dependent contribution (the $\fFR$'s) and a target-dependent contribution (the $\MFR$'s). To make this separation manifest we define the target dependent self-energy as
\begin{align}
    \hat{\Pi}_{\ell m}(\bm{p}_1, \bm{p}_2) \equiv -\sum_{IF} \left[ \frac{ \MFR_{IF, \ell}(\bm{p}_1) \, \MFR_{FI, m}(-\bm{p}_2) }{\omega - E_F + E_I + i \epsilon} - \frac{ \MFR_{IF, m}(-\bm{p}_2) \, \MFR_{FI, \ell}(\bm{p}_1) }{\omega + E_F - E_I - i \epsilon} \right] \, ,
    \label{eq:Pi_hat_definition}
\end{align}
so the Feynman rule and diagram for a loop becomes
\begin{itemize}
    \item \textbf{Scalar Loop Diagram (Type I)}: For each single electron loop diagram with $\phi$ external legs and incoming momentum $\bm{p}_1$ and outgoing momentum $\bm{p}_2$, insert
    \vspace{1em}
    \begin{align*}
        \begin{gathered}
            \vspace{-0.6em}
            \begin{fmffile}{loop_scalar}
    \begin{fmfgraph*}(120,75)
        \fmfleft{i} \fmfright{f}
        \fmf{dashes,label=$\mArrowS{0.4cm}{p}{1}$,label.side=left}{i,m1}
        \fmf{plain,left,tension=.4}{m1,m2}
        \fmf{plain,left,tension=.4}{m2,m1}
        \fmf{dashes,label=$\mArrowS{0.4cm}{p}{2}$,label.side=left}{m2,f}
        \fmfv{decor.shape=circle,decor.filled=hatched,decor.size=0.09w}{m1}
        \fmfv{decor.shape=circle,decor.filled=hatched,decor.size=0.09w}{m2}
        \fmflabel{$\phi$}{i}
        \fmflabel{$\phi$}{f}
        \fmfforce{0.5*vloc(__m1)+0.5*vloc(__m2) shifted (-2mm,8mm)}{al}
        \fmfforce{0.5*vloc(__m1)+0.5*vloc(__m2) shifted (2mm,8mm)}{ar}
        \fmf{phantom_arrow}{al,ar}
        \fmfforce{0.5*vloc(__m1)+0.5*vloc(__m2) shifted (-2mm,-8mm)}{al1}
        \fmfforce{0.5*vloc(__m1)+0.5*vloc(__m2) shifted (2mm,-8mm)}{ar1}
        \fmf{phantom_arrow}{ar1,al1}
    \end{fmfgraph*}
\end{fmffile}
        \end{gathered} \hspace{5em} \Longrightarrow \hspace{4em} i \sum_{\ell m} \fFR_{\phi, \ell}(\bm{p}_1) \; \hat{\Pi}_{\ell m}(\bm{p}_1,\bm{p}_2) \; \fFR_{\phi, m}(-\bm{p}_2)\, .
    \end{align*}
\end{itemize}

An important subtlety is that in general $\bm{p}_1$ does not have to equal $\bm{p}_2$, because momentum does not have to be conserved inside the target medium. However, most previous studies do not consider this ``off-diagonal" momentum non-conserving contribution to the self-energies, and make the approximation
~\cite{Mitridate:2021ctr,Knapen:2021run} 
\begin{align}
    \hat{\Pi}_{\ell m}(\bm{p}_1, \bm{p}_2) \approx \hat{\Pi}_{\ell m}(\bm{p}_1, \bm{p}_1) \, \delta_{\bm{p}_1\bm{p}_2} \, . 
    \label{eq:Pi_hat_approximation}
\end{align}
We will adopt this approximation when computing observables below. See Ref.~\cite{Knapen:2021run} for additional discussion.

\vspace{1em}
\begin{center}
    \textit{ \normalsize Scalar-Vector Loop Diagram (Type II) }
\end{center}
\vspace{1em}

The final loop diagram we compute is for a second loop topology, which we will refer to as a ``type II" loop diagram
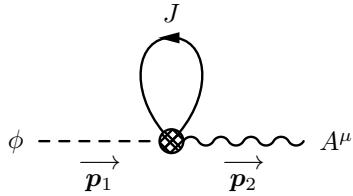
\vspace{2em}
\begin{align*}
    \begin{fmffile}{loop2_warmup}
    \begin{fmfgraph*}(100,75)
        \fmfleft{i} \fmfright{v}
        \fmf{dashes,label=$\overset{\ds \longrightarrow}{\ds \vec{p}_1}$}{i,m1}
        \fmf{plain,right,tension=.85,label=$J$}{m1,m1}
        \fmf{photon,label=$\overset{\ds \longrightarrow}{\ds \vec{p}_2}$}{m1,v}
        \fmflabel{$\phi$}{i}
        \fmflabel{$A^\mu$}{v}
        \fmfv{decor.shape=circle,decor.filled=hatched,decor.size=0.09w}{m1}
        \fmffreeze 
        \fmfset{arrow_len}{3.5mm}
        \fmfforce{vloc(__m1) shifted (-1.75mm,13.75mm)}{al}
        \fmfforce{vloc(__m1) shifted (1.75mm,13.75mm)}{ar}
        \fmf{phantom_arrow}{ar,al}
    \end{fmfgraph*}
\end{fmffile}
\end{align*}
\vspace{-3em}

\noindent Since the Feynman rule for the four-point vertex was already derived in Sec.~\ref{subsec:4_pt_vertex}, we apply the previously developed Feynman rules based on \Eq{eq:general_4_pt_L_int} to compute the matrix element
\begin{align}
    i \mathcal{M}_\text{NR}^\mu & = \left( -1 \right) \left( \sum_J \int \frac{dE}{2 \pi} \right) \left( i \sum_\ell \left[ \fFR_{\phi A, \ell}(\bm{p}_1, \bm{p}_2)\right]^\mu \MFR_{JJ, \ell}(\bm{p}_1 - \bm{p}_2) \right) \left( \frac{i}{E - E_J + i \epsilon_J} \right) \nonumber \\ 
    & = i \sum_\ell \left[ f_{\phi A, \ell}(\bm{p}_1, \bm{p}_2) \right]^\mu  \sum_J \theta(- E_J) \MFR_{JJ, \ell}(\bm{p}_1 - \bm{p}_2)  \, ,
\end{align}
where we have used the Cauchy integral formula to integrate over energy. As in previous sections, the self-energy matrix element cleanly separates into a model-dependent contribution (the $\fFR$'s) and a target-dependent contribution (the sum over $\MFR$'s). To make this factorization manifest we define
\begin{align}
    \hat{\Pi}_\ell(\vec{p}_1 - \vec{p}_2) \equiv \sum_J \theta(- E_J) \MFR_{JJ, \ell}(\bm{p}_1 - \bm{p}_2) = \sum_I \int d^3 {\bm x} \, e^{i \left( \vec{p}_1 - \vec{p}_2 \right) \cdot \vec{x}} \, \psi^\dagger_I({\bm x}) \, \ONRBasis_\ell \, \psi_I({\bm x}),
    \label{eq:Pi_hat_1}
\end{align}
where the sum over $I$ is over filled states in the target ($E_I < 0$).

This NR matrix element expression is most commonly evaluated in the limit where $\bm{p}_1 = \bm{p}_2$, where it simplifies to
\begin{align}
    i \mathcal{M}_\text{NR}^\mu(\bm{p}_1, \bm{p}_1) & \approx i \sum_\ell \left[ \fFR_{\phi A, \ell}(\bm{p}_1, \bm{p}_1) \right]^\mu \hat{\Pi}_{\ell}(0)~.
\end{align}
Note that $\hat{\Pi}_\ell(0)$ is just the expectation value of the operator $\ONRBasis_\ell$ over the target volume. For example, for $\ONRBasis_1 = 1$, $\hat{\Pi}_1(0) = \sum_I =  N_e$, where $N_e$ is the total number of electrons in the target. In terms of $\hat{\Pi}_\ell(\vec{p}_1 - \vec{p}_2)$ the general type-II loop Feynman rule is
\begin{itemize}
    \item \textbf{Scalar-Vector Loop Diagram (Type II)}: For every type II loop connecting an incoming scalar $\phi$ with momentum $\vec{p}_1$, and an outgoing vector $V$ with momentum $\bm{p}_2$ insert,
    \vspace{1em}
    \begin{align*}
        \begin{gathered}
            \vspace{-0.6em}
            \begin{fmffile}{loop2_scalar_vector}
    \begin{fmfgraph*}(100,75)
        \fmfleft{i} \fmfright{v}
        \fmf{dashes,label=$\overset{\ds \longrightarrow}{\ds \vec{p}_1}$}{i,m1}
        \fmf{plain,right,tension=.85}{m1,m1}
        \fmf{photon,label=$\overset{\ds \longrightarrow}{\ds \vec{p}_2}$}{m1,v}
        \fmflabel{$\phi$}{i}
        \fmflabel{$A^\mu$}{v}
    \fmfv{decor.shape=circle,decor.filled=hatched,decor.size=0.09w}{m1}
    \fmffreeze 
    \fmfset{arrow_len}{3.5mm}
    \fmfforce{vloc(__m1) shifted (-1.75mm,13.75mm)}{al}
    \fmfforce{vloc(__m1) shifted (1.75mm,13.75mm)}{ar}
    \fmf{phantom_arrow}{ar,al}
    \end{fmfgraph*}
\end{fmffile}
        \end{gathered} \hspace{5em} \Longrightarrow \hspace{4em} i \sum_{\ell = 1}^8 \left[ \fFR_{\phi A,\sumIndex}(\bm{p}_1,\bm{p}_2) \right]^{\mu} \; \hat{\Pi}_\ell(\vec{p}_1 - \vec{p}_2) \, .
    \end{align*}
\end{itemize}

\subsection{Summary}
\label{subsec:feynman_rules_summary}

We now collect all of the Feynman rules derived in Secs.~\ref{subsec:3_pt_vertex}~-~\ref{subsec:4_pt_vertex}, and reference the section in which each one is derived. Wherever possible, we factorize the Feynman rules in to a model-dependent contribution (denoted by expressions involving $\fFR$ vertex factors) which is directly a function of the NR EFT coefficients introduced in Sec.~\ref{sec:NR_DM_electron_interaction}, and target-dependent factors
\begin{align}
    \MFR_{IJ, \ell}(\qVec) & \equiv \int d^3 {\bm x} \, e^{i \bm{q} \cdot {\bm x}} \, \psi_J^\dagger({\bm x}) \, \ONRBasis_\ell \, \psi_I({\bm x}) \\ 
    \hat{\Pi}_{\ell m}(\bm{p}_1, \bm{p}_2) & \equiv -\sum_{IF}\left[ \frac{ \MFR_{IF, \ell}(\bm{p}_1) \, \MFR_{FI, m}(-\bm{p}_2) }{\omega - E_F + E_I + i \epsilon} - \frac{ \MFR_{IF, m}(-\bm{p}_2) \, \MFR_{FI, \ell}(\bm{p}_1) }{\omega + E_F - E_I - i \epsilon} \right] \\ 
    \hat{\Pi}_\ell(\vec{p}_1 - \vec{p}_2) & \equiv \sum_I \int d^3 {\bm x} \, e^{i ( \vec{p}_1 - \vec{p}_2 ) \cdot \vec{x}} \, \psi^\dagger_I({\bm x}) \, \ONRBasis_\ell \, \psi_I({\bm x}) \, ,
\end{align}
(typically denoted with a ``$\;\hat{\phantom{o}}\;$" symbol) which are directly a function of the target electronic structure. $I$ indexes filled states with $E_I < 0$, and $F$ indexes unfilled states with $E_F > 0$. This list of Feynman rules is not exhaustive; these examples are chosen as a representative set useful in computing the observables discussed in Secs.~\ref{sec:absorption},~\ref{sec:scattering}, and~\ref{sec:dark_thomson_scattering}.
\begin{itemize}
    \item \textbf{Three-point Scalar Vertex} (Sec.~\ref{subsec:3_pt_vertex}): At each three-point vertex between an incoming particle $\phi$ with momentum $\qVec$ and electronic states $J, K$ insert,
        \vspace{1em}
        \begin{align*}
            \begin{gathered}
                \begin{fmffile}{3pt_scalar}
    \begin{fmfgraph*}(100,75)
        \fmfleft{i} \fmfright{f1,f2}
        \fmf{dashes,label=$\mArrow{0.4cm}{q}$,label.side=left}{i,v}
        \fmf{fermion}{v,f2}
        \fmf{fermion}{f1,v}
        \fmfv{decor.shape=circle,decor.filled=hatched,decor.size=0.125w}{v}
        \fmflabel{$\phi$}{i}
        \fmflabel{$J$}{f1}
        \fmflabel{$K$}{f2}
    \end{fmfgraph*}
\end{fmffile}
            \end{gathered} \quad\quad\quad \Longrightarrow \quad\quad\quad i \sum^8_{\ell = 1} \fFR_{\phi, \ell}(\qVec) \; \MFR_{JK, \ell}(\qVec)      
        \end{align*}
        \vspace{1em}
    \item \textbf{Three-point Vector Vertex} (Sec.~\ref{subsec:3_pt_vertex}): At each three-point vertex between an incoming vector particle $V$ with momentum $\qVec$ and electronic states $J, K$ insert
        \vspace{1em}
        \begin{align*}
            \begin{gathered}
                \begin{fmffile}{3pt_vector}
    \begin{fmfgraph*}(100,75)
        \fmfleft{i} \fmfright{f1,f2}
        \fmf{photon,label=$\mArrow{0.4cm}{q}$,label.side=left}{i,v}
        \fmf{fermion}{v,f2}
        \fmf{fermion}{f1,v}
        \fmfv{decor.shape=circle,decor.filled=hatched,decor.size=0.125w}{v}
        \fmflabel{$V^\mu$}{i}
        \fmflabel{$J$}{f1}
        \fmflabel{$K$}{f2}
    \end{fmfgraph*}
\end{fmffile}
            \end{gathered} \quad\quad\quad \Longrightarrow \quad\quad\quad i \sum^8_{\ell = 1} \left[ \fFR_{V, \ell}(\qVec) \right]^{\mu} \, \MFR_{JK, \ell}(\qVec).   
        \end{align*}
        \vspace{0.5em}
    \item \textbf{Four-Point Scalar-Vector Vertex} (Sec.~\ref{subsec:4_pt_vertex}): At each four-point vertex between an incoming scalar particle $\phi$ with momentum $\bm{p}_1$, an outgoing vector particle $A$ with momentum $\bm{p}_2$, and electronic states $J, K$, insert
        \vspace{1.5em}
        \begin{align*}
            \begin{gathered}
                \begin{fmffile}{4pt_scalar_vector}
    \begin{fmfgraph*}(100,75)
        \fmfleft{i1,i2} \fmfright{f1,f2}
        \fmf{dashes,label=$\rotatebox{-45}{ $\xrightarrow{\makebox[0.4cm]{}}$ }$,label.dist=-0.08w}{i2,v}
        \fmf{fermion}{i1,v}
        \fmf{fermion}{v,f1}
        \fmf{photon,label=$\rotatebox{45}{ $\xrightarrow{\makebox[0.4cm]{}}$ }$,label.dist=-0.08w,label.side=right}{v,f2}
        \fmflabel{$\phi$}{i2}
        \fmflabel{$A^\mu$}{f2}
        \fmflabel{$J$}{i1}
        \fmflabel{$K$}{f1}
        \fmfv{decor.shape=circle,decor.filled=hatched,decor.size=0.125w}{v}
        \fmffreeze
        \fmfforce{0.5*vloc(__v)+0.5*vloc(__i2) shifted (-0.75mm,-4mm)}{l}
        \fmflabel{$\vec{p}_1$}{l}
        \fmfforce{0.5*vloc(__v)+0.5*vloc(__f2) shifted (+1mm,-4mm)}{l2}
        \fmflabel{$\vec{p}_2$}{l2}
    \end{fmfgraph*}
\end{fmffile}
            \end{gathered} \quad\quad\quad \Longrightarrow \quad\quad\quad i \sum_{\ell=1}^8 \left[ f_{\phi A, \ell}(\bm{p}_1, \bm{p}_2) \right]^{\mu} \MFR_{JK, \ell}(\bm{p}_1 - \bm{p}_2) \, ,
        \end{align*}
        \vspace{0.5em}
    \item \textbf{Four-Point Vector-Vector Vertex} (Sec.~\ref{subsec:4_pt_vertex}): At each four-point vertex between an incoming vector particle $V$ with momentum $\bm{p}_1$, an outgoing photon $A$ with momentum $\bm{p}_2$, and electronic states $J, K$, insert
        \vspace{1.5em}
        \begin{align*}
            \begin{gathered}
                \begin{fmffile}{4pt_vector_vector}
    \begin{fmfgraph*}(100,75)
        \fmfleft{i1,i2} \fmfright{f1,f2}
        \fmf{photon,label=$\rotatebox{-45}{ $\xrightarrow{\makebox[0.4cm]{}}$ }$,label.dist=-0.08w}{i2,v}
        \fmf{fermion}{i1,v}
        \fmf{fermion}{v,f1}
        \fmf{photon,label=$\rotatebox{45}{ $\xrightarrow{\makebox[0.4cm]{}}$ }$,label.dist=-0.08w,label.side=right}{v,f2}
        \fmflabel{$V^\mu$}{i2}
        \fmflabel{$A^\nu$}{f2}
        \fmflabel{$J$}{i1}
        \fmflabel{$K$}{f1}
        \fmfv{decor.shape=circle,decor.filled=hatched,decor.size=0.125w}{v}
        \fmffreeze
        \fmfforce{0.5*vloc(__v)+0.5*vloc(__i2) shifted (-0.75mm,-4mm)}{l}
        \fmflabel{$\vec{p}_1$}{l}
        \fmfforce{0.5*vloc(__v)+0.5*vloc(__f2) shifted (+1mm,-4mm)}{l2}
        \fmflabel{$\vec{p}_2$}{l2}
    \end{fmfgraph*}
\end{fmffile}
            \end{gathered} \quad\quad\quad \Longrightarrow \quad\quad\quad i \sum_{\ell=1}^8 \left[ f_{V A, \ell}(\bm{p}_1, \bm{p}_2) \right]^{\mu \nu} \MFR_{JK, \ell}(\bm{p}_1 - \bm{p}_2) \, 
        \end{align*}
        \vspace{0.5em}
    \item \textbf{NR Electron Propagator} (Sec.~\ref{subsec:nr_electron_propagator}): For every internal electron line labelled by $J$, insert the 
    propagator from \Eq{eq:NR_electron_propagator1} 
        \vspace{-1em}
            \begin{align*}
                \begin{gathered}
                    \vspace{-0.6em}
                    \begin{fmffile}{e_prop}
    \begin{fmfgraph*}(100,75)
        \fmfleft{i} \fmfright{f}
        \fmf{fermion,label=$J$,label.dist=-0.15w}{i,f}
    \end{fmfgraph*}
\end{fmffile}
                \end{gathered} \quad\quad\quad \Longrightarrow \quad\quad\quad \frac{i}{\omega - E_{J} + i \epsilon_J}~~~,~~~\epsilon_J = \epsilon \, \text{sign}\left( E_J \right) \, .
            \end{align*}
        \vspace{-1em}
    \item Conserve energy at each vertex (Sec.~\ref{subsec:nr_matrix_element_def}).
    \item For each undetermined energy $E$, insert factors of $\displaystyle \left( \int \frac{dE}{ 2\pi} \right)$ and perform the corresponding integral (Sec.~\ref{subsec:nr_matrix_element_def}).
    \item Sum over all internal electron lines labelled by $J,K,\dots$ by inserting $\displaystyle\sum_{~~JK \dots}$ (Sec.~\ref{subsec:loop_diagram}).
    \item Multiply the resulting expression by $-i$ (Sec.~\ref{subsec:nr_matrix_element_def}).
\end{itemize}

Using these rules one can compute more complicated diagrams. For usability we provide Feynman rules for a few of the composite loop diagrams.

\begin{itemize}
    \item \textbf{Scalar Loop Diagram (Type I)} (Sec.~\ref{subsec:loop_diagram}): For each single electron loop diagram between $\phi$ states  with incoming and outgoing momenta $\bm{p}_1$ and $\bm{p}_2$, respectively, insert
        \vspace{1em}
        \begin{align*}
            \begin{gathered}
                \vspace{-0.6em}
                \begin{fmffile}{loop_scalar}
    \begin{fmfgraph*}(120,75)
        \fmfleft{i} \fmfright{f}
        \fmf{dashes,label=$\mArrowS{0.4cm}{p}{1}$,label.side=left}{i,m1}
        \fmf{plain,left,tension=.4}{m1,m2}
        \fmf{plain,left,tension=.4}{m2,m1}
        \fmf{dashes,label=$\mArrowS{0.4cm}{p}{2}$,label.side=left}{m2,f}
        \fmfv{decor.shape=circle,decor.filled=hatched,decor.size=0.09w}{m1}
        \fmfv{decor.shape=circle,decor.filled=hatched,decor.size=0.09w}{m2}
        \fmflabel{$\phi$}{i}
        \fmflabel{$\phi$}{f}
        \fmfforce{0.5*vloc(__m1)+0.5*vloc(__m2) shifted (-2mm,8mm)}{al}
        \fmfforce{0.5*vloc(__m1)+0.5*vloc(__m2) shifted (2mm,8mm)}{ar}
        \fmf{phantom_arrow}{al,ar}
        \fmfforce{0.5*vloc(__m1)+0.5*vloc(__m2) shifted (-2mm,-8mm)}{al1}
        \fmfforce{0.5*vloc(__m1)+0.5*vloc(__m2) shifted (2mm,-8mm)}{ar1}
        \fmf{phantom_arrow}{ar1,al1}
    \end{fmfgraph*}
\end{fmffile}
            \end{gathered} \hspace{5em} \Longrightarrow \hspace{4em} i \sum_{\ell m} \fFR_{\phi, \ell}(\bm{p}_1) \; \hat{\Pi}_{\ell m}(\bm{p}_1,\bm{p}_2) \; \fFR_{\phi, m}(-\bm{p}_2)\, .
        \end{align*}
    \item \textbf{Scalar-Vector Loop Diagram (Type II)} (Sec.~\ref{subsec:loop_diagram}): For every type II loop connecting an incoming scalar $\phi$ with momentum $\vec{p}_1$, and an outgoing vector $V$ with momentum $\bm{p}_2$ insert,
        \vspace{2em}
        \begin{align*}
            \begin{gathered}
                \vspace{-0.6em}
                \begin{fmffile}{loop2_scalar_vector}
    \begin{fmfgraph*}(100,75)
        \fmfleft{i} \fmfright{v}
        \fmf{dashes,label=$\overset{\ds \longrightarrow}{\ds \vec{p}_1}$}{i,m1}
        \fmf{plain,right,tension=.85}{m1,m1}
        \fmf{photon,label=$\overset{\ds \longrightarrow}{\ds \vec{p}_2}$}{m1,v}
        \fmflabel{$\phi$}{i}
        \fmflabel{$A^\mu$}{v}
    \fmfv{decor.shape=circle,decor.filled=hatched,decor.size=0.09w}{m1}
    \fmffreeze 
    \fmfset{arrow_len}{3.5mm}
    \fmfforce{vloc(__m1) shifted (-1.75mm,13.75mm)}{al}
    \fmfforce{vloc(__m1) shifted (1.75mm,13.75mm)}{ar}
    \fmf{phantom_arrow}{ar,al}
    \end{fmfgraph*}
\end{fmffile}
            \end{gathered} \hspace{5em} \Longrightarrow \hspace{4em} i \sum_{\ell} \left[ \fFR_{\phi A,\sumIndex}(\bm{p}_1,\bm{p}_2) \right]^{\mu} \; \hat{\Pi}_\ell(\vec{p}_1 - \vec{p}_2) \, .
        \end{align*}
\end{itemize}

These loop diagrams are typically evaluated in the $\vec{p}_1 = \vec{p}_2$ limit. To simplify the notation later, when $\vec{p}_1 = \vec{p}_2$ we write self-energies as, $\Pi(\vec{p}_1) \equiv \Pi(\vec{p}_1, \vec{p}_1)$, $\hat{\Pi}_{\ell m}(\vec{p}_1) \equiv \hat{\Pi}_{\ell m}(\vec{p}_1, \vec{p}_1)$, and $\hat{\Pi}_\ell(0) \equiv \hat{\Pi}_{\ell}$.


\section{Absorption}
\label{sec:absorption}

\begin{figure}[ht!]
    \input{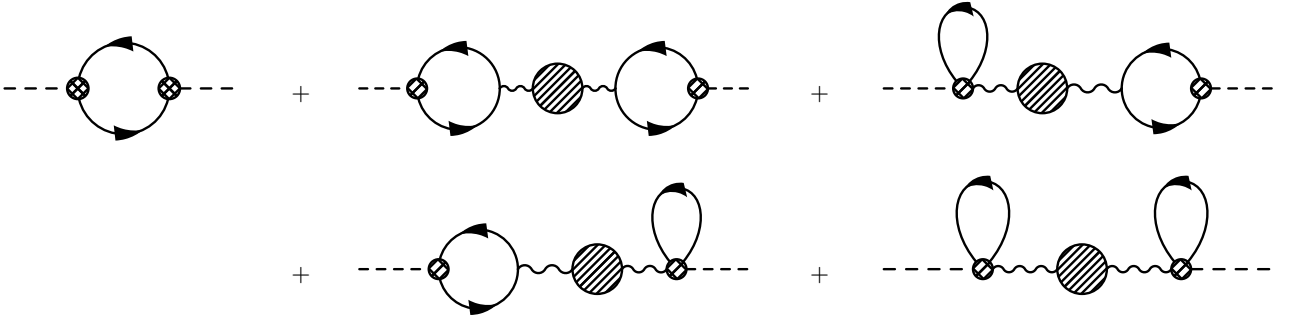}
    \caption{Example Feynman diagrams for DM (dashed lines) absorption into electronic excitations. The vertex blobs indicate a vertex whose Feynman rules are derived from the DM-electron NR EFT discussed in Secs.~\ref{sec:NR_DM_electron_interaction} and~\ref{sec:feynman_rules}. The Feynman rule coefficients for specific DM models can be found in the tables in App.~\ref{app:summary_tables}. The photon propagator with a shaded internal circle represents a sum over all one particle irreducible (1PI) electron loop diagrams, as discussed in detail in Appendix \ref{app:in_medium_photon_propagator}. Diagrams involving a photon propagators \textit{screen} the DM-electron interactions.}
    \label{fig:absorption_feynman_diagram}
\end{figure}

Unstable bosonic DM particles can be absorbed in target materials to yield electronic excitations. In this section we calculate general absorption rates for spin-0 and spin-1 particles whose electron interactions have arbitrary Lorentz structure. As in previous sections, we find clean factorization between model-dependent expressions that characterize the DM-electron interaction Lagrangian and target-dependent 
expressions that characterize properties of the target material. Previous literature on bosonic DM absorption for specific targets and DM models can be found in Refs.~\cite{Mitridate:2021ctr,Chen:2022pyd,Berlin:2023ppd,Berlin:2023ubt,Krnjaic:2023nxe}. Throughout this section, we make frequent use of the resummed in-medium photon propagator:
\be
    G_{AA}^{\mu\nu} = i \sum_\lambda \frac{\epsilon_\lambda^\mu \epsilon_\lambda^
    \nu}{\omega^2 - \bm{q}^2 - \Pi^\lambda_{AA}} \, ~~,~~
    \Pi_{AA}^\lambda \equiv -\epsilon^\lambda_{\mu} \, \Pi_{AA}^{\mu \nu} \, \epsilon^\lambda_{\nu}~~,
\ee
where $\omega$ and $\bm q$ are the photon energy and momentum, respectively, 
$\Pi_{AA}^{\mu\nu}$ is the photon self energy tensor,\footnote{After this point, $\Pi_{AA}^{\mu \nu}$ will be referred to as $\Pi_{AA}^{\text{UV}, \mu \nu}$ since it is computed with UV matrix elements defined in Eq.~\eqref{eq:matrix_element_QFT}, as opposed to the NR matrix elements defined in Eq.~\eqref{eq:matrix_element}. For further clarification see the discussion before Eq.~\eqref{eq:self_energy_vol_factor}.}
$\epsilon_\mu^\lambda$ is a polarization vector, and the sum is over photon polarization states $\lambda$. A detailed discussion of $G_{AA}^{\mu\nu}$ and its relation to the material dielectric function can be found in Appendix \ref{app:in_medium_photon_propagator}.

\subsection{Spin-0 Dark Matter}
\label{subsec:abs_scalar}

The absorption rate per incoming spin-0 particle $\phi$ with momentum $\bm{p}$ and energy $\omega^2 = \bm{p}^2 + m_\phi^2$ is related to the imaginary part of the particles self-energy via the optical theorem
\begin{align}
    \Gamma_{\phi}(\bm{p}) & \approx - \frac{1}{m_\phi} \text{Im} \left[ \, \Pi_{\phi\phi}^\text{UV}(\bm{p}) \, \right] \, ,
    ~~
    \label{eq:scalar_optical_theorem}
\end{align}
where we have assumed the incoming $\phi$ is NR, such that $\omega \approx m_\phi$. The total absorption rate per unit of detector exposure (detector mass $\times$ observation time) is then found by multiplying $\Gamma_\phi$ by the number of $\phi$ particles in the detector $N_\phi = \rho_\phi \vol / m_\phi$, and dividing by the detector mass, $\rho_T \vol$, to yield 
\begin{align}
    R_\phi = - \frac{\rho_\phi}{\rho_T \, m_\phi^2} \text{Im} \left[ \, \Pi_{\phi\phi}^\text{UV}(\bm{p}) \, \right] \, ,
\end{align}
where $\rho_\phi$ is the local DM mass density and $\rho_T$ is the target mass density.

The self-energy $\Pi_{\phi \phi}^\text{UV}$ is calculated by evaluating the series of diagrams shown in Fig.~\ref{fig:absorption_feynman_diagram}, where the dashed lines represent $\phi$ particles.\footnote{Although we are agnostic about the Lorentz structure of the high-energy $\phi$-electron coupling as in Sec.~\ref{subsec:example_high_to_low_match}, we restrict to interactions linear in $\phi$, so there are no $\phi$ self interactions that contribute to the sum in Fig.~\ref{fig:absorption_feynman_diagram}.} However there is one subtlety: the self-energy in the optical theorem relationship in Eq.~\eqref{eq:scalar_optical_theorem} is defined in terms of UV matrix elements (Eq.~\eqref{eq:matrix_element_QFT}), and therefore has a different mass dimension than the NR self-energy one would compute with the Feynman rules in Sec.~\ref{sec:feynman_rules}. The difference is a factor of $\vol$, which can be seen by equating
 Eqs.~\eqref{eq:matrix_element_QFT} and~\eqref{eq:matrix_element} to yield
\begin{align}
    (2 \pi)^3 \delta^{3}(0) \, \Pi_{\phi \phi}^\text{UV}(\vec{p}) = \vol \, \Pi_{\phi \phi}^\text{UV}(\vec{p}) = \Pi_{\phi \phi}^\text{NR}(\vec{p}) \, ,
    \label{eq:self_energy_vol_factor}
\end{align}
 where $\Pi_{\phi \phi}^\text{NR}$ is the $\phi$ self-energy computed using the NR Feynman rules from Sec.~\ref{sec:feynman_rules}. For simplicity we will drop the ``NR" superscript when the self-energies are defined in terms of NR matrix elements.

With this subtlety in mind, the first diagram in the upper left of Fig.~\ref{fig:absorption_feynman_diagram} has been computed in detail in Sec.~\ref{subsec:loop_diagram}, and its contribution to the self-energy is
\begin{align}
    -i \Pi_{\phi \phi}(\bm{p}) \supset i \sum_{\ell m} \fFR_{\phi, \ell}(\bm{p}) \, \hat{\Pi}_{\ell m}(\bm{p}) \fFR_{\phi, m}(-\bm{p}) \, ,
    \label{eq:phi_phi_self_energy_first_term}
\end{align}
which is just the expression in \Eq{eq:se_2} with the replacements  $\bm{p}_1 = \bm{p}_2  \equiv \bm p$. The additional minus sign on the left hand side of Eq.~\eqref{eq:phi_phi_self_energy_first_term} is added because the scalar particle self-energy is defined as the negative of the associated Feynman diagram~\cite{Peskin:1995ev}.\footnote{This is to ensure the scalar propagator resums as
\begin{align*}
    G_{\phi \phi} & = \frac{i}{q^2 - m_\phi^2} + \left( \frac{i}{q^2 - m_\phi^2} \right) \left( -i \Pi_{\phi \phi}^\text{UV} \right) \left( \frac{i}{q^2 - m_\phi^2} \right) + \cdots = \frac{i}{q^2 - m_\phi^2 - \Pi_{\phi \phi}^\text{UV}} \, .
\end{align*}
}

The other four diagrams in Fig.~\ref{fig:absorption_feynman_diagram}  contain in-medium photon propagators and, as we will see in Sec.~\ref{subsec:screening_effects}, these {\it screen} the DM interaction. While each can be computed individually it is convenient to split the calculation into pieces. The sum of these screening diagrams factorize as
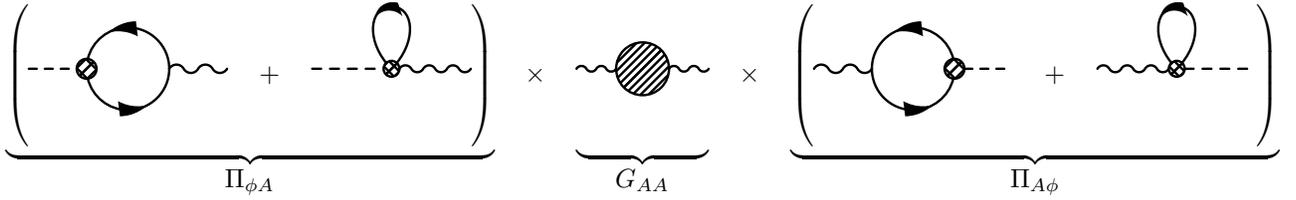
\vspace{2em}
\begin{fmffile}{abs_grouped}
    \begin{align*}
        \underbrace{\left( 
            \begin{gathered}
                \begin{fmfgraph*}(75,50)
                    \fmfleft{i} \fmfright{f}
                    \fmf{dashes}{i,m1}
                    \fmf{fermion,right,tension=.35}{m1,m2}
                    \fmf{fermion,right,tension=.35}{m2,m1}
                    \fmf{photon}{m2,f}
                    \fmfv{decor.shape=circle,decor.filled=hatched,decor.size=0.1w}{m1}
                \end{fmfgraph*}
            \end{gathered} \quad + \quad
            \begin{gathered}
                \begin{fmfgraph*}(60,50)
                    \fmfleft{i} \fmfright{f}
                    \fmf{dashes}{i,m1}
                    \fmf{fermion,right,tension=.85}{m1,m1}
                    \fmf{photon}{m1,f}
                    \fmfv{decor.shape=circle,decor.filled=hatched,decor.size=0.1w}{m1}
                \end{fmfgraph*}
            \end{gathered} 
        \right)}_{ \textstyle \Pi_{\phi A} } \quad \times \quad 
        \underbrace{\begin{gathered}
            \begin{fmfgraph*}(50,50)
                \fmfleft{i} \fmfright{f}
                \fmf{photon}{i,v}
                \fmf{photon}{v,f}
                \fmfblob{.4w}{v}
            \end{fmfgraph*}
        \end{gathered}}_{ \textstyle G_{AA} } \quad \times \quad
        \underbrace{\left( 
            \begin{gathered}
                \begin{fmfgraph*}(75,50)
                    \fmfleft{i} \fmfright{f}
                    \fmf{photon}{i,m1}
                    \fmf{fermion,right,tension=.35}{m1,m2}
                    \fmf{fermion,right,tension=.35}{m2,m1}
                    \fmf{dashes}{m2,f}
                    \fmfv{decor.shape=circle,decor.filled=hatched,decor.size=0.1w}{m2}
                \end{fmfgraph*}
            \end{gathered} \quad + \quad
            \begin{gathered}
                \begin{fmfgraph*}(60,50)
                    \fmfleft{i} \fmfright{f}
                    \fmf{photon}{i,m1}
                    \fmf{fermion,right,tension=.85}{m1,m1}
                    \fmf{dashes}{m1,f}
                    \fmfv{decor.shape=circle,decor.filled=hatched,decor.size=0.1w}{m1}
                \end{fmfgraph*}
            \end{gathered} 
        \right)}_{ \textstyle \Pi_{A \phi} }
    \end{align*}
\end{fmffile}

\noindent where the in-medium photon propagator $G_{AA}$ is computed in detail in App.~\ref{app:in_medium_photon_propagator}.
Using the Feynman rules in Sec.~\ref{subsec:feynman_rules_summary}, the off-diagonal $\Pi_{\phi A}$ and $\Pi_{A \phi}$ contributions are  
\begin{align}
    i \Pi^\mu_{\phi A}(\bm{p}, \bm{p}') & =  i \sum_{\ell m} \fFR_{\phi, \ell}(\bm{p}) \, \hat{\Pi}_{\ell m}(\bm{p}, \bm{p}') \, [\fFR_{A, m}(-\bm{p}')]^\mu + i \sum_{\ell} [\fFR_{\phi A, \ell}(\bm{p}, \bm{p}')]^\mu \; \hat{\Pi}_{\ell}(\vec{p} - \vec{p}') 
    \label{eq:off-diagonal-vector-scalar-Pi1}
    \\ 
    i \Pi^\mu_{A \phi}(\bm{p}', \bm{p}) & = i \sum_{\ell m} [\fFR_{A, \ell}(\bm{p}')]^\mu \, \hat{\Pi}_{\ell m}(\bm{p}', \bm{p}) \, \fFR_{\phi, m}(-\bm{p}) + i \sum_{\ell} [\fFR_{A \phi, \ell}(\bm{p}', \bm{p})]^\mu \; \hat{\Pi}_{\ell}(\vec{p}' - \vec{p}) \, ,
    \label{eq:off-diagonal-vector-scalar-Pi2}
\end{align}
and $\bm{p}'$ is the undetermined momentum flowing through the photon line. The total self-energy can then be written 
\begin{align}
    -i \Pi_{\phi \phi}(\bm{p}) = i \sum_{\ell m} \fFR_{\phi, \ell}(\bm{p}) \, \hat{\Pi}_{\ell m}(\bm{p}) \fFR_{\phi, m}(-\bm{p}) + \left( \int \frac{d^3 \bm{p}'}{(2 \pi)^3} \right) \left[ i \Pi^\mu_{\phi A}(\bm{p}, \bm{p}') \right] \left[ G_{AA}(\bm{p}') \right]_{\mu\nu} \! \left[ i \Pi^\nu_{A \phi}(\bm{p}', \bm{p}) \right]~,
\end{align}
where the first term is from the diagonal contribution in \Eq{eq:phi_phi_self_energy_first_term} and in the second term we have added the integral over $\bm{p}'$ to sum over all undetermined intermediate photon states. With the approximation in Eq.~\eqref{eq:Pi_hat_approximation}, the $\bm{p}'$ integral can be removed, which introduces a factor of $\vol$ since $\delta_{\bm{p}\bm{p}'} = (2 \pi)^3 \delta^{3}(\bm{p} - \bm{p}') / \vol$. Therefore, the total self-energy simplifies to yield
\begin{align}
    -\Pi_{\phi \phi}(\bm{p}) & = \sum_{\ell m} \fFR_{\phi, \ell}(\bm{p}) \, \hat{\Pi}_{\ell m}(\bm{p}) \fFR_{\phi, m}(-\bm{p}) + \frac{i}{\vol} \, \Pi^\mu_{\phi A}(\bm{p}) \, \left[  G_{AA}(\bm{p}) \right]_{\mu\nu}   \Pi^\nu_{A \phi}(\bm{p}) \, .
    \label{eq:final_scalar_se}
\end{align}
In typical targets, $\Pi_{\phi \phi}^\text{UV}$ is $\vol$ independent, such that the absorption rate in Eq.~\eqref{eq:scalar_optical_theorem} is also $\vol$ independent. Therefore, by Eq.~\eqref{eq:self_energy_vol_factor}, $\Pi_{\phi \phi}$ and $\hat{\Pi}_{\ell m}$ must be linear in $\vol$. This $\vol$ factor will arise from explicitly evaluating $\hat{\Pi}_{\ell m}$ in Eq.~\eqref{eq:Pi_hat_definition} through either $\hat{\mathcal{M}}_{JK, \ell}$ or the state sums.

\subsection{Spin-1 Dark Matter}
\label{subsec:abs_vector}

We now consider absorption of a dark spin-1 particle, $V$, with mass $m_V$. For a fixed polarization $\lambda$, the absorption per incoming particle is 
\begin{align}
    \Gamma_V^\lambda(\bm{p}) \approx - \frac{1}{m_V \vol} \text{Im} \left[ \Pi_{VV}^{\lambda}(\bm{p}) \right]~~~,~~~\Pi_{VV}^{\lambda} = -\epsilon^\lambda_\mu \, \Pi_{VV}^{\mu \nu} \, \epsilon^\lambda_\nu \, ,
\end{align}
where we have assumed NR kinematics for the incoming $V$, written $\Gamma_{V}^\lambda$ in terms of the NR self-energies (related to the usual self-energies used in the optical theorem via Eq.~\eqref{eq:self_energy_vol_factor}), and projected the self-energy using real polarization vectors satisfying
\begin{align}
    \epsilon^\mu_L & = \frac{1}{\sqrt{p^2}} (|\bm{p}|, \omega \hat{\bm{p}})~~~,~~~\epsilon^\mu_{\pm} = (0, \hat{\bm{p}}_\pm)~~~,~~~p^\mu \epsilon_\mu^\lambda = 0~~~,~~~\sum_{\lambda} \epsilon^\mu_{\lambda} \epsilon^\nu_{\lambda} = - \eta^{\mu \nu} + \frac{p^\mu p^\nu}{p^2}~~~,~~~ \epsilon^\mu_\lambda \epsilon_{\lambda' \mu} = - \delta_{\lambda \lambda'} ~,
    \label{eq:polarization_vector_properties}
\end{align}
where $\hat{\bm{p}}_\pm$ are any two vectors chosen to be mutually orthonormal to $\hat{\bm{p}}$, $p^\mu = (\omega, \bm{p})$ is the incoming four-momentum, and $p^2 = m_V^2$.

The total absorption rate averaged over the polarization components is then\footnote{This is not appropriate when the incoming DM modes have different abundances. For example, the sun produces primarily longitudinal dark photons~\cite{An:2013yua,An:2013yfc}, which must be taken into account when averaging Eq.~\eqref{eq:abs_rate_vector}.}
\begin{align}
    R_V & = - \frac{\rho_V}{\rho_T \, m_V^2 \vol} \left( \frac{1}{3} \sum_{\lambda} \right) \text{Im} \left[ \Pi_{VV}^{\lambda}(\bm{p}) \right] = \frac{\rho_V}{3 \rho_T \, m_V^2 \vol} \left( - \eta_{\mu \nu} + \frac{p_\mu p_\nu}{m_V^2} \right) \text{Im} \left[ \Pi_{VV}^{\mu \nu}(\bm{p}) \right]\, ,
    \label{eq:abs_rate_vector}
\end{align}
where $\rho_V$ is the mass density of $V$, and we have replaced the polarization sum with Eq.~\eqref{eq:polarization_vector_properties}. The self-energy $\Pi_{VV}^{\mu \nu}$ calculation is analogous to the previously considered scalar $\Pi_{\phi \phi}$, except now the self-energies have additional Lorentz indices matching those of the corresponding field $V$. Explicitly including these Lorentz indices, the relevant off-diagonal self-energies are
\begin{align}
    i \Pi_{V A}^{\mu \nu}(\bm{p}) & = i \sum_{\ell m} \left[ \fFR_{V, \ell}(\bm{p})\right]^{\mu} \, \hat{\Pi}_{\ell m}(\bm{p}) \, \left[ \fFR_{A, m}(-\bm{p}) \right]^{\nu} + i \sum_{\ell} \left[\fFR_{V A, \ell}(\bm{p}) \right]^{\mu \nu} \; \hat{\Pi}_{\ell} \\ 
    i \Pi_{A V}^{\mu \nu}(\bm{p}) & = i \sum_{\ell m} \left[ \fFR_{A, 
    \ell}(\bm{p}) \right]^{\mu} \, \hat{\Pi}_{\ell m}(\bm{p}) \, \left[ \fFR_{V, m}(-\bm{p})\right]^{\nu} + i \sum_{\ell} \left[\fFR_{A V, \ell}( \bm{p}) \right]^{\mu \nu} \; \hat{\Pi}_{\ell} ,
    \label{eq:final_vector_se}
\end{align}
and the diagonal self-energy can be written
\be
\Pi_{VV}^{\mu \nu}(\bm{p})  = \sum_{\ell m} \left[\fFR_{V, \ell}(\bm{p})\right]^{\mu} \, \hat{\Pi}_{\ell m}( \bm{p}) \left[ \fFR_{V, \ell}(-\bm{p})\right]^{\nu} + \frac{i}{\vol} \, \Pi_{V A}^{\mu \rho}(\bm{p}) \,  \left[ G_{AA} (\bm{p})\right]_{\rho \sigma}  \Pi_{A V}^{\sigma \nu}(\bm{p}) \, ,
\label{eq:Pi_VV}
\ee
so the final $\Pi_{VV}$ would be identical in form to the scalar analogue in Eq.~\eqref{eq:final_scalar_se} if the Lorentz indices were left implicit. As noted in Sec.~\ref{subsec:abs_scalar}, once $\hat \Pi_{\ell m}$ is evaluated explicitly in typical targets both the $\hat{\Pi}$'s and $\Pi$'s will be linear in $\vol$, leaving the absorption rate in Eq.~\eqref{eq:abs_rate_vector} $\vol$ independent.

\subsection{Screening Effects}
\label{subsec:screening_effects}

Above it was mentioned that the four diagrams in Fig.~\ref{fig:absorption_feynman_diagram} containing the photon ``screen" the DM interaction with electrons and  suppress the absorption rate. To justify this, here we investigate DM absorption in two specific DM models -- one of these exhibits screening, and the other does not. 

\vspace{1em}
\begin{center}
    \textit{ \normalsize  An Example With Screening }
\end{center}
\vspace{1em}

First, consider the scenario where the $V$ is a kinetically mixed dark photon, whose mass basis UV interaction is 
\begin{align}
    \LUV_\text{int} = - \kappa e V_\mu \bar{\Psi} \gamma^\mu \Psi \, . 
\end{align}
To simplify the calculation of the absorption rate in Eq.~\eqref{eq:abs_rate_vector} note that $\Pi_{VV}^{\mu \nu}$ satisfies the Ward Identity (WI): $p_\mu \Pi_{VV}^{\mu \nu} = \Pi_{VV}^{\mu \nu} p_\nu = 0$, where $p_\mu = (\omega, \bm p)$ is the $V$ momentum. This eliminates the contribution from contracting $\Pi_{VV}^{\mu \nu}$ with $p_\mu p_\nu / m_V^2$ in Eq.~\eqref{eq:abs_rate_vector}. Furthermore, the WI demands that for absorption kinematics, $|\vec{p}| \ll \omega$, the temporal components of the $V$ self-energy are negligible since
\be
 \Pi^{00}_{VV} = \frac{\vec{p}^i \, \Pi_{VV}^{ij} \, \vec{p}^j }{\omega^2}~~.
\ee
Thus, in the absorption limit, the absorption rate in Eq.~\eqref{eq:abs_rate_vector} reduces to 
\begin{align}
    R_V \approx \frac{\rho_V}{3 \rho_T \, m_V^2 \vol} \text{Im} \left[ \Pi_{VV}^{ii}(\bm{p}) \right] \, ,
    \label{eq:R_V_screen}
\end{align}
where we have dropped terms of order $|\bm p|/m_V$. To compute $\Pi_{VV}^{ii}$ with Eq.~\eqref{eq:Pi_VV} we can use the NR Feynman rule coefficients from Table~\ref{tab:vector_summary_table} and replace $g_v \rightarrow -\kappa e$
\begin{align}
    \left[\fFR_{V, 2} \right]^{i, j} = i \kappa e \, \delta^{ij}~~~,~~~\left[ f_{V A, 1} \right]^{ij} = - \frac{\kappa e^2}{m_e} \, \delta^{ij} \, .
\end{align}
These expressions are identical to the leading order NR Feynman rule coefficients of NR QED (App.~\ref{app:summary_tables}) with the replacement $\kappa \rightarrow 1$, and therefore we can use the relationships in App.~\ref{app:in_medium_photon_propagator} to relate each of the terms in Eq.~\eqref{eq:Pi_VV} to the dielectric function $\varepsilon(\omega)$ (assuming an isotropic target for simplicity)
\begin{align}
    & \text{Im} \left[ \left[ f_{V, 2} \right]^{i} \hat{\Pi}_{22} \left[ f_{V, 2} \right]^{j} \right] = \vol \, \kappa^2 \, \text{Im}\left[ \Pi_{AA}^{\text{UV}, ij} \right] = \vol \, \kappa^2 \omega^2 \,  \text{Im} \left[ \varepsilon(\omega) \right] \, \delta^{ij} \nonumber \\
    & \Pi_{VA}^{ij} = \Pi_{AV}^{ij} = - \vol \, \kappa \, \omega^2 \, [1 - \varepsilon(\omega)] \, \delta^{ij} ~~,~~G_{AA}^{ij} = \frac{i \delta^{ij}}{\omega^2 \, \varepsilon(\omega)} . \, ~~~~~~~
    \label{eq:dielectric_relationship}
\end{align}
Substituting these expressions with $\omega \approx m_V$ into Eq.~\eqref{eq:Pi_VV}, and then Eq.~\eqref{eq:R_V_screen},  leads to the familiar expression for the absorption rate of dark photon DM~\cite{An:2013yua,An:2014twa,Hochberg:2016sqx,Hochberg:2016ajh,Mitridate:2021ctr}
\begin{align}
    R_V = \kappa^2 \frac{\rho_V}{\rho_T} \text{Im} \left[ \frac{-1}{\varepsilon(m_V)} \right] = \kappa^2 \frac{\rho_V}{\rho_T} \frac{\text{Im}\left[ \varepsilon(m_V) \right]}{|\varepsilon(m_V)|^2} \, ,
\end{align}
which is \textit{screened} due to the appearance of $1/|\varepsilon|^2$, where $|\varepsilon| > 1$ in typical materials.

\vspace{1em}
\begin{center}
    \textit{ \normalsize  An Example Without Screening }
\end{center}
\vspace{1em}

We now consider a DM model which will not be screened. Consider the interactions of an axion-like particle $a$ with mass $m_a$
\begin{align}
    \LUV_\text{int} = g_{ae} ( \partial_\mu a ) \bar{\Psi} \gamma^\mu \gamma^5 \Psi \, .
\end{align}
Since this is a momentum dependent interaction, in the absorption limit, $|\bm{p}| \ll \omega$, the zero Lorentz component dominates,
and we have 
\begin{align}
    \LUV_\text{int} \approx g_{ae} ( \partial_t a ) \bar{\Psi} \gamma^0 \gamma^5 \Psi \, ,
\end{align}
for which the dark operator $(\partial_t a)$ has the same electronic interactions as the spin-1 $V_0$ state coupled to the $\bar \Psi \gamma^0 \gamma^5 \Psi$ current
discussed in 
Sec.~\ref{sec:NR_DM_electron_interaction}. Therefore the NR EFT interaction Lagrangian coefficients ($C$) for $\partial_t a$ are identical to those of $V^0$ in Table \ref{tab:vector_coefficients} (with $g_a \rightarrow g_{ae}$), and the corresponding Feynman rule coefficient $f$ is
\begin{align}
    \left[ C_{\partial_t a, 5} \right]^{ij} = - i g_{ae} \, \delta^{ij}~~~,~~~\left[ \fFR_{a, 5} \right]^{ij} = - g_{ae} \, \omega \, \delta^{ij} \, .
\end{align}
Using the self-energy from Eq.~\eqref{eq:final_scalar_se} and the relationship between the in-medium
propagator and material dielectric in simple targets from Eq.~\eqref{eq:dielectric_relationship}, the pseudoscalar self-energy is
\begin{align}
    -\Pi_{aa} =  f_{a, 5}(m_a) \, \hat{\Pi}_{55} \, f_{a, 5}(-m_a) - \frac{1}{\vol} \frac{\Pi_{a A}^i \Pi_{Aa}^i}{m_a^2 \, \varepsilon(m_a)} \, .
    \label{eq:pi_aa}
\end{align}
From Eq.~\eqref{eq:pi_aa} we see that there can only be screening if $\Pi_{a A} \neq 0$, i.e., if the pseudoscalar mixes with the photon. Since the axion-like particle only couples to the the fifth NR basis operator, and the photon dominantly couples to the second (App.~\ref{app:summary_tables}), $\Pi_{aA}$ is only non-zero if $\hat{\Pi}_{25} \neq 0$. Let us focus on the calculation of $\hat{\Pi}_{25}$ for a target where spin is a good quantum number such that $I = \{ b, s \}$, and the wavefunctions may be decomposed as 
\begin{align}
    \psi_{b s}(\vec{x}) = \psi_b(\vec{x}) \, \xi_s~~~,~~~\xi_\uparrow = \begin{pmatrix}
        1 \\ 0 
    \end{pmatrix}~~~,~~~\xi_\downarrow = \begin{pmatrix}
        0 \\ 1
    \end{pmatrix} \, ,
\end{align}
and the energy levels are spin-degenerate, $E_{b, s} = E_b$. In this limit, for electronic transitions $I = \{b,s\}  \to F = \{b^\prime,s^\prime\}$, we have 
\be
    \left[ \hat{\Pi}_{25} \right]^{ijk} \!\! = - \frac{1}{m_e^2}  \sum_{bb^\prime} \left[ 
      \frac{
    \left( \displaystyle \int d^3 \vec{y} e^{i \bm{p} \cdot \vec{y}} \psi^*_{b^\prime} \nabla^i \psi_b \right)
    \left( \displaystyle\int d^3 \vec{x} e^{-i \bm{p} \cdot \vec{x}} \psi^*_b \nabla^k \psi_{b^\prime} \right)
    \left( \displaystyle \sum_{ss'} \xi_{s'}^\dagger \xi_s \xi_s^\dagger \sigma^j \xi_{s'} \right)}{m_a - E_b + E_{b^\prime} + i \epsilon} - \left( b, s \leftrightarrow b^\prime, s'\, , \epsilon \rightarrow - \epsilon \right) \right]  ,~~~~~~~~
\ee
where the sum over spins satisfies $\sum_{s} \xi_s \xi_{s}^\dagger = 1$, so this expression simplifies to yield
\be
\left[ \hat{\Pi}_{25} \right]^{ijk}  \propto \sum_{s'} \xi_{s'}^\dagger \bm{\sigma} \xi_{s'} =\text{Tr}\left[ \bm{\sigma} \right] =  0 \, ,
\ee
and therefore there is \textit{no screening}, which follows generically from inserting a spin-independent and spin-dependent operator in $\hat{\Pi}_{\ell m}$. 

Using similar arguments, we arrive at the general conclusion: at leading order in simple targets, 
\textit{there is \textbf{no screening} if the dominant NR EFT interaction is spin-dependent} --  i.e., if the dark fields couple dominantly to $\ONRBasis_4, \ONRBasis_5, \ONRBasis_6$,  or $\ONRBasis_8$ from Eqs.~\eqref{eq:ONR_basis2} and ~\eqref{eq:ONR_basis3}.


\section{Scattering}
\label{sec:scattering}

\begin{figure}[ht!]
    \input{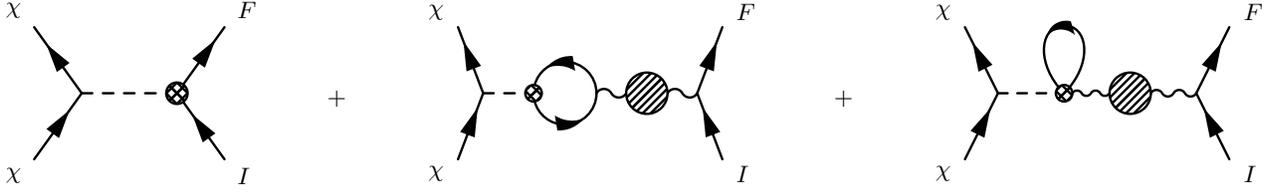}
    \caption{Feynman diagrams for fermion DM-electron scattering. The first diagram on the left is the tree level process, while the second two \textit{screen} the interactions. As in Fig.~\ref{fig:absorption_feynman_diagram}, the vertex blobs indicate a vertex whose Feynman rules are derived from the DM-electron NR EFT discussed in Secs.~\ref{sec:NR_DM_electron_interaction} and~\ref{sec:feynman_rules}. The Feynman rules for specific models can be found in App.~\ref{app:summary_tables}. The photon propagator includes a resummation of all 1PI diagrams and is discussed in detail in App.~\ref{app:in_medium_photon_propagator}.}
    \label{fig:scattering_feynman_diagram}
\end{figure}

\subsection{General Formalism}

In addition to the absorption process discussed in Sec.~\ref{sec:absorption}, DM may scatter off a target, inducing transitions between the filled and unfilled electronic states. The probability for an interaction to occur given an initial state $| \mathcal{I} \rangle$ and final state $| \mathcal{F} \rangle$, over some time period $T$ due to a transfer matrix $\mathcal{T}$ is~\cite{Peskin:1995ev,Schwartz}
\begin{align}
    \mathcal{P}_{\mathcal{I} \rightarrow \mathcal{F}} = \frac{\left| \langle \mathcal{F} | \, i \mathcal{T} \, | \mathcal{I} \rangle \right|^2}{\langle \mathcal{I} | \mathcal{I} \rangle \langle \mathcal{F} | \mathcal{F} \rangle} \, ,
\end{align}
and the corresponding interaction rate is
\begin{align}
\label{eq:Gamma_IF}
    \Gamma_{\mathcal{I} \rightarrow \mathcal{F}} \equiv \frac{{\cal P}_{{\cal I} \to {\cal F}}}{T} \, .
\end{align}
For the scattering process of interest here, the initial state is an incoming DM particle $\chi$ with momentum $\vec{p}$, and spin $s$, and an electron in a filled state $| I \rangle$. The final state is a DM particle with momentum $\vec{p}'$ and spin $s'$, and an electron in an unfilled state $| F \rangle$
\begin{align}
    | \mathcal{I} \rangle = | \vec{p}, s \rangle \otimes | I \rangle~~~,~~~| \mathcal{F} \rangle = | \vec{p}', s' \rangle \otimes | F \rangle \, .
    \label{eq:scattering_states}
\end{align}
The total scattering rate per incoming DM particle is the sum over the final state quantum numbers, $\vec{p}'$, $F$, $s'$, and initial, filled electron states $I$
\begin{align}
    \Gamma(\vec{p}, s) & = \sum_{\vec{p}'} \sum_I \sum_F \sum_{s'} \Gamma_{\mathcal{I} \rightarrow \mathcal{F}} = \frac{1}{T} \sum_{\vec{p}'} \sum_{IF} \sum_{s'} \frac{\left| \langle \mathcal{F} | \, i \mathcal{T} \, | \mathcal{I} \rangle \right|^2}{\langle \mathcal{I} | \mathcal{I} \rangle \langle \mathcal{F} | \mathcal{F} \rangle} ~.
    \label{eq:scatter_rate_1}
\end{align}
By inserting the (relativistic) DM state normalization factors~\cite{Schwartz} 
\be
    \langle \vec{p}, s | \vec{p}, s \rangle = 2 E_{\vec{p}} (2 \pi)^3 \delta^3(0) = 2 E_{\vec{p}} \vol~~~,~~~\langle \vec{p}', s' | \vec{p}', s' \rangle = 2 E_{\vec{p}'} (2 \pi)^3 \delta^3(0) = 2 E_{\vec{p}'} \vol \, ,
\ee
and using the fact that, for the continuum index $\bm p^\prime$, the sum can be written as an integral
\begin{align}
    \sum_{\vec{p}'} = \vol \int \frac{d^3 \vec{p}'}{(2 \pi)^3} \, ,
\end{align}
 the general formula from Eq.~\eqref{eq:scatter_rate_1} becomes
\begin{align}
    \Gamma(\vec{p}, s) & = \frac{1}{T} \frac{1}{2 E_{\vec{p}} \vol} \left( \int \frac{d^3 \vec{p}'}{(2 \pi)^3} \frac{1}{2 E_{\vec{p}'}} \right) \sum_{IF} \sum_{s'} \frac{\left| \langle \mathcal{F} | \, i \mathcal{T} \, | \mathcal{I} \rangle \right|^2}{\langle I | I \rangle \langle F | F \rangle} \, ,
    \label{eq:scatter_rate_2}
\end{align}
which represents the physical scattering rate per incoming DM particle in a target of volume $\vol$. 

\vspace{1em}
\begin{center}
    \textit{ \normalsize Electron State Normalization }
\end{center}
\vspace{1em}

The electron state normalization factors in Eq.~\eqref{eq:scatter_rate_2} can be simplified further. Using the electron state normalization factors discussed in Sec.~\ref{subsec:nr_electron_field_quantization}, if $I$ is a discrete index, then
\be
\langle I | I \rangle = 1~~, ~~\frac{1}{\langle I | I \rangle} \sum_I = \sum_I~~,
\ee
and if $I$ is a continuous index (e.g., electron spatial momentum $\vec{k}$), then 
\be
\langle I | I \rangle =  \langle   \vec{k} | \vec{k} \rangle = \vol
 ~~,~~ 
  ~~\frac{1}{\langle I | I \rangle} \sum_I =
 \frac{1}{\langle \vec{k} | \vec{k} \rangle} \sum_{\vec{k}} = \int \frac{d^3\bm{k} }{(2 \pi)^3}~,
\ee
where the electron states are normalized
according to \Eq{eq:electron-state-normalization}. Both cases here are neatly handled by replacing
\begin{align}
    \sum_I \frac{1}{\langle I | I \rangle} \rightarrow \sum_I \, ,
\end{align}
for both $I$ and $F$, and \textit{interpreting the sum on the right-hand side as a sum when $I$ is discrete, and an integral when $I$ is continuous}. This is consistent with the convention introduced in Sec.~\ref{subsec:nr_electron_field_quantization}, where sums over discrete/continuous indices are summed/integrated over. With this convention for the electron state sums, the scattering rate in Eq.~\eqref{eq:scatter_rate_2} can be written
\begin{align}
    \Gamma(\vec{p}, s) & = \frac{1}{T} \frac{1}{2 E_{\vec{p}} \vol} \left( \int \frac{d^3 \vec{p}'}{(2 \pi)^3} \frac{1}{2 E_{\vec{p}'}} \right) \sum_{IF} \sum_{s'} \left| \langle \mathcal{F} | \, i \mathcal{T} \, | \mathcal{I} \rangle \right|^2 \, ,
    \label{eq:scatter_rate_3}
\end{align}
where the sums over $I, F$ become integrals for the continuous indices. 

\subsection{NR Dark Matter Scattering}

We now replace $\mathcal{T}$ in Eq.~\eqref{eq:scatter_rate_3} with $\mathcal{T}_\text{NR}$ from Eq.~\eqref{eq:matrix_element}, noting that
\be
\left| \langle \mathcal{F} | \, i \mathcal{T}_\text{NR} \, | \mathcal{I} \rangle \right|^2 = 
     \left[ 2 \pi \delta\left({\textstyle \sum E} \right) \right]
     \left[ 2 \pi \delta(0) \right] |\mathcal{M}_{\rm NR} |^2 = T \left[2 \pi \delta\left({\textstyle \sum E} \right) \right] |\mathcal{M}_{\rm NR} |^2~,
 \label{eq:T_NR_sub}
\ee
where we have used
\be
2 \pi \delta(\omega) = \lim_{T\to \infty} \int_{-T/2}^{T/2} dt \, e^{-i \omega t} ~~,~~ 2 \pi \delta(0) \to T\, .
\ee
Substituting Eq.~\eqref{eq:T_NR_sub} in to Eq.~\eqref{eq:scatter_rate_3}, while taking the NR limit of the energy normalization factors $E_{\vec{p}} \approx E_{\vec{p}'} \approx m_\chi$ and trading the $\vec{p}' \equiv \vec{p} - \vec{q}$ integral for an integral over momentum transfer $\vec{q}$, the total scattering rate per incoming DM particle becomes\footnote{
    To connect back to the usual QFT free-particle limit, consider the case where the electrons are free particles: $I = \{ \vec{k}, \sigma \}$, $F = \{ \vec{k}', \sigma' \}$, and we average over incoming fermion spins, $\sigma, s$. In this limit Eq.~\eqref{eq:gamma_NR_DM} becomes,
    \begin{align*}
        \ds \Gamma(\vec{p}) = \frac{1}{2 m_\chi} \, \int \overline{|\mathcal{M}|^2} \, (2 \pi)^4 \delta^4(\textstyle{\sum} p) \ds \, \left[ \frac{d^3 \mathbf{k}}{(2 \pi)^3} \frac{1}{2 m_e} \right]  \,  d \Pi_f~~~,~~~ \ds d\Pi_f = \left[ 
        \frac{d^3 \vec{p}'}{(2 \pi)^3} \frac{1}{2 m_\chi} \right] \left[ 
        \frac{d^3 \vec{k}'}{(2 \pi)^3} \frac{1}{2 m_e} \right] \, ,
    \end{align*}
    where $\overline{|\mathcal{M}|^2}$ is the usual QFT spin-averaged matrix element, we have added factors of $2 m_e$ to match the usual QFT state normalization for electrons, and note that $\mathcal{M}_\text{NR} = \mathcal{M} \, (2 \pi)^3 \delta^3(\textstyle\sum \vec{p})$. Here the integral over $\bm k$ arises from the replacement $\sum_I \to \sum_{\sigma} \int d^3 {\bm k}/(2\pi)^3$.
}
\begin{align}
    \Gamma(\vec{p}, s) = \frac{2 \pi}{4 \, m_\chi^2 \vol} \sum_{IF} \sum_{s'} \int \frac{d^3\bm{q}}{(2 \pi)^3} \, \delta\left( \textstyle \sum E \right) \; |\mathcal{M}_\text{NR}|^2 \, .
    \label{eq:gamma_NR_DM}
\end{align}

Since we are interested in NR DM scattering, their momentum and energy are typically approximated in terms of their velocity $\vec{v}$, $\vec{p} \approx m_\chi \vec{v}, E_{\vec{p}} \approx m_\chi + m_\chi \vec{v}^2 / 2$. Therefore the argument of the energy conserving delta function becomes
\begin{align}
    \sum E \implies \left( \frac{|\vec{p} - \bm{q}|^2}{2 m_\chi} + E_F \right) - \left( \frac{\vec{p}^2}{2 m_\chi} + E_I\right) = - \bm{q} \cdot \vec{v} + \frac{\bm{q}^2}{2 m_\chi} + E_F - E_I \, .
    \label{eq:scattering_energy_conservation}
\end{align}
The average scattering rate per incoming DM particle $\bar{\Gamma}$ is found by averaging over the incoming DM velocity distribution function $f_{\chi}(\vec{v})$ and spin $s$ to obtain 
\begin{align}
    \bar{\Gamma} = \frac{2 \pi }{4 \, m_\chi^2 \vol} \sum_{IF} \left( \frac{1}{2} \sum_{ss'} \right) \int \frac{d^3\bm{q}}{(2 \pi)^3} \, \int d^3 \vec{v} \, f_{\chi}(\vec{v} + \vec{v}_e) \, \delta\left( \textstyle \sum E \right) \; |\mathcal{M}_\text{NR}|^2 \, ,
\end{align}
where we have boosted to the Earth frame with the velocity of the Earth in the galactic frame, $\vec{v}_e$, and we model  $f_\chi$ as a truncated Maxwell-Boltzmann distribution
\begin{align}
    f_{\chi}(\vec{v}; v_0, v_\text{esc}) & = \frac{1}{N_0} e^{- \vec{v}^2 / v_0^2} \, \Theta(v_\text{esc} - v)~~~,~~~N_0 = \pi^{3/2} v_0^3 \left[ \text{erf}\left( \frac{v_\text{esc}}{v_0} \right) - \frac{2}{\sqrt{\pi}} \frac{v_\text{esc}}{v_0} e^{- v_\text{esc}^2 / v_0^2} \right] \label{eq:N_0} \, .
\end{align}
where $v_0$ is the velocity dispersion and $v_\text{esc}$ is the escape velocity \cite{Freese:2012xd,Baxter:2021pqo}.
Finally, the total scattering rate $R$ (per unit of detector exposure) is found by multiplying the average scattering rate $\bar{\Gamma}$ by the number of DM particles inside the detector $N_\chi = (\rho_\chi / m_\chi) \vol$, and dividing by the mass of the detector $\rho_T \vol$ to yield the general scattering rate
\begin{align}
    R = \frac{\rho_\chi}{\rho_T m_\chi} \, \frac{2 \pi}{4 \, m_\chi^2 \vol} \sum_{IF} \left( \frac{1}{2} \sum_{ss'} \right) \int \frac{d^3\bm{q}}{(2 \pi)^3} \, \int d^3 \vec{v} \, f_{\chi}(\vec{v} + \vec{v}_e) \, \delta\left( \textstyle \sum E \right) \; |\mathcal{M}_\text{NR}|^2 \, .
    \label{eq:general_rate_formula}
\end{align}
From this expression, one simply needs to evaluate the matrix element $\mathcal{M}_\text{NR}$ using the NR EFT Feynman rules introduced in Sec.~\ref{sec:feynman_rules}. The leading order scattering diagrams are given in Fig.~\ref{fig:scattering_feynman_diagram}. 

Before computing the NR matrix elements, there is another useful manipulation which separates the DM-mediator interactions from the mediator-electron interactions. Notice that for all the diagrams in Fig.~\ref{fig:scattering_feynman_diagram} the DM-mediator interaction contribution is identical. We can factorize these sub-diagrams as
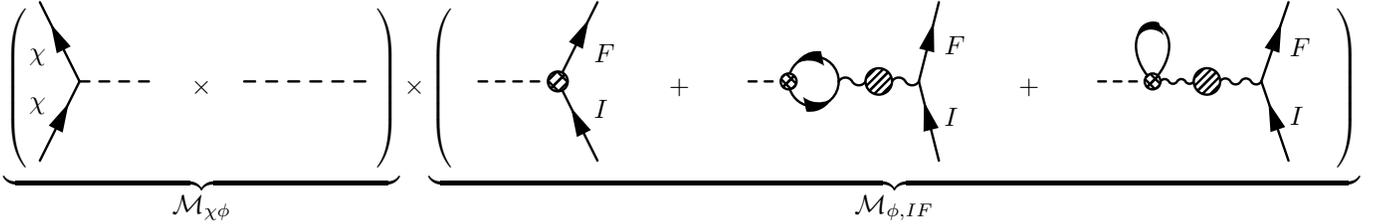
\begin{fmffile}{scatter_grouped}
    \begin{align*}
        & \underbrace{\left( \begin{gathered}
                \begin{fmfgraph*}(50,60)
                    \fmfleft{i1,i2} \fmfright{f}
                    \fmf{dashes}{v1,f}
                    \fmf{fermion,label=$\chi$,label.side=left}{v1,i2}
                    \fmf{fermion,label=$\chi$,label.side=left}{i1,v1}
                \end{fmfgraph*}
            \end{gathered} \quad \times \quad
            \begin{gathered}
                \begin{fmfgraph*}(50,60)
                    \fmfleft{i} \fmfright{f}
                    \fmf{dashes}{i,f}
                \end{fmfgraph*}
            \end{gathered}
            \right)}_{ \textstyle \mathcal{M}_{\chi\phi} } \times \underbrace{\left( \quad
                \begin{gathered}
                    \begin{fmfgraph*}(50,60)
                        \fmfleft{i} \fmfright{f1,f2}
                        \fmf{dashes}{i,v}
                        \fmf{fermion,label=$ I $,label.side=right}{f1,v}
                        \fmf{fermion,label=$ F $}{v,f2}
                        \fmfv{decor.shape=circle,decor.filled=hatched,decor.size=0.15w}{v}
                    \end{fmfgraph*}
                \end{gathered} \quad \quad + \quad \quad
                \begin{gathered}
                    \begin{fmfgraph*}(80,70)
                        \fmfleft{i} \fmfright{f1,f2}
                        \fmf{dashes}{i,v2}
                        \fmf{fermion,right,tension=.4}{v2,v3}
                        \fmf{fermion,right,tension=.4}{v3,v2}
                        \fmf{photon}{v3,v4}
                        \fmf{photon}{v4,v5}
                        \fmf{fermion,label=$ F $,label.side=right}{v5,f2}
                        \fmf{fermion,label=$ I $,label.side=right}{f1,v5}            
                        \fmfv{decor.shape=circle,decor.filled=hatched,decor.size=0.075w}{v2}
                        \fmfblob{.125w}{v4,v4}
                    \end{fmfgraph*}
                \end{gathered}
                \quad\quad + \quad\quad
                \begin{gathered}
                    \begin{fmfgraph*}(80,60)
                        \fmfleft{i} \fmfright{f1,f2}
                        \fmf{dashes}{i,v2}
                        \fmf{fermion,right,tension=.65}{v2,v2}
                        \fmf{photon}{v2,v4}
                        \fmf{photon}{v4,v5}
                        \fmf{fermion,label=$ F $,label.side=right}{v5,f2}
                        \fmf{fermion,label=$ I $,label.side=right}{f1,v5}            
                        \fmfv{decor.shape=circle,decor.filled=hatched,decor.size=0.075w}{v2}
                        \fmfblob{.125w}{v4,v4}
                    \end{fmfgraph*}
                \end{gathered} \quad
            \right)}_{ \textstyle \mathcal{M}_{\phi, I F} }
    \end{align*}
\end{fmffile}

\noindent where, due to the split, $\mathcal{M}_{\chi\phi}$ and $\mathcal{M}_{\phi, IF}$ have Lorentz indices that match those of the mediator. With this decomposition, the total matrix element is a product of two different matrix elements
\begin{align}
    i \mathcal{M}_\text{NR} = \left[ i \mathcal{M}_{\chi \phi}(\bm{q}, \vec{v}) \right] \,\left[ i \mathcal{M}_{\phi, I F}(\bm{q}) \right]~~~,~~~i \mathcal{M}_\text{NR} = \left[ i \mathcal{M}_{\chi V}^\mu(\bm{q}, \vec{v}) \right] \! \left[ i \left[ \mathcal{M}_{V, I F}(\bm{q}) \right]_\mu \right]\, ,
    \label{eq:matrix_element_split}
\end{align}
where the NR matrix element on the left applies for scalar mediators, and the one on the right applies for vector mediators. This decomposition is beneficial because $\vec{v}$ and $s,s'$ \textit{only} appear on the DM-mediator side of the calculation. Note that  $\mathcal{M}_{\chi \phi}$ and $\mathcal{M}_{\chi V}$ are the familiar QFT matrix elements, which can then be expanded in $\vec{v}$ and (analytically) integrated over the DM velocity distribution. The resulting ``DM form factor" can then be used \textit{independently} of how the mediator couples to the NR electrons. Defining this form factor for scalar and vector mediators as
\begin{align}
    \mathcal{F}_{\chi \phi}(\bm{q}, \omega) & \equiv \frac{2 \pi}{4 m_\chi^2} \, \frac{1}{2} \sum_{ss'} \, \int d^3 \vec{v} \, f_{\chi}(\vec{v} + \vec{v}_e) \, \delta\left( \omega - \bm{q} \cdot \vec{v} + \frac{\bm{q}^2}{2 m_\chi} \right) \, \left[\mathcal{M}_{\chi \phi} \right]_{ss'}^* \left[ \mathcal{M}_{\chi \phi} \right]_{ss'} \nonumber \\
    \mathcal{F}_{\chi V}^{\mu \nu}(\bm{q}, \omega) & \equiv \frac{2 \pi}{4 m_\chi^2} \, \frac{1}{2} \sum_{ss'} \, \int d^3 \vec{v} \, f_{\chi}(\vec{v} + \vec{v}_e) \, \delta\left( \omega - \bm{q} \cdot \vec{v} + \frac{\bm{q}^2}{2 m_\chi} \right) \, \left[\mathcal{M}_{\chi V} \right]_{ss'}^{\mu, *} \left[ \mathcal{M}_{\chi V} \right]_{ss'}^\nu \, ,
    \label{eq:DM_form_factor}
\end{align}
the rate in Eq.~\eqref{eq:general_rate_formula} simplifies to
\begin{align}
    R & = \frac{\rho_\chi }{\rho_T m_\chi \vol} \sum_{IF}  \int \frac{d^3\bm{q}}{(2 \pi)^3} \, \mathcal{F}_{\chi \phi}(\bm{q}, E_F - E_I) \mathcal{M}_{\phi, I F}^* \, \mathcal{M}_{\phi , I F} \nonumber \\
    R & = \frac{\rho_\chi}{\rho_T m_\chi \vol} \sum_{IF}  \int \frac{d^3\bm{q}}{(2 \pi)^3} \, \mathcal{F}_{\chi V}^{\mu \nu}(\bm{q}, E_F - E_I) \left[ \mathcal{M}_{V, I F}\right]^*_{\mu} \, \left[ \mathcal{M}_{V, I F} \right]_{\nu} \, ,
    \label{eq:final_rate_formula_pre}
\end{align}
for scalar and vector mediators, respectively. Using Eq.~\eqref{eq:DM_form_factor} the $\mathcal{F}_{\chi \phi}$ and $\mathcal{F}_{\chi V}$ form factors will be explicitly computed for a variety of DM models in Sec.~\ref{subsec:scatter_scalar_mediator}.

\subsection{Screening Effects And In-Medium Coupling Coefficients}

We now turn to the calculation of the matrix element $\mathcal{M}_{\phi, IF}$ and $\mathcal{M}_{V, IF}$ in Eq.~\eqref{eq:matrix_element_split}. The first term in the diagrammatic expansion is straightforwardly computed with the Feynman rules discussed in Sec.~\ref{sec:feynman_rules}. The next two can be written in terms of $\Pi_{\phi A}$ and $\Pi_{VA}$, the mediator photon mixing self-energies introduced in Secs.~\ref{subsec:abs_scalar}~\ref{subsec:abs_vector}, and $G_{AA}$, the in-medium photon propagator discussed in detail in App.~\ref{app:in_medium_photon_propagator}. In total, the matrix elements are
\begin{align}
    i \mathcal{M}_{\phi, I F} & = i \sum_{\ell} \fFR_{\phi, \ell} \, \MFR_{IF, \ell} + \left( \int \frac{d^3 \bm{q}'}{(2 \pi)^3} \right) \left( i  \Pi_{\phi A}^\mu(\bm{q}, \bm{q}') \right) \left( \left[ G_{AA}(\bm{q}') \right]_{\mu \nu} \right) \left( i \sum_{\ell} \left[ \fFR_{A, \ell} \right]^{\nu} \, \MFR_{IF, \ell} \right) \nonumber \\ 
    i \mathcal{M}_{V, I F}^\mu & = i \sum_{\ell} [ \fFR_{V, \ell} ]^\mu \, \MFR_{IF, \ell} + \left( \int \frac{d^3 \bm{q}'}{(2 \pi)^3} \right) \left( i \Pi_{V A}^{\mu \alpha}(\bm{q}, \bm{q}') \right) \left( \left[ G_{AA}(\bm{q}') \right]_{\alpha \beta} \right) \left( i \sum_{\ell} \left[ \fFR_{A, \ell} \right]^{\beta} \, \MFR_{IF, \ell} \right) \, ,
    \label{eq:m_scatter_general_pre}
\end{align}
where the $\bm{q}'$ integral arises from summing over the intermediate photon states. These expressions can be further simplified using the $\bm q' \approx \bm q$ approximation from Eq.~\eqref{eq:Pi_hat_approximation}
\begin{align}
    i \mathcal{M}_{\phi, I F} & = i \sum_{\ell} \left( \fFR_{\phi, \ell} + \frac{i}{\vol} \left[ \Pi_{\phi A}^\mu(\bm{q}) \right] \left[ G_{AA}(\bm{q}) \right]_{\mu \nu} [\fFR_{A, \ell}]^\nu \right) \MFR_{IF, \ell} \equiv i \sum_{\ell} g_{\phi, \ell} \, \MFR_{IF, \ell} \nonumber \\ 
    i \mathcal{M}_{V, I F}^\mu & = i \sum_{\ell} \left( [\fFR_{V, \ell}]^\mu + \frac{i}{\vol} \, \left[ \Pi_{V A}^{\mu\alpha}(\bm{q}) \right] \left[ G_{AA}(\bm{q}) \right]_{\alpha \beta} [\fFR_{A, \ell}]^\beta \right) \MFR_{IF, \ell} \equiv i \sum_{\ell} g_{V, \ell}^\mu \, \MFR_{IF, \ell}\, ,
    \label{eq:m_scatter_general}
\end{align}
where we have defined  \textit{in-medium} coupling coefficients
\begin{align}
 g_{\phi, \ell}(\bm{q}) & \equiv \fFR_{\phi, \ell}(\bm{q}) + \frac{i}{\vol} \left[ \Pi_{\phi A}^\mu(\bm{q}) \right] \, \left[ G_{AA}(\bm{q}) \right]_{\mu \nu} \, [\fFR_{A, \ell}(\bm{q})]^\nu \nonumber \\ 
  [g_{V, \ell}(\bm{q}) ]^\mu  & \equiv [\fFR_{\phi, \ell}(\bm{q})]^\mu + \frac{i}{\vol} \left[ \Pi_{V A}^{\mu \alpha}(\bm{q}) \right] \, \left[ G_{AA}(\bm{q}) \right]_{\alpha \beta} \, [\fFR_{A, \ell}(\bm{q})]^\beta\, .
\end{align}
If the mediator mixes with the photon, then $\Pi_{\phi A} \neq 0$ or $\Pi_{VA} \neq 0$ and the scattering rate may be \textit{screened}, thereby reducing $g_{\phi, \ell}$ ($g_{V, \ell}$) from its tree level value, $\fFR_{\phi, \ell}$ ($\fFR_{V, \ell}$). 

For example, consider the scenario where the vector mediator $V$ is a kinetically mixed dark photon (see \cite{Fabbrichesi:2020wbt} for a review) with 
electron couplings 
\be
{\cal L}_{\rm int} \supset -\kappa e V_\mu \bar \Psi \gamma^\mu \Psi,
\ee
where $\kappa\ll 1$ is a mixing parameter and the Feynman coefficients are all proportional to the photon couplings $\fFR_V = \kappa f_A$.  Assuming that scattering is dominated by the zero components of $\Pi_{VA} = \kappa \, \Pi_{AA}$ and $G_{AA}$, the in-medium coupling is 
\begin{align}
    [ g_{V, \ell} ]^0 \approx \kappa \, [f_{A, \ell}]^0 \left[ 1 + \frac{i}{\vol} \Pi_{AA}^{00} G^{AA}_{00}  \right]  = \kappa \, [f_{A, \ell}]^0 \left[ 1 + i\Pi_{AA}^{\text{UV}, 00} G^{AA}_{00}  \right] \, ,
\end{align}
where we have used $\Pi_{AA}^{00} = \vol \Pi_{AA}^{\text{UV}, 00}$ in Eq.~\eqref{eq:self_energy_vol_factor} to make the cancellation of the $\vol$ factor transparent. Using the results in App.~\ref{app:in_medium_photon_propagator} we can relate $\Pi_{AA}^{\text{UV},00}$ and $G_{AA}^{00}$ to the target dielectric tensor $\bm{\varepsilon}$
\begin{align}
    \Pi_{AA}^{\text{UV}, 00} = - \vec{q}^2 \, (1 - \hat{\vec{q}} \cdot \bm{\varepsilon} \cdot \hat{\vec{q}})~~~,~~~G_{AA}^{00} = \frac{i}{\vec{q} \cdot \bm{\varepsilon} \cdot \vec{q}} \, .
\end{align}
where $\hat{\vec{q}} \equiv \bm q /|\bm q|$, and $\bm q \bm \cdot \bm \varepsilon \cdot \bm q \equiv q^i \varepsilon^{ij} q^j$. Thus, for the dark photon mediated model, the effective coupling is 
\begin{align}
    [ g_{V, \ell} ]^0 \approx \frac{\kappa \, [ \fFR_{A, \ell} ]^0 }{\hat{\vec{q}} \cdot \bm{\varepsilon} \cdot \hat{\vec{q}}} \, ,
\end{align}
so the interaction is \textit{screened} by the dielectric.

In terms of the in-medium coupling coefficients, the DM-electron scattering rate for scalar and vector mediated models are
\begin{align}
    R & = \frac{\rho_\chi}{\rho_T m_\chi \vol} \, \sum_{\ell m} \sum_{IF}  \int \frac{d^3\bm{q}}{(2 \pi)^3} \, \mathcal{F}_{\chi\phi}(\bm{q}, E_F - E_I) \, g_{\phi, \ell}^* \, g_{\phi, m} \, \MFR_{I F, \ell}^* \, \MFR_{I F, m} \nonumber \\
    R & = \frac{\rho_\chi}{\rho_T m_\chi \vol} \, \sum_{\ell m} \sum_{IF}  \int \frac{d^3\bm{q}}{(2 \pi)^3} \, \mathcal{F}_{\chi V}^{\mu \nu}(\bm{q}, E_F - E_I) \, [g_{V, \ell}^*]_\mu \, [ g_{V, m} ]_\nu \, \MFR_{I F, \ell}^* \, \MFR_{I F, m}\, .
    \label{eq:final_rate_formula}
\end{align}
To summarize the meaning of each term in this expression: $\mathcal{F}_{\chi \phi}$ and $\mathcal{F}_{\chi V}$ from Eq.~\eqref{eq:DM_form_factor} are ``DM form factors" and only dependent on the physics of the dark sector. $g_{\phi, \ell}$ and $g_{V, \ell}$ from Eq.~\eqref{eq:m_scatter_general} are ``in-medium coupling coefficients," which depend on how the mediator couples to electrons in the UV, and may be screened via mixing with the photon. Lastly, $\MFR_{IF, \ell}$  from Eq.~\eqref{eq:M_hat_definition} are the target-dependent transition matrix elements for each NR basis operator, $\ONRBasis_\ell$. We now compute the DM form factors $\mathcal{F}_{\chi \phi}$, $\mathcal{F}_{\chi V}$ for a variety of fermionic DM models.

\subsection{Dark Matter Form Factors}
\label{subsec:scatter_scalar_mediator}

\vspace{1em}
\begin{center}
    \textit{ \normalsize Scalar Mediated Models}
\end{center}
\vspace{1em}

For a fermionic DM particle $\chi$, the most general renormalizable scalar couplings are 
\begin{align}
    \LUV \supset \left( y_{\chi s} \, \bar{\chi} \chi + i y_{\chi p} \, \bar{\chi} \gamma^5 \chi \right) \phi \, ,
\end{align}
where the squared, spin averaged sub-amplitude can be computed using familiar QFT Feynman rules 
\begin{align}
    \frac{1}{2} \sum_{ss'} \mathcal{M}_{\chi \phi, ss'}^* \mathcal{M}_{\chi \phi, ss'} \approx
     \frac{  4y_{\chi s}^2  m_\chi^2   + y_{\chi p}^2  \bm{q}^2  }{(\vec{q}^2 + m_\phi^2 )^2  } ,
     \label{eq:squared_sub_amplitude}
\end{align}
and we have approximated $q^2 \approx - \vec{q}^2$ and  $p \cdot p'  \approx m_\chi^2 + \bm{q}^2 / 2$ in the NR limit.
The DM form factor from Eq.~\eqref{eq:DM_form_factor} can now be written
\begin{align}
    \mathcal{F}_{\chi \phi} =  \frac{1}{4 m_\chi^2} \, \left[  \frac{  4y_{\chi s}^2  m_\chi^2   + y_{\chi p}^2  \bm{q}^2  }{(\vec{q}^2 + m_\phi^2 )^2  } \right] \, K_0(\bm{q}, \omega, \vec{v}_e)\;  \, ,
    \label{eq:general_scalar_mediator_DM_form_factor}
\end{align}
where $K_0$ is the \textit{kinematic function} introduced in Ref.~\cite{Trickle:2019nya,Coskuner:2021qxo,Trickle:2020oki}
\be
\label{eq:K0_func}
    K_0(\bm{q}, \omega, \vec{v}_e)  \equiv 2 \pi \int d^3 \vec{v} f_{\chi}(\vec{v} + \vec{v}_e) \, \delta\left( \omega - \bm{q} \cdot \vec{v} + \frac{\bm{q}^2}{2 m_\chi} \right) = \frac{2 \pi^2}{N_0 |\bm{q}|}  \left( e^{-v_-^2 / v_0^2} - e^{-v_\text{esc}^2 / v_0^2} \right) ,
\ee
where $N_0$ is the velocity profile normalization factor defined in Eq.~\eqref{eq:N_0} 
and we have defined the velocities
 \be
    v_- \equiv \text{min} \left( v_\text{esc}, v_* \right)~~~,~~~v_*  \equiv \frac{1}{|\bm{q}|} \left( \bm{q} \cdot \vec{v}_e + \frac{\bm{q}^2}{2 m_\chi} + \omega \right)~~.
\ee
 Note that the squared sub-amplitude
in \Eq{eq:squared_sub_amplitude} only depends on the norm $\bm q^2$, so it can be taken outside of the velocity integral in 
\Eq{eq:K0_func}.

\vspace{1em}
\begin{center}
    \textit{ \normalsize Vector Mediated Models}
\end{center}
\vspace{1em}

We now consider DM $\chi$ coupled to a vector mediator $V_\mu$ with the most general dimension-four interactions
\begin{align}
    \LUV \supset \left( g_{\chi v} \, \bar{\chi} \gamma^\mu \chi + g_{\chi a} \, \bar{\chi} \gamma^{\mu} \gamma^5 \chi \right) V_\mu  ,
\end{align}
%
and, as in Sec.~\ref{subsec:scatter_scalar_mediator}, we first calculate the DM-mediator sub-amplitude $\mathcal{M}_{\chi V}^\mu$ using 
conventional QFT Feynman rules:
\begin{align}
    i \mathcal{M}_{\chi V}^\mu = \left( \frac{i}{q^2 - m_V^2} \right) \left( -\eta^{\mu \nu} + \frac{q^\mu q^\nu}{m_V^2} \right) \bar{u}_{s'}(\vec{p} - \bm{q}) \left[ g_{\chi v} \gamma_{\nu}  + i g_{\chi a} \gamma_\nu \gamma^5 \right] u_s(\vec{p}) \, .
\end{align}
Squaring and spin-averaging yields 
\begin{align}
    \frac{1}{2} \sum_{ss'} \mathcal{M}_{\chi V}^{* \mu} \mathcal{M}_{\chi V}^{\nu} 
    &\approx \left( \frac{1}{\vec{q}^2 + m_V^2} \right)^2 \Bigg\{ \; g_{\chi v}^2 \left[ - \bm{q}^2 \eta^{\mu \nu} + 2 \left( p^\mu {p'}^{\nu} + p^\nu {p'}^{\mu} \right) \right] + 4 i \, g_{\chi v} g_{\chi a} \epsilon^{\mu \nu \rho \lambda} p_\rho p'_\lambda \nonumber \\
    & \hspace{-5em} \hspace{7.5em} + g_{\chi a}^2 \left[ -(4 m_\chi^2 + \bm{q}^2) \eta^{\mu \nu} + 2 ( p^\mu {p'}^\nu + {p'}^\mu p^\nu ) + \frac{4 m_\chi^2 (2 m_V^2 + \bm{q}^2)}{m_V^4} q^\mu q^\nu \right] \Bigg\} \, ,
\end{align}
where, again, $p$ and $p^\prime$ are respectively the incoming and outgoing DM four momenta. Expanding to second order in $|\vec{q}|$ and $|\vec{p}|$ for each component gives
\be
\frac{1}{2} \sum_{ss'} \mathcal{M}_{\chi V}^{* \mu} \mathcal{M}_{\chi V}^{\nu} &\approx& \frac{1}{ (\bm{q}^2 + m_V^2)^2}  \Bigg\{
      g_{\chi v}^2 
        \begin{pmatrix} 
            4 m_\chi^2 & 2 m_\chi (2 p^i - q^i) \\ 
            2 m_\chi (2 p^i - q^i) & 4 p^i p^j - 2 (p^i q^j + q^i p^j) + \bm{q}^2 \delta^{ij}
        \end{pmatrix} 
        + 4 i g_{\chi v} g_{\chi a}  
        \begin{pmatrix}
            0 & -\epsilon^{i j k} p^j q^k \\ 
            \epsilon^{ijk} p^j q^k & m_\chi \epsilon^{ijk} q^k  
        \end{pmatrix}
        \nonumber \\[2ex]
   && +  g_{\chi a}^2 \left[
        \begin{pmatrix} 
            4 \vec{p}^2 - 4 \bm{q} \cdot \vec{p} + \bm{q}^2 & 2 m_\chi (2 p^i - q^i) \\
            2 m_\chi (2 p^i - q^i) & 4 m_\chi^2 \delta^{ij}
        \end{pmatrix}
        + \frac{4 m_\chi^2 (2 m_V^2 + \bm{q}^2)}{m_V^4} 
        \begin{pmatrix} 
             0 & 0 \\
             0 & q^i q^j 
        \end{pmatrix}
        \right]
    \Bigg\}~~,
    \label{eq:vector_matrix_element_sq_approx}
\ee
where we have condensed the $\mu \nu$ components to a matrix; the $00$ component is in the upper left, the $0i$, $i0$ components are in the upper right and lower left, respectively, the $ij$ components are in the lower right, and we have separated the anomalous contribution to the axial-vector current.

Substituting Eq.~\eqref{eq:vector_matrix_element_sq_approx} in to Eq.~\eqref{eq:DM_form_factor} we can compute the DM form factor, $\mathcal{F}^{\mu \nu}_{\chi V}$
\be
    \mathcal{F}_{\chi V}^{\mu \nu} &=& \frac{1}{(\bm{q}^2 + m_V^2)^2}
      \left\{ g_{\chi v}^2 
        \begin{pmatrix} 
            K_0 & \displaystyle K_1^i - \frac{q^i}{2 m_\chi} K_0 \\ 
            \displaystyle K_1^i - \frac{q^i}{2 m_\chi} K_0 & \displaystyle K_2^{ij} - \left( K^i_1 \frac{q^j}{2 m_\chi} + \frac{q^i}{2 m_\chi} K^j_1 \right) + \frac{\bm{q}^2}{4 m_\chi^2} \delta^{ij} \, K_0
        \end{pmatrix}  \right. \nonumber \\
        && \quad \quad \quad  \quad \quad \quad +  g_{\chi a}^2 
        \left[
        \begin{pmatrix} 
            \displaystyle K_2^{ii} - \frac{q^i}{m_\chi} K_1^i + \bm{q}^2 K_0 & \displaystyle K_1^i - \frac{q^i}{2 m_\chi} K_0 \\
            \displaystyle K_1^i - \frac{q^i}{2 m_\chi} K_0 &  \delta^{ij} \, K_0
        \end{pmatrix}
        + \frac{2 m_V^2 + \bm{q}^2}{m_V^4} 
        \begin{pmatrix} 
             0 & 0 \\
             0 & \displaystyle q^i q^j K_0 
        \end{pmatrix}
        \right] 
        \nonumber \\
    && \left. \quad \quad  \quad  \quad \quad  \quad +  2 i g_{\chi v} g_{\chi a}
        \begin{pmatrix} 
            0 & \displaystyle - \epsilon^{ijk} K_1^j \frac{q^k}{2 m_\chi} \\ 
            \displaystyle \epsilon^{ijk} K_1^j \frac{q^k}{2 m_\chi} & \displaystyle  \epsilon^{ijk} \frac{q^k}{2 m_\chi} \, K_0 
        \end{pmatrix} 
    \right\}~,
\ee
where the velocity integrals have been rewritten in terms the generalized kinematic functions
$K_{1}^{i}$ and $K_{2}^{ij}$, first derived in Ref.~\cite{Trickle:2020oki}, where $K^i_1$ satisfies
\begin{align}
     K_1^i(\bm{q}, \omega, \vec{v}_e) & = 2 \pi \int d^3 \vec{v} \, v^i \, f_{\chi}(\vec{v} + \vec{v}_e) \, \delta \left( \omega - \bm{q} \cdot \vec{v} + \frac{\bm{q}^2}{2 m_\chi} \right) =  \left( v_* \hat{q}^i - v_e^i \right) K_0(\bm{q}, \omega, \vec{v}_e) \, ,
\end{align}
and the $K_2^{ij}$ function can be written  
\be
    K_2^{ij}(\bm{q}, \omega, \vec{v}_e) &=& 2 \pi \int d^3 \vec{v} \, v^i v^j \, f_{\chi}(\vec{v} + \vec{v}_e) \, \delta \left( \omega - \bm{q} \cdot \vec{v} + \frac{\bm{q}^2}{2 m_\chi} \right) \nonumber \\ 
    &=& v_*^2 \hat{q}^i \hat{q}^j \, K_0(\bm{q}, \omega, \vec{v}_e) + \left( \delta^{ij} - \hat{q}^i \hat{q}^j \right) \left[ \frac{\pi^2 v_0^2}{|\bm{q}| N_0} \left( v_0^2 e^{- v_-^2 / v_0^2 } - (v_0^2 - v_-^2 + v_\text{esc}^2) e^{- v_\text{esc}^2 / v_0^2} \right) \right] \nonumber \\ 
    && \quad\quad - \left[ v_e^i \, K_1^j(\bm{q}, \omega, \vec{v}_e) + v_e^j \, K_1^i(\bm{q}, \omega, \vec{v}_e) \right] + v_e^i \, v_e^j \, K_0(\bm{q}, \omega, \vec{v}_e) \, .
\ee
%


\section{Dark Thomson Scattering}
\label{sec:dark_thomson_scattering}

\begin{figure}[t!]
    \centering
    \input{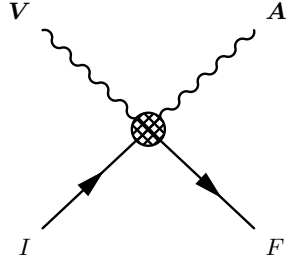}
    \vspace{1em}
    \caption{An example Feynman diagram for the dark Thomson scattering process.}
    \label{fig:thomson_feynman_diagram}
\end{figure}

The last process we consider is ``dark Thomson scattering," where an incoming DM vector particle $V$ inelastically scatters off an electron and converts into a photon, as depicted Fig.~\ref{fig:thomson_feynman_diagram}.
This process is the low-energy limit of dark Compton scattering considered in Ref.~\cite{Hochberg:2021zrf}. The interaction rate per incoming $V$ particle from \Eq{eq:Gamma_IF} can be written in terms of ${\cal M}_\text{NR}$ as
\begin{align}
    \Gamma_{\mathcal{I} \rightarrow \mathcal{F}} = \frac{2 \pi}{\langle \mathcal{I} | \mathcal{I} \rangle \langle \mathcal{F} | \mathcal{F} \rangle} \delta\left( \textstyle \sum E \right) \, |\mathcal{M}_{\text{NR}}|^2  \, .
\end{align}
Using the Feynman rules developed in Sec.~\ref{sec:feynman_rules} we can evaluate the NR matrix element for incoming and outgoing states given by,
\begin{align}
    | \mathcal{I} \rangle = | \vec{p}, \lambda \rangle \otimes | I \rangle~~~,~~~| \mathcal{F} \rangle = | \vec{p} - \vec{q}, \lambda' \rangle \otimes | F \rangle \, ,
\end{align}
where $\lambda$ is the polarization of the incoming DM particle and $\lambda'$ is the polarization of the outgoing photon. The total scattering rate of a polarization $\lambda$, per detector exposure, $R_\lambda$, is then found by summing over the electron and photon states, multiplying by the total number of DM particles in the detector, $N_V = ( \rho_V / m_V ) \vol$, and dividing by the detector mass, $\rho_T \vol$,
\begin{align}
    R_{\lambda} = \frac{\rho_V}{\rho_T m_V} \frac{2 \pi}{4 m_V \omega \vol} \sum_{IF} \sum_{\lambda'} \int \frac{d^3 \bm{q}}{(2 \pi)^3} \, \delta(\textstyle \sum E) \, |\mathcal{M}_\text{NR}|^2 \, .
\end{align} 
where $\omega = |\vec{p} - \vec{q}|$ is the energy of the outgoing photon and, as in Sec.~\ref{sec:scattering}, the sums over $I, F$ become integrals when the indices are continuous. Lastly, averaging over the incoming DM polarization gives the expected total scattering rate,
\begin{align}
    R = \frac{\rho_V}{\rho_T m_V} \frac{2 \pi}{4 m_V \omega \vol} \sum_{IF} \left( \frac{1}{3} \sum_{\lambda \lambda'} \right) \int \frac{d^3 \bm{q}}{(2 \pi)^3} \, \delta(\textstyle \sum E) \, |\mathcal{M}_\text{NR}|^2 \, .
    \label{eq:dt_rate_1}
\end{align} 
The NR matrix element for the dark Thomson process can then be computed with the Feynman rules from Sec.~\ref{sec:feynman_rules},
\begin{align}
    i \mathcal{M}_\text{NR} =  \bm{\epsilon}_{V, i}^{\lambda} \bm{\epsilon}_{A, j}^{\lambda'} \left( i \sum_{\sumIndex} \left[ \fFR_{V A} \right]^{ij} \MFR_{IF, \sumIndex} \right) \, ,
    \label{eq:thomson_matrix_element}
\end{align}
where $\bm{\epsilon}_{V}, \bm{\epsilon}_A$ are the DM and photon polarization vectors, respectively, and we have ignored the zero component contribution because it is higher order.

\vspace{1em}
\begin{center}
    \textit{ \normalsize A Dark Thomson Scattering Example}
\end{center}
\vspace{1em}

To provide an example calculation we consider dark Thomson scattering in a simple model and target. As our model we take the kinetically mixed dark photon ($g_v = - \kappa e$ in the representative UV Lagrangian in Sec.~\ref{sec:NR_DM_electron_interaction}). In this model the dominant Feynman rule coefficient contributing to dark Thomson scattering is
\begin{align}
    \left[ f_{V A, 1} \right]^{ij} = -\frac{\kappa \, e^2}{m_e} \delta^{ij} \, .
\end{align}

For the target we assume that the electrons are free ($\Phi = 0$), and the initial state is $N_e = n_e \vol$ electrons with momentum much less than $m_V$. This is equivalent to the approximations made in Ref.~\cite{Hochberg:2021zrf}, and the final scattering rate will match Ref.~\cite{Hochberg:2021zrf} in the limit of $m_V \ll m_e$. Therefore the electron states are indexed by momentum and spin quantum numbers,
\begin{align}
    |I \rangle & = | \vec{k}, s \rangle~~~,~~~\sum_I \rightarrow \sum_s \int \frac{d^3 \vec{k}}{(2 \pi)^3}~~~,~~~|F \rangle = | \vec{k}', s' \rangle~~~,~~~\sum_F \rightarrow \sum_{s'} \int \frac{d^3 \vec{k}'}{(2 \pi)^3} \, .
\end{align}
The relevant transition matrix element is $\MFR_{\vec{k}s \, \vec{k}'s',1}$
\begin{align}
    \MFR_{\vec{k}s \, \vec{k}'s', 1} = \delta_{ss'} (2 \pi)^3  \delta^3(\vec{q} - \vec{k}' + \vec{k}) \, ,
\end{align}
and upon substitution the rate in Eq.~\eqref{eq:dt_rate_1} becomes
\begin{align}
    R & = \frac{\rho_V}{\rho_T m_V} \frac{\kappa^2 e^4}{m_e^2} \frac{2 \pi}{4 m_V^2} \left( \frac{1}{3} \sum_{\lambda \lambda'} |\bm{\epsilon}_{V, \lambda} \cdot \bm{\epsilon}_{A, \lambda'} |^2 \right) \left( \sum_{ss'} \delta_{ss'} \right)\int \frac{d^3 \vec{k}}{(2 \pi)^3} \frac{d^3 \vec{k}'}{(2 \pi)^3} \delta(m_V - |\bm{k}'|) \, .
    \label{eq:dt_rate_2}
\end{align}
This can be further simplified using the polarization sum relationships from Eq.~\eqref{eq:polarization_vector_properties}
\begin{align}
       \sum_{\lambda} \epsilon_{V, \lambda}^i \epsilon_{V, \lambda}^j~\approx~\delta^{ij}~~~\implies~~~\sum_{\lambda \lambda'} |\bm{\epsilon}_{V, \lambda} \cdot \bm{\epsilon}_{A, \lambda'} |^2 \approx \sum_{\lambda'}  \bm{\epsilon}_{A, \lambda'} \cdot \bm{\epsilon}_{A, \lambda'} = 2 \, ,
\end{align}
which simplifies Eq.~\eqref{eq:dt_rate_2} to yield
\begin{align}
    R = \frac{n_e}{\rho_T} \frac{\rho_V}{m_V} \frac{e^4 \kappa^2}{6 \pi m_e} \, ,
    \label{eq:DT_simple_rate}
\end{align}
noting that the sum over initial electron states is simply their number density $n_e = 2 \int d^3 \vec{k} / (2 \pi)^3$.


\section{Conclusions}
\label{sec:conclusions}

Understanding DM-electron interactions in a variety of targets is necessary to maximize the discovery potential of current, and future, electron based direct detection experiments. 
In this work we formulate the NR EFT of DM-electron interactions as a tool to compute any DM-electron observable, in any target. The EFT is developed from a ``top-down" perspective, where the starting point is a Lagrangian defining DM-electron interactions at energies above the electron mass $m_e$.
The high-energy theory is then matched to a low-energy theory describing the interactions of the DM and NR electron field, which satisfies the Schr\"odinger equation (including corrections to an arbitrary order in $1/m_e$). The mapping between the high and low-energy electron fields, discussed in Sec.~\ref{sec:NRQED}, can be found by ``integrating out the positron" field which is done with the \FWspell method~\cite{PhysRev.78.29,PhysRev.87.688,drell,Gardestig:2007mk,Smith:2023htu} and maps QED to NR QED~\cite{Paz:2015uga}. In Sec.~\ref{sec:NR_DM_electron_interaction} this mapping is then applied to the interaction operators in the high-energy theory to find the NR EFT Lagrangian of DM-electron interactions.

While the NR EFT dictates the structure of the DM-electron interactions, it is still one step removed from the observables one wants to compute, e.g., absorption or scattering rates. Typically in relativistic QFT this connection is made by using Feynman rules and diagrams to compute a matrix element, which is then used to compute the observable. In Sec.~\ref{sec:feynman_rules} we develop analogous Feynman rules for the NR EFT. This allows any process to be computed diagrammatically which has the benefit of composability: more complicated diagrams and observables can be constructed from a few primitive vertices and propagators. These Feynman rules are then used to compute the DM absorption, scattering, and dark Thomson scattering rates for a wide variety of DM models in Secs.~\ref{sec:absorption},~\ref{sec:scattering}, and~\ref{sec:dark_thomson_scattering}, respectively. 

A major benefit of the formalism developed here is that the interaction rates of a given DM model can be directly applied to any electronic target. This is because the target electronic structure enters only through the evaluation of the Feynman rules. Furthermore we identified a basis of eight NR operators, Eqs.~\eqref{eq:ONR_basis1}~-~\eqref{eq:ONR_basis3}, which can determine the electronic response of any DM model to order $m_e^{-2}$, assuming the electron is in only a background electric potential. Extending the existing codes for DM-electron interaction rates in crystals (\textsf{EXCEED-DM}~\cite{Griffin:2021znd,Trickle:2022fwt}, \textsf{QCDark}~\cite{Dreyer:2023ovn}, \textsf{QEDark}~\cite{Essig:2015cda}, \textsf{QEDark-EFT}~\cite{Catena:2021qsr}), or atomic targets (\textsf{DarkART}~\cite{Catena:2019gfa}) to compute these eight transition matrix elements in Eq.~\eqref{eq:M_hat_definition}, in any kinematic regime, would allow them to compute all the DM-electron interaction rates discussed here. 

There are other interesting DM induced processes which deposit an NR amount of energy and momentum to the electron degrees of freedom, and therefore could be similarly generalized with the results of this paper. For example, relativistic absorption (e.g., absorption of solar axions~\cite{vanBibber:1988ge,Moriyama:1995bz,Redondo:2013wwa}), or even those which involve some nuclear degrees of freedom, e.g., the Migdal effect~\cite{Ibe:2017yqa,Dolan:2017xbu,Baxter:2019pnz,Essig:2019xkx,Liu:2020pat,Berghaus:2022pbu,Adams:2022zvg} or ``fermionic absorption"~\cite{Dror:2019dib,PandaX:2022osq,Ge:2022ius,Cox:2023cjw}. Given a term in the NR EFT which describes these interactions, similar Feynman rules may be analogously derived for those processes, and therefore generalized to any target. Additionally, an advantage of using an EFT for only the electron degrees of freedom is that the usual tools from relativistic QFT can be used to handle any other degrees of freedom. For example, if DM couples to quarks, and therefore only interacts with electrons via a loop~\cite{Diamond:2023fsm}, the DM-quark and mixing diagrams can be computed as usual, and one simply needs to replace the electron vertices with the Feynman rules developed here.


\acknowledgments

We would like to thank Kevin Zhou for helpful discussion.
This research was supported by Fermilab which is operated by the Fermi Research Alliance, LLC under Contract DE-AC02-07CH11359 with the U.S.~Department of Energy.  Part of this work was completed at the Aspen Center for Physics, which is supported by National Science Foundation grant PHY-2210452. This work is supported by the U.S. Department of Energy, Office of Science, Office of Workforce Development for Teachers and Scientists, Office of Science Graduate Student Research (SCGSR) program. The SCGSR program is administered by the Oak Ridge Institute for Science and Education for the DOE under contract number DE‐SC0014664.

\appendix

\section{NR Interaction Lagrangian Summary Tables}
\label{app:summary_tables}

\SetTblrInner{rowsep=5pt}
\begin{table}[ht!]
    \begin{tblr}{width=\textwidth,colspec={ X[-1,l] X[1,l] }} 
        \toprule
            \SetCell[c=2]{l}{ \textbf{High-Energy Interaction Lagrangian} } \\ 
            \phantom{\quad} & $\displaystyle \LUV_\text{int} = y_s \, \phi \, \bar{\Psi} \Psi$ \\
            [1ex]\midrule
            \SetCell[c=2]{l}{ \textbf{NR EFT Interaction Lagrangian} } \\
            & $\ds \LNR_\text{int} = \sum_{\ell} \left[ C_{\phi, \ell} \phi + C_{(\nabla \phi), \ell} \frac{(\nabla \phi)}{m_e}  + C_{(\nabla \nabla \phi), \ell} \frac{(\nabla \nabla \phi)}{m_e^2} \right.$ \\
            & $ \ds \hspace{7em} \left. + \, C_{\phi \Av, \ell} \frac{\phi \Av}{m_e} + C_{(\nabla \phi) \Av, \ell} \frac{(\nabla \phi) \Av}{m_e^2} + C_{\phi (\nabla \Av), \ell} \frac{\phi (\nabla \Av)}{m_e^2} + C_{\phi \Av \Av, \ell} \frac{\phi \Av \Av}{m_e^2} \right] \left[ \psi^\dagger \ONRBasis_\ell \psi \right]$ \\[1ex]
            & $\ds C_{\phi, 1} = y_s~~~,~~~ \left[C_{\phi,3}\right]^{ij} = \frac{y_s}{2} \delta^{ij} ~~~,~~~\left[ C_{(\nabla \phi), 2} \right]^{ij} = \frac{y_s}{2} \delta^{i,j}~~~,~~~\left[ C_{(\nabla \phi), 5} \right]^{i,jk} = \frac{-i y_s}{4} \epsilon^{ijk}~~~,~~~\left[ C_{(\nabla \nabla \phi), 1} \right]^{ij} = \frac{y_s}{8} \delta^{ij}$ \\
            & $\ds \left[ C_{\phi \Av, 2} \right]^{i,j} = - i e y_s \delta^{ij}~~~,~~~\left[ C_{(\nabla \phi) \Av, 1} \right]^{ij} = \frac{-ie y_s}{2} \delta^{ij}~~~,~~~\left[ C_{(\nabla \phi) \Av, 4} \right]^{ij,k} = \frac{e y_s}{4} \epsilon^{ijk}~~~,~~~\left[ C_{\phi (\nabla \Av), 1} \right]^{ij} = \frac{-ie y_s}{2} \delta^{ij}$\\
            & $\ds \left[ C_{\phi (\nabla \Av), 4} \right]^{ij,k} = \frac{e y_s}{2} \epsilon^{ijk} ~~~,~~~ \left[ C_{\phi \Av \Av} \right]^{ij} = - \frac{e^2 y_s}{2} \delta^{ij}$\\[1ex]
            \midrule
            \SetCell[c=2]{l}{ \textbf{NR Feynman Rules} } & \\ 
            & \vtop{\vspace{-1em}
                    \begin{align*}
                        \begin{gathered}
                            \scalebox{0.8}{\begin{fmffile}{3pt_scalar}
    \begin{fmfgraph*}(100,75)
        \fmfleft{i} \fmfright{f1,f2}
        \fmf{dashes,label=$\mArrow{0.4cm}{q}$,label.side=left}{i,v}
        \fmf{fermion}{v,f2}
        \fmf{fermion}{f1,v}
        \fmfv{decor.shape=circle,decor.filled=hatched,decor.size=0.125w}{v}
        \fmflabel{$\phi$}{i}
        \fmflabel{$J$}{f1}
        \fmflabel{$K$}{f2}
    \end{fmfgraph*}
\end{fmffile}}
                        \end{gathered} \quad\quad\quad \Longrightarrow \quad\quad\quad i \sum_{\ell = 1}^8 f_{\phi, \ell}(\bm{q}) \; \hat{\mathcal{M}}_{JK, \ell}(\bm{q})  \, .
                    \end{align*}
                    \vspace{-1em}
                } \\
            & \vtop{
                \begin{align*}
                    f_{\phi, 1} & = y_s \left(1 - \frac{\vec{q}^2}{8 m_e^2} \right)~~~,~~~\left[ f_{\phi, 2} \right]^i = \frac{i y_s}{2} \frac{q^i}{m_e}~~~,~~~\left[ f_{\phi, 3} \right]^{ij} = \frac{ y_s}{2} \delta^{ij} ~~~,~~~\left[ f_{\phi, 5} \right]^{ij} = \frac{y_s}{4} \epsilon^{ijk} \frac{q^k}{m_e} 
                \end{align*}
            } \\
            \cmidrule{2}
            & \vtop{ 
                    \begin{align*}
                        \begin{gathered}
                            \scalebox{0.8}{\begin{fmffile}{4pt_scalar_vector}
    \begin{fmfgraph*}(100,75)
        \fmfleft{i1,i2} \fmfright{f1,f2}
        \fmf{dashes,label=$\rotatebox{-45}{ $\xrightarrow{\makebox[0.4cm]{}}$ }$,label.dist=-0.08w}{i2,v}
        \fmf{fermion}{i1,v}
        \fmf{fermion}{v,f1}
        \fmf{photon,label=$\rotatebox{45}{ $\xrightarrow{\makebox[0.4cm]{}}$ }$,label.dist=-0.08w,label.side=right}{v,f2}
        \fmflabel{$\phi$}{i2}
        \fmflabel{$A^\mu$}{f2}
        \fmflabel{$J$}{i1}
        \fmflabel{$K$}{f1}
        \fmfv{decor.shape=circle,decor.filled=hatched,decor.size=0.125w}{v}
        \fmffreeze
        \fmfforce{0.5*vloc(__v)+0.5*vloc(__i2) shifted (-0.75mm,-4mm)}{l}
        \fmflabel{$\vec{p}_1$}{l}
        \fmfforce{0.5*vloc(__v)+0.5*vloc(__f2) shifted (+1mm,-4mm)}{l2}
        \fmflabel{$\vec{p}_2$}{l2}
    \end{fmfgraph*}
\end{fmffile}}
                        \end{gathered} \quad\quad\quad \Longrightarrow \quad\quad\quad i \sum_{\ell = 1}^8 \left[ f_{\phi A, \ell} (\vec{p}_1, \vec{p}_2) \right]^\mu \hat{\mathcal{M}}_{JK, \ell}(\vec{p}_1 - \vec{p}_2) \, . 
                    \end{align*}
                    \vspace{-1em}
            } \\
            & \vtop{
                \begin{align*}
                    \left[ f_{\phi A, 1} \right]^i = - \frac{e y_s}{2 m_e^2} (p_1^i - p_2^i)~~~,~~~\left[ f_{\phi A, 2} \right]^{i,j} = \frac{i e y_s}{m_e} \delta^{ij}~~~,~~~\left[ f_{\phi A, 4} \right]^{i,j} = - \frac{i e y_s}{4m_e^2} \epsilon^{ijk} (p_1^k - 2 p_2^k)
                \end{align*}
            } \\
        \bottomrule
    \end{tblr}
    \caption{Summary of the NR EFT Interaction Lagrangian and Feynman rules generated by the UV interaction Lagrangian $\LUV_\text{int} = y_s \phi \bar{\Psi} \Psi$. See Sec.~\ref{sec:feynman_rules} to see how to find the $f$'s in terms of the $C$'s.}
    \label{tab:scalar_summary_table}
\end{table}

\clearpage

\SetTblrInner{rowsep=5pt}
\begin{table}[ht!]
    \begin{tblr}{width=\textwidth,colspec={ X[-1,l] X[1,l] }} 
        \toprule
            \SetCell[c=2]{l}{ \textbf{High-Energy Interaction Lagrangian} } \\ 
            \phantom{\quad} & $\ds \LUV_\text{int} = i y_p \, \phi \, \bar{\Psi} \gamma^5 \Psi$ \\
            [1ex]\midrule
            \SetCell[c=2]{l}{ \textbf{NR EFT Interaction Lagrangian} } \\
            & $\ds \LNR_\text{int} = \sum_{\ell} \left[ C_{\phi, \ell} \phi + C_{(\nabla \phi), \ell} \frac{(\nabla \phi)}{m_e} + C_{\phi (\partial_t \Av),\ell} \frac{\phi (\partial_t \Av)}{m_e^2} + C_{\phi (\nabla A^0),\ell} \frac{\phi (\nabla A^0)}{m_e^2} \right] \left[ \psi^\dagger \ONRBasis_{\ell} \psi \right]$ \\[1ex]
            & $\ds \left[ C_{\phi, 8} \right]^{ij} = - \frac{y_p}{2} \delta^{ij}~~~,~~~\left[ C_{(\nabla \phi), 4} \right]^{i,j} = - \frac{y_p}{2} \delta^{ij}~~~,~~~\left[ C_{\phi (\partial_t \Av), 4} \right]^{i,j} = - \frac{e y_p}{2} \delta^{ij}~~~,~~~\left[ C_{\phi (\nabla A^0), 4} \right]^{i,j} = - \frac{e y_p}{2} \delta^{ij}$ \\[1ex]
            \midrule
            \SetCell[c=2]{l}{ \textbf{NR Feynman Rules} } & \\ 
            & \vtop{\vspace{-1em}
                    \begin{align*}
                        \begin{gathered}
                            \scalebox{0.8}{\begin{fmffile}{3pt_scalar}
    \begin{fmfgraph*}(100,75)
        \fmfleft{i} \fmfright{f1,f2}
        \fmf{dashes,label=$\mArrow{0.4cm}{q}$,label.side=left}{i,v}
        \fmf{fermion}{v,f2}
        \fmf{fermion}{f1,v}
        \fmfv{decor.shape=circle,decor.filled=hatched,decor.size=0.125w}{v}
        \fmflabel{$\phi$}{i}
        \fmflabel{$J$}{f1}
        \fmflabel{$K$}{f2}
    \end{fmfgraph*}
\end{fmffile}}
                        \end{gathered} \quad\quad\quad \Longrightarrow \quad\quad\quad i \sum_{\ell = 1}^8 f_{\phi, \ell}(\bm{q}) \; \hat{\mathcal{M}}_{JK, \ell}(\bm{q})  \, .
                    \end{align*}
                    \vspace{-1em}
                } \\
            & \vtop{
                \begin{align*}
                    \left[ f_{\phi, 4} \right]^{i} = - \frac{i y_p}{2} \frac{q^i}{m_e}~~~,~~~\left[ f_{\phi, 8} \right]^{ij} = - \frac{y_p}{2} \delta^{ij}
                \end{align*}
            } \\
            \cmidrule{2}
            & \vtop{ 
                    \begin{align*}
                        \begin{gathered}
                            \scalebox{0.8}{\begin{fmffile}{4pt_scalar_vector}
    \begin{fmfgraph*}(100,75)
        \fmfleft{i1,i2} \fmfright{f1,f2}
        \fmf{dashes,label=$\rotatebox{-45}{ $\xrightarrow{\makebox[0.4cm]{}}$ }$,label.dist=-0.08w}{i2,v}
        \fmf{fermion}{i1,v}
        \fmf{fermion}{v,f1}
        \fmf{photon,label=$\rotatebox{45}{ $\xrightarrow{\makebox[0.4cm]{}}$ }$,label.dist=-0.08w,label.side=right}{v,f2}
        \fmflabel{$\phi$}{i2}
        \fmflabel{$A^\mu$}{f2}
        \fmflabel{$J$}{i1}
        \fmflabel{$K$}{f1}
        \fmfv{decor.shape=circle,decor.filled=hatched,decor.size=0.125w}{v}
        \fmffreeze
        \fmfforce{0.5*vloc(__v)+0.5*vloc(__i2) shifted (-0.75mm,-4mm)}{l}
        \fmflabel{$\vec{p}_1$}{l}
        \fmfforce{0.5*vloc(__v)+0.5*vloc(__f2) shifted (+1mm,-4mm)}{l2}
        \fmflabel{$\vec{p}_2$}{l2}
    \end{fmfgraph*}
\end{fmffile}}
                        \end{gathered} \quad\quad\quad \Longrightarrow \quad\quad\quad i \sum_{\ell = 1}^8 \left[ f_{\phi A, \ell} (\vec{p}_1, \vec{p}_2) \right]^\mu \hat{\mathcal{M}}_{JK, \ell}(\vec{p}_1 - \vec{p}_2) \, . 
                    \end{align*}
                    \vspace{-1em}
            } \\
            & \vtop{
                \begin{align*}
                    \left[ f_{\phi A, 4} \right]^{0, i} = \frac{i e y_p}{2 m_e} \frac{p_2^i}{m_e}~~~,~~~\left[ f_{\phi A, 4} \right]^{i,j} = \frac{i e y_p}{2 m_e} \frac{\omega_2}{m_e} \delta^{ij} 
                \end{align*}
            } \\
        \bottomrule
    \end{tblr}
    \caption{Summary of the NR EFT Interaction Lagrangian and Feynman rules generated by the UV interaction Lagrangian $\LUV_\text{int} = i y_p \phi \bar{\Psi} \gamma^5 \Psi$. See Sec.~\ref{sec:feynman_rules} to see how to find the $f$'s in terms of the $C$'s.}
    \label{tab:ps_summary_table}
\end{table}

\clearpage

\SetTblrInner{rowsep=5pt}
\begin{table}[ht!]
    \begin{tblr}{width=\textwidth,colspec={ X[-1,l] X[1,l] }} 
        \toprule
            \SetCell[c=2]{l}{ \textbf{High-Energy Interaction Lagrangian} } \\ 
            \phantom{\quad} & $\ds \LUV_\text{int} = g_v \, V_\mu \, \bar{\Psi} \gamma^\mu \Psi$ \\
            [1ex]\midrule
            \SetCell[c=2]{l}{ \textbf{NR EFT Interaction Lagrangian} } \\
            & $\ds \LNR_\text{int} = \sum_\ell \left[ C_{V^0, \ell} V^0 + C_{(\nabla V^0), \ell} \frac{( \nabla V^0 )}{m_e} + C_{(\nabla \nabla V^0), \ell} \frac{(\nabla \nabla V^0)}{m_e^2} + C_{(\nabla V^0 \Av), \ell} \frac{(\nabla V^0) \Av}{m_e^2} \right.$ \\
            & $ \ds \left. \hspace{7.5em} + \, C_{\Vv, \ell} \Vv + C_{(\nabla \Vv), \ell} \frac{(\nabla \Vv)}{m_e} + C_{\Vv \Av, \ell} \frac{\Vv \Av}{m_e} + C_{\Vv (\partial_t \Av), \ell} \frac{ \Vv (\partial_t \Av)}{m_e^2} + C_{\Vv (\nabla A^0), \ell} \frac{\Vv (\nabla A^0)}{m_e^2} \right] \left[ \psi^\dagger \ONRBasis_{\ell}  \psi \right]$ \\[1ex]
            & $\ds C_{V^0, 1} = g_v~~~,~~~\left[ C_{(\nabla V^0), 5} \right]^{i, jk} = \frac{i g_v}{4} \epsilon^{ijk}~~~,~~~\left[ C_{(\nabla \nabla V^0), 1} \right]^{ij} = \frac{g_v}{8} \delta^{ij}~~~,~~~\left[ C_{(\nabla V^0) \Av, 4} \right]^{ij, k} = - \frac{e g_v}{4} \epsilon^{ijk}$ \\
            & $\ds \left[ C_{(\nabla \Vv), 1} \right]^{i,j} = \frac{i g_v}{2} \delta^{ij} ~~~,~~~ \left[ C_{(\nabla \Vv), 4} \right]^{ij,k} = - \frac{g_v}{2} \epsilon^{ijk} ~~~,~~~ \left[ C_{\Vv, 2} \right]^{i,j} = i g_v \delta^{ij}~~~,~~~\left[ C_{\Vv, 8} \right]^{i,jk} = - \frac{g_v}{2} \epsilon^{ijk}$\\
            & $ \ds \left[ C_{\Vv \Av, 1} \right]^{ij} = e g_v \delta^{ij}~~~,~~~\left[ C_{\Vv (\partial_t \Av),4} \right]^{ij,k} = \frac{e g_v}{2} \epsilon^{ijk} ~~~,~~~\left[ C_{\Vv (\nabla A^0),4} \right]^{ij,k} = \frac{e g_v}{2} \epsilon^{ijk}$ \\
            [1ex]\midrule
            \SetCell[c=2]{l}{ \textbf{NR Feynman Rules} } \\ 
            & \vtop{\vspace{-1em} 
                    \begin{align*}
                         \begin{gathered}
                            \scalebox{0.8}{\begin{fmffile}{3pt_vector}
    \begin{fmfgraph*}(100,75)
        \fmfleft{i} \fmfright{f1,f2}
        \fmf{photon,label=$\mArrow{0.4cm}{q}$,label.side=left}{i,v}
        \fmf{fermion}{v,f2}
        \fmf{fermion}{f1,v}
        \fmfv{decor.shape=circle,decor.filled=hatched,decor.size=0.125w}{v}
        \fmflabel{$V^\mu$}{i}
        \fmflabel{$J$}{f1}
        \fmflabel{$K$}{f2}
    \end{fmfgraph*}
\end{fmffile}}
                        \end{gathered} \quad\quad\quad \Longrightarrow \quad\quad\quad i \sum_{\ell} \left[ f_{V, \ell}(\bm{q}) \right]^\mu \hat{\mathcal{M}}_{JK, \ell}(\bm{q})  \, .
                    \end{align*}
                    \vspace{-1em} 
                } \\
            & \vtop{
                \begin{align*}
                    & f_{V, 1}^0 = g_v \left( 1 - \frac{\vec{q}^2}{8 m_e^2} \right)~~~,~~~\left[ f_{V, 5} \right]^{0,ij} = - \frac{g_v}{4} \epsilon^{ijk} \frac{q^k}{m_e} \\ 
                    & \left[ f_{V, 1} \right]^{i} = \frac{g_v}{2m_e} q^i~~~,~~~\left[ f_{V, 2} \right]^{i, j} = -i g_v \delta^{ij}~~~,~~~\left[ f_{V, 4} \right]^{i, j} = \frac{i g_v}{2m_e} \epsilon^{ijk} q^k~~~,~~~\left[ f_{V, 8} \right]^{i, jk} =  \frac{g_v}{2} \epsilon^{ijk} 
                \end{align*}
            } \\
            \cmidrule{2}
            & \vtop{ 
                    \begin{align*}
                        \begin{gathered}
                            \scalebox{0.8}{\begin{fmffile}{4pt_vector_vector}
    \begin{fmfgraph*}(100,75)
        \fmfleft{i1,i2} \fmfright{f1,f2}
        \fmf{photon,label=$\rotatebox{-45}{ $\xrightarrow{\makebox[0.4cm]{}}$ }$,label.dist=-0.08w}{i2,v}
        \fmf{fermion}{i1,v}
        \fmf{fermion}{v,f1}
        \fmf{photon,label=$\rotatebox{45}{ $\xrightarrow{\makebox[0.4cm]{}}$ }$,label.dist=-0.08w,label.side=right}{v,f2}
        \fmflabel{$V^\mu$}{i2}
        \fmflabel{$A^\nu$}{f2}
        \fmflabel{$J$}{i1}
        \fmflabel{$K$}{f1}
        \fmfv{decor.shape=circle,decor.filled=hatched,decor.size=0.125w}{v}
        \fmffreeze
        \fmfforce{0.5*vloc(__v)+0.5*vloc(__i2) shifted (-0.75mm,-4mm)}{l}
        \fmflabel{$\vec{p}_1$}{l}
        \fmfforce{0.5*vloc(__v)+0.5*vloc(__f2) shifted (+1mm,-4mm)}{l2}
        \fmflabel{$\vec{p}_2$}{l2}
    \end{fmfgraph*}
\end{fmffile}}
                        \end{gathered} \quad\quad\quad \Longrightarrow \quad\quad\quad i \sum_{\ell} \left[ f_{V A, \ell} (\vec{p}_1, \vec{p}_2) \right]^{\mu \nu} \hat{\mathcal{M}}_{JK, \ell}(\vec{p}_1 - \vec{p}_2) \, . 
                    \end{align*}
            } \\
            & \vtop{
                \begin{align*}
                    & \left[ f_{V A, 4} \right]^{0i, j} = \frac{i e g_v}{4 m_e} \epsilon^{ijk} \frac{p_1^k}{m_e}~~~,~~~ \left[ f_{VA, 4} \right]^{i0, j} = -\frac{i e g_v}{2 m_e} \epsilon^{ijk} \frac{p_2^k}{m_e} \\
                    & \left[ f_{VA, 1} \right]^{ij} = \frac{e g_v}{m_e} \delta^{ij}~~~,~~~\left[ f_{VA, 4} \right]^{ij, k} = \frac{i e g_v}{2 m_e} \frac{\omega_2}{m_e} \epsilon^{ijk}
                \end{align*}
            } \\
        \bottomrule
    \end{tblr}
    \caption{Summary of the NR EFT Interaction Lagrangian and Feynman rules generated by the UV interaction Lagrangian $\LUV_\text{int} = g_v V_\mu \bar{\Psi} \gamma^\mu \Psi$. See Sec.~\ref{sec:feynman_rules} to see how to find the $f$'s in terms of the $C$'s.}
    \label{tab:vector_summary_table}
\end{table}

\clearpage 

\SetTblrInner{rowsep=5pt}
\begin{table}[ht!]
    \begin{tblr}{width=\textwidth,colspec={ X[-1,l] X[1,l] }} 
        \toprule
            \SetCell[c=2]{l}{ \textbf{High-Energy Interaction Lagrangian} } \\ 
            \phantom{\quad} & $\ds \LUV_\text{int} = g_a \, V_\mu \, \bar{\Psi} \gamma^\mu \gamma^5 \Psi$ \\
            \midrule
            \SetCell[c=2]{l}{ \textbf{NR EFT Interaction Lagrangian} } \\
            & $\ds \LNR_\text{int} = \sum_{\ell} \left[ C_{V^0, \ell} V^0 + C_{(\nabla V^0), \ell} \frac{(\nabla V^0)}{m_e} + C_{V^0 \Av, \ell} \frac{V^0 \Av}{m_e} + C_{\Vv, \ell} \Vv + C_{(\nabla \Vv), \ell} \frac{(\nabla \Vv)}{m_e} + C_{\Vv \Av, \ell} \frac{\Vv \Av}{m_e} \right.$ \\
            & $\ds \left. \hspace{6em} + \, C_{(\nabla \nabla \Vv), \ell} \frac{(\nabla \nabla \Vv)}{m_e^2} + C_{(\nabla \Vv) \Av, \ell} \frac{(\nabla \Vv) \Av}{m_e^2} + C_{\Vv (\nabla \Av), \ell} \frac{\Vv (\nabla \Av)}{m_e^2} + C_{\Vv \Av \Av, \ell} \frac{\Vv \Av \Av}{m_e^2} \right] \left[ \psi^\dagger \ONRBasis_{\ell} \psi \right]$ \\[1ex]
            & $\ds [C_{V^0, 5}]^{ij} = -i g_a \delta^{ij}~~~,~~~[C_{(\nabla V^0), 4}]^{i, j} = - \frac{i g_a}{2} \delta^{ij}~~~,~~~[C_{V^0 \Av,4}]^{i, j} = - e g_a \delta^{ij}$ \\
            & $\ds[C_{\Vv, 4}]^{i, j} = - g_a \delta^{ij}~~~,~~~[C_{\Vv, 6}]^{i,jkl} = \frac{g_a}{2} H_1^{ijkl}~~~,~~~[C_{\nabla \Vv, 2}]^{ij,k} = \frac{i g_a}{4} \epsilon^{ijk}~~~,~~~[C_{\nabla \Vv, 5}]^{ij,kl} = \frac{g_a}{4} H_2^{ijkl}$ \\
            & $\ds [C_{\Vv \Av, 5}]^{ij, kl} = - \frac{i e g_a}{2} H_2^{ijkl}~~~,~~~[C_{(\nabla \nabla \Vv), 4}]^{ijk,l} = - \frac{g_a}{8} \delta^{ij} \delta^{kl}~~~,~~~[C_{(\nabla \Vv) \Av, 1}]^{ijk} = \frac{e g_a}{4} \epsilon^{ijk}$ \\
            & $ \ds [C_{(\nabla \Vv) \Av, 4}]^{ijk,l} = -\frac{i e g_a}{4} H_2^{ijlk}~~~,~~~[C_{\Vv (\nabla \Av), 1}]^{ijk} = - \frac{e g_a}{2} \epsilon^{ijk}~~~,~~~[C_{\Vv (\nabla \Av), 4}]^{ijk,l} = -\frac{i e g_a}{4} H_2^{ijkl}$\\
            & $\ds [C_{\Vv\Av\Av, 4}]^{ijk,l} = - \frac{e^2g_a}{2} H_1^{ijlk}$ \\
            [1ex]\midrule
            \SetCell[c=2]{l}{ \textbf{NR Feynman Rules} } \\ 
            & \vtop{\vspace{-2em} 
                    \begin{align*}
                         \begin{gathered}
                            \scalebox{0.8}{\begin{fmffile}{3pt_vector}
    \begin{fmfgraph*}(100,75)
        \fmfleft{i} \fmfright{f1,f2}
        \fmf{photon,label=$\mArrow{0.4cm}{q}$,label.side=left}{i,v}
        \fmf{fermion}{v,f2}
        \fmf{fermion}{f1,v}
        \fmfv{decor.shape=circle,decor.filled=hatched,decor.size=0.125w}{v}
        \fmflabel{$V^\mu$}{i}
        \fmflabel{$J$}{f1}
        \fmflabel{$K$}{f2}
    \end{fmfgraph*}
\end{fmffile}}
                        \end{gathered} \quad\quad\quad \Longrightarrow \quad\quad\quad i \sum_{\ell} \left[ f_{V, \ell}(\bm{q}) \right]^{\mu} \hat{\mathcal{M}}_{JK, \ell}(\bm{q})  \, .
                    \end{align*}
                } \\
            & \vtop{\vspace{-2.5em}
                \begin{align*}
                    & \left[ f_{V, 4} \right]^{0, i} = \frac{g_a}{2} \frac{q^i}{m_e}~~~,~~~\left[ f_{V, 5} \right]^{0, ij} = - i g_a \delta^{ij} \\ 
                    & \left[ f_{V, 2} \right]^{i,j} = \frac{g_a}{4} \epsilon^{ijk} \frac{q^k}{m_e}~~~,~~~[ f_{V, 4} ]^{i,j} = g_a \delta^{ij} \left(1 - \frac{\vec{q}^2}{8 m_e^2} \right)~~~,~~~[f_{V,5} ]^{i, jk} = - \frac{i g_a}{4} H_2^{lijk} \frac{q^l}{m_e}~~~,~~~[f_{V, 6}]^{i, jkl} = -\frac{g_a}{2} H_1^{ijkl} 
                \end{align*}
            } \\
            \cmidrule{2}
            & \vtop{ 
                    \begin{align*}
                        \begin{gathered}
                            \scalebox{0.8}{\begin{fmffile}{4pt_vector_vector}
    \begin{fmfgraph*}(100,75)
        \fmfleft{i1,i2} \fmfright{f1,f2}
        \fmf{photon,label=$\rotatebox{-45}{ $\xrightarrow{\makebox[0.4cm]{}}$ }$,label.dist=-0.08w}{i2,v}
        \fmf{fermion}{i1,v}
        \fmf{fermion}{v,f1}
        \fmf{photon,label=$\rotatebox{45}{ $\xrightarrow{\makebox[0.4cm]{}}$ }$,label.dist=-0.08w,label.side=right}{v,f2}
        \fmflabel{$V^\mu$}{i2}
        \fmflabel{$A^\nu$}{f2}
        \fmflabel{$J$}{i1}
        \fmflabel{$K$}{f1}
        \fmfv{decor.shape=circle,decor.filled=hatched,decor.size=0.125w}{v}
        \fmffreeze
        \fmfforce{0.5*vloc(__v)+0.5*vloc(__i2) shifted (-0.75mm,-4mm)}{l}
        \fmflabel{$\vec{p}_1$}{l}
        \fmfforce{0.5*vloc(__v)+0.5*vloc(__f2) shifted (+1mm,-4mm)}{l2}
        \fmflabel{$\vec{p}_2$}{l2}
    \end{fmfgraph*}
\end{fmffile}}
                        \end{gathered} \quad\quad\quad \Longrightarrow \quad\quad\quad i \sum_{\ell} \left[ f_{V A, \ell}(\vec{p}_1, \vec{p}_2) \right]^{\mu \nu} \hat{\mathcal{M}}_{JK, \ell}(\vec{p}_1 - \vec{p}_2) \, . 
                    \end{align*}  
            } \\
            & \vtop{\vspace{-2.5em}
                \begin{align*}
                    & \left[ f_{VA, 4} \right]^{0i,j} = \frac{e g_a}{m_e} \delta^{ij} \\ 
                    & \left[ f_{VA, 1} \right]^{ij} = \frac{i e g_a}{4 m_e^2} \epsilon^{ijk} (p_1^k - p_2^k)~~~,~~~[f_{VA, 4}]^{ij, k} =  \frac{e g_a}{4 m_e^2} \left( p_1^l - p_2^l \right) H_2^{iljk}~~~,~~~[f_{VA, 5}]^{ij, kl} = - \frac{i e g_a}{2 m_e} H_2^{ijlk} 
                \end{align*}
            } \\
        \bottomrule
    \end{tblr}
    \caption{Summary of the NR EFT Interaction Lagrangian and Feynman rules generated by the UV interaction Lagrangian $\LUV_\text{int} = g_a V_\mu \bar{\Psi} \gamma^\mu \gamma^5 \Psi$. See Sec.~\ref{sec:feynman_rules} to see how to find the $f$'s in terms of the $C$'s. Furthermore for readability we have defined, $H_1^{ijkl} \equiv \delta^{ik} \delta^{jl} - \delta^{ij} \delta^{kl}$, $H^{ijkl}_2 \equiv \delta^{ij} \delta^{kl} + \delta^{ik} \delta^{jl} - 2 \delta^{il} \delta^{jk}$.}
    \label{tab:axial_vector_summary_table}
\end{table}

\clearpage

\SetTblrInner{rowsep=5pt}
\begin{table*}
    \begin{tblr}{width=\textwidth,colspec={ X[-1,l] X[1,l] }} 
        \toprule
            \SetCell[c=2]{l}{ \textbf{NR QED Interaction Lagrangian} } \\
            \phantom{\quad} & $\ds \LNR_\text{int} = \sum_{\ell} \left[ C_{A_0, \ell} A^0 + C_{\Av, \ell} \Av + C_{(\nabla A^0), \ell} \frac{(\nabla A^0)}{m_e} + C_{(\partial_t \Av), \ell} \frac{(\partial_t \Av)}{m_e} + C_{(\nabla \Av), \ell} \frac{(\nabla \Av)}{m_e} + C_{\Av \Av, \ell} \frac{\Av \Av}{m_e} \right.$ \\
            & $\ds \left. \hspace{7.5em} + \, C_{(\partial_t \nabla \Av), \ell} \frac{(\partial_t \nabla \Av)}{m_e^2} + C_{(\nabla \nabla A^0), \ell} \frac{(\nabla \nabla A^0)}{m_e^2} + C_{(\nabla A^0) \Av, \ell} \frac{(\nabla A^0) \Av}{m_e^2} \,\right] \left[ \psi^\dagger \ONRBasis_{\ell} \psi \right] $ \\[1ex]
            & $\ds C_{A_0, 1} = -e~~~,~~~[C_{\Av, 2}]^{i,j} = - i e\delta^{ij}~~~,~~~[C_{\Av, 8}]^{i, jk} = \frac{e}{4} \epsilon^{ijk}~~~,~~~[C_{(\nabla A^0),5}]^{i, jk} = - \frac{ie}{4} \epsilon^{ijk}~~~,~~~[C_{(\partial_t \Av), 5}]= - \frac{ie}{4} \epsilon^{ijk}$ \\
            & $\ds [C_{(\nabla \Av), 1}]^{ij} = - \frac{ie}{2} \delta^{ij} ~~~,~~~[C_{(\nabla \Av), 4}]^{ij,k} =\frac{e}{2} \epsilon^{ijk}~~~,~~~[C_{\Av \Av, 1}]^{ij} = - \frac{e^2}{2} \delta^{ij}~~~,~~~[C_{(\partial_t \nabla \Av), 1}]^{ij} = - \frac{e}{8} \delta^{ij}$ \\
            & $\ds [C_{(\partial_t \nabla \Av), 4}]^{ij,k} = - \frac{i e}{8} \epsilon^{ijk}~~~,~~~[C_{(\nabla \nabla A^0), 1}]^{ij} = - \frac{e}{8} \delta^{ij}~~~,~~~[C_{(\nabla A^0) \Av, 4}]^{ij, k} = \frac{e^2}{4} \epsilon^{ijk}$ \\
            [1ex]\midrule
            \SetCell[c=2]{l}{ \textbf{NR Feynman Rules} } \\ 
            & \vtop{ 
                    \begin{align*}
                        \begin{gathered}
                            \scalebox{0.8}{\begin{fmffile}{3pt_A}
    \begin{fmfgraph*}(100,75)
        \fmfleft{i} \fmfright{f1,f2}
        \fmf{photon,label=$\mArrow{0.4cm}{q}$,label.side=left}{i,v}
        \fmf{fermion}{v,f2}
        \fmf{fermion}{f1,v}
        \fmfv{decor.shape=circle,decor.filled=hatched,decor.size=0.125w}{v}
        \fmflabel{$A^\mu$}{i}
        \fmflabel{$J$}{f1}
        \fmflabel{$K$}{f2}
    \end{fmfgraph*}
\end{fmffile}}
                        \end{gathered} \quad\quad\quad \Longrightarrow \quad\quad\quad i \sum_{\ell = 1}^8 \left[ f_{A, \ell}(\bm{q})\right]^\mu \hat{\mathcal{M}}_{JK, \ell}(\bm{q})  \, .
                    \end{align*}
                } \\
            & \vtop{
                \begin{align*}
                    & \left[ f_{A, 1} \right]^0 = -e \left(1 - \frac{\vec{q}^2}{8 m_e^2} \right)~~~,~~~\left[ f_{A, 5}\right]^{0,ij} = \frac{e}{4} \epsilon^{ijk} \frac{q^k}{m_e} \\ 
                    & \left[ f_{A, 1} \right]^i = \frac{e}{2} \frac{q^i}{m_e} \left( -1 + \frac{\omega}{4 m_e} \right)~~~,~~~[f_{A, 2}]^{i, j} = i e \delta^{ij}~~~,~~~\left[ f_{A, 4} \right]^{i, j} = \frac{i e}{2} \epsilon^{ijk} \frac{q^k}{m_e} \left( -1 + \frac{\omega}{4 m_e} \right)~~~,~~~[f_{A, 5}]^{i, jk} = \frac{e\omega}{4 m_e} \epsilon^{ijk} \\ 
                    & \left[ f_{A, 8} \right]^{i, jk} = - \frac{e}{4} \epsilon^{ijk}
                \end{align*}
            } \\
            \cmidrule{2}
            & \vtop{ 
                    \begin{align*}
                        \begin{gathered}
                            \scalebox{0.8}{\begin{fmffile}{4pt_A_A}
    \begin{fmfgraph*}(100,75)
        \fmfleft{i1,i2} \fmfright{f1,f2}
        \fmf{photon,label=$\rotatebox{-45}{ $\xrightarrow{\makebox[0.4cm]{}}$ }$,label.dist=-0.08w}{i2,v}
        \fmf{fermion}{i1,v}
        \fmf{fermion}{v,f1}
        \fmf{photon,label=$\rotatebox{45}{ $\xrightarrow{\makebox[0.4cm]{}}$ }$,label.dist=-0.08w,label.side=right}{v,f2}
        \fmflabel{$A^\mu$}{i2}
        \fmflabel{$A^\nu$}{f2}
        \fmflabel{$J$}{i1}
        \fmflabel{$K$}{f1}
        \fmfv{decor.shape=circle,decor.filled=hatched,decor.size=0.125w}{v}
        \fmffreeze
        \fmfforce{0.5*vloc(__v)+0.5*vloc(__i2) shifted (-0.75mm,-4mm)}{l}
        \fmflabel{$\vec{p}_1$}{l}
        \fmfforce{0.5*vloc(__v)+0.5*vloc(__f2) shifted (+1mm,-4mm)}{l2}
        \fmflabel{$\vec{p}_2$}{l2}
    \end{fmfgraph*}
\end{fmffile}}
                        \end{gathered} \quad\quad\quad \Longrightarrow \quad\quad\quad i \sum_{\ell = 1}^8 \left[ f_{A A, \ell}(\vec{p}_1, \vec{p}_2) \right]^{\mu \nu}  \hat{\mathcal{M}}_{JK, \ell}(\vec{p}_1 - \vec{p}_2) \, . 
                    \end{align*}  
            } \\
            & \vtop{
                \begin{align*}
                    \left[ f_{A A, 1} \right]^{ij} = -\frac{e^2}{m_e} \delta^{ij}~~~,~~~[f_{AA, 4}]^{0i, j} = - \frac{i e^2}{4 m_e} \epsilon^{ijk} \frac{p_1^k}{m_e}~~~,~~~[f_{AA, 4}]^{i0, j} = \frac{i e^2}{4 m_e} \epsilon^{ijk} \frac{p_2^k}{m_e}
                \end{align*}
            } \\
        \bottomrule
    \end{tblr}
    \label{tab:QED_summary_table}
    \caption{Feynman rules for NR QED to order $m_e^{-2}$. The NR QED Lagrangian is derived in Sec.~\ref{sec:NRQED}. Note that four-point amplitude Feynman coefficients with identical fields is multiplied by a symmetry factor of two.}
\end{table*}

\clearpage

\vspace{1em}
\begin{center}
    \textit{ \normalsize Dipole Interaction  NR Limit }
\end{center}
\vspace{1em}
The high-energy/UV magnetic dipole interaction may be decomposed in to dark electric and magnetic fields ($\vec{E}', \vec{B}'$, respectively) as
\begin{align}
    \LUV_\text{int} = \frac{d_M}{2} \, V_{\mu\nu} \, \bar{\Psi} \sigma^{\mu \nu} \Psi = i d_M \, {E'}^i \, \bar{\Psi} \gamma^0 \gamma^i \Psi - \frac{i d_M}{2} \, {B'}^k \, \bar{\Psi} \left[ \epsilon^{kij} \gamma^i \gamma^j \right] \Psi \, ,
\end{align}
where $V_{0i} = - {E'}_i$, $V_{ij} = \epsilon_{ijk} {B'}^k$, and $\sigma^{\mu \nu} = \frac{i}{2} [\gamma^\mu, \gamma^\nu]$. The antisymmetric field strength tensor is written in terms of these two vector fields, and we construct the NR Lagrangian in terms of these coefficients



\begin{align}
    \LNR_\text{int} = \frac{1}{m_e} 
    \sum_{\ell = 1}^{8} &\left[ C_{\bm E', \ell} \bm E' + C_{(\nabla \bm E'), \ell} \frac{(\nabla \bm E')}{m_e} + C_{\bm E' \Av, \ell} \frac{\bm E' \Av}{m_e}  + C_{\bm E' (\nabla A^0), \ell} \frac{\bm E' (\nabla A^0)}{m_e^2} + C_{\bm E' (\partial_t \Av), \ell} \frac{\bm E' (\partial_t \Av)}{m_e^2} \right.
    \nonumber \\
    & \left. + \, C_{\bm B', \ell} \bm B' + C_{(\nabla \bm B'), \ell} \frac{\nabla \bm B'}{m_e} + C_{(\nabla \nabla \bm B'), \ell} \frac{(\nabla \nabla \bm B')}{m_e^2} + C_{\bm B' \Av, \ell} \frac{\vec B' \bm A}{m_e} + C_{(\nabla \bm B') \bm A, \ell} \frac{(\nabla \bm B') \Av}{m_e^2} \right. \nonumber
    \\
    &\left. +\, C_{\vec B' (\nabla \Av)} \frac{\vec B' (\nabla \Av)}{m_e^2}
    + C_{\bm B' \Av \Av, \ell} \frac{\bm B' \Av \Av}{m_e^2} \right] \left[ \psi^\dag \ONRBasis_\ell \psi \right]\, .
\end{align}
Where the coefficients are given by
\begin{align}
    & \left[ C_{\vec{E}', 5} \right]^{i, jk} = - i d_M m_e \epsilon^{ijk}~~,~~\left[ C_{\vec{E}', 7} \right]^{i, j} = - \frac{d_M m_e}{2} \delta^{ij}~~,~~\left[ C_{\nabla\vec{E}', 1} \right]^{ij} = - \frac{1}{2} d_M m_e \delta^{ij}~~,~~\left[ C_{\nabla\vec{E}', 4} \right]^{ij,k} = - \frac 12 i d_M m_e \epsilon^{ijk} \nonumber \\ 
    &\left[ C_{\vec{E}' \Av, 4} \right]^{ij,k} =  e d_M m_e \epsilon^{ijk}~~,~~\left[ C_{\vec{E}' (\partial_t \Av), 1}\right]^{ij} = - \frac 12 e d_M m_e \delta^{ij}~~,~~\left[ C_{\vec{E}' (\nabla A^0),1} \right]^{i,j} = - \frac 12 e d_M m_e \delta^{ij} 
\end{align}
\begin{align}
    & \left[ C_{\vec{B}', 4} \right]^{i,j} = - d_M m_e \delta^{ij}~~~,~~~\left[ C_{\vec{B}', 6} \right]^{i,jkl} = - \frac 12 d_M m_e \delta^{ik} \delta^{jl}~~~,~~~\left[ C_{\nabla \vec{B}', 2} \right]^{ij,k} = - \frac 14 i d_M m_e \epsilon^{ijk}
    \nonumber\\
    & \left[ C_{\nabla \vec{B}', 5} \right]^{ij,kl} = - \frac 14 d_M m_e (\delta^{ik} \delta^{jl} + \delta^{ij} \delta^{kl})~~~,~~~\left[ C_{\nabla \nabla \vec{B}', 4} \right]^{ijk,l} = - \frac 18 d_M m_e \delta^{ij} \delta^{kl}
    \nonumber\\
    & \left[ C_{\vec{B}' \bm A, 5} \right]^{ij,kl} =  \frac 12 i e d_M m_e (\delta^{il} \delta^{jk} + \delta^{ij} \delta^{kl})~~~,~~~\left[ C_{(\nabla \vec{B}') \bm A, 1} \right]^{ijk} = - \frac 14 e d_M m_e \epsilon^{ijk}
    \nonumber\\
    & \left[ C_{(\nabla \vec{B}') \bm A, 4} \right]^{ijk,l} =  \frac 14 i e d_M m_e (\delta^{il} \delta^{jk} + \delta^{ij} \delta^{kl})~~~,~~~\left[ C_{\vec{B}' (\nabla \Av), 4} \right]^{ijk,l} = \frac 14 i e d_M m_e (\delta^{ik} \delta^{jl} + \delta^{ij} \delta^{kl})
    \nonumber \\
    & \left[ C_{\vec{B}' \Av \Av, 4} \right]^{ijk,l} =  \frac 12 e^2 d_M m_e \delta^{ik} \delta^{jl} 
\end{align}
Since $\vec E'$ is related to $V^0$ and $V^i$, we can write the coefficients of the expansion of the NR Lagrangian in terms of those in the $\vec E', \, \vec B'$ expansion. For instance, 
%
\begin{align}
    C_{(\nabla V^0)\cdots, \ell} &= C_{(\partial_t \vec V)\cdots, \ell} =  -C_{\bm E\cdots} 
    \\
    C_{(\nabla \vec V)\cdots, \ell}^{ij\cdots} &= \varepsilon^{ijk} \left[ C_{\vec B \cdots, \ell} \right]^{k\cdots}
\end{align}
Where the ellipses indicate the possibility of additional photon terms. See Table~\ref{tab:dipole_summary_table} for the Feynman rule coefficients. 

\clearpage

\begin{table}[ht!]
    \begin{tblr}{width=\textwidth,colspec={ X[-1,l] X[1,l] }} 
        \toprule
            \SetCell[c=2]{l}{ \textbf{High-Energy Interaction Lagrangian} } \\ 
            \phantom{\quad} & $\displaystyle \LUV_\text{int} = \frac{d_M}{2} V_{\mu \nu} \bar \Psi \sigma^{\mu \nu} \Psi $ \\
            [1ex]\midrule
            \SetCell[c=2]{l}{ \textbf{NR Feynman Rules} } & \\ 
            & \vtop{\vspace{-1em}
                    \begin{align*}
                        \begin{gathered}
                            \scalebox{0.8}{\begin{fmffile}{3pt_vector}
    \begin{fmfgraph*}(100,75)
        \fmfleft{i} \fmfright{f1,f2}
        \fmf{photon,label=$\mArrow{0.4cm}{q}$,label.side=left}{i,v}
        \fmf{fermion}{v,f2}
        \fmf{fermion}{f1,v}
        \fmfv{decor.shape=circle,decor.filled=hatched,decor.size=0.125w}{v}
        \fmflabel{$V^\mu$}{i}
        \fmflabel{$J$}{f1}
        \fmflabel{$K$}{f2}
    \end{fmfgraph*}
\end{fmffile}}
                        \end{gathered} \quad\quad\quad \Longrightarrow \quad\quad\quad i \sum_{\ell = 1}^8 f_{V, \ell}(\bm{q}) \; \hat{\mathcal{M}}_{JK, \ell}(\bm{q})  \, .
                    \end{align*}
                    \vspace{-1em}
                } \\
            & \vtop{
                \begin{align*}
                    &\left[ f_{V, 1} \right]^0 = -\frac{d_M \vec q^2}{2 m_e}
                    ~~~,~~~ \left[ f_{V, 5} \right]^{0,ij} = -d_M \epsilon^{ijk} q^k ~~~,~~~ \left[ f_{V, 7} \right]^{0,i} = \frac{i d_M}{2} q^i 
                    \\
                    &\left[ f_{V, 1} \right]^{i}  = - d_M \frac{\omega q^i}{2 m_e} ~~~,~~~ \left[ f_{V, 2} \right]^{i}  = - \frac{i d_M}{4 m_e} (q^i q^j - \delta^{ij} q^2) ~~~,~~~ \left[ f_{V, 4} \right]^{i,j}  =  i d_M \epsilon^{ijk} q^k \left(1 - \frac{\omega}{2m_e} - \frac{\vec q^2}{8 m_e^2} \right)
                    \\ 
                    &\left[ f_{V, 5} \right]^{i,jk}  = - d_M \left( \omega \epsilon^{ijk} + \frac{q^j \epsilon^{ikm} q^m}{4 m_e }  \right) ~~~,~~~ \left[ f_{V, 6} \right]^{i,jkl} = \frac{i d_M}{4} q^m \left( \delta^{jl} \epsilon^{ikm} + \delta^{jk} \epsilon^{ilm} \right) ~~~,~~~ \left[ f_{V, 7} \right]^{i,j}  = \frac{i d_M \omega} {2}\delta^{ij}
                \end{align*}
            } \\
            \cmidrule{2}
            & \vtop{ 
                    \begin{align*}
                        \begin{gathered}
                            \scalebox{0.8}{\begin{fmffile}{4pt_vector_vector}
    \begin{fmfgraph*}(100,75)
        \fmfleft{i1,i2} \fmfright{f1,f2}
        \fmf{photon,label=$\rotatebox{-45}{ $\xrightarrow{\makebox[0.4cm]{}}$ }$,label.dist=-0.08w}{i2,v}
        \fmf{fermion}{i1,v}
        \fmf{fermion}{v,f1}
        \fmf{photon,label=$\rotatebox{45}{ $\xrightarrow{\makebox[0.4cm]{}}$ }$,label.dist=-0.08w,label.side=right}{v,f2}
        \fmflabel{$V^\mu$}{i2}
        \fmflabel{$A^\nu$}{f2}
        \fmflabel{$J$}{i1}
        \fmflabel{$K$}{f1}
        \fmfv{decor.shape=circle,decor.filled=hatched,decor.size=0.125w}{v}
        \fmffreeze
        \fmfforce{0.5*vloc(__v)+0.5*vloc(__i2) shifted (-0.75mm,-4mm)}{l}
        \fmflabel{$\vec{p}_1$}{l}
        \fmfforce{0.5*vloc(__v)+0.5*vloc(__f2) shifted (+1mm,-4mm)}{l2}
        \fmflabel{$\vec{p}_2$}{l2}
    \end{fmfgraph*}
\end{fmffile}}
                        \end{gathered} \quad\quad\quad \Longrightarrow \quad\quad\quad i \sum_{\ell = 1}^8 \left[ f_{V A, \ell} (\vec{p}_1, \vec{p}_2) \right]^{\mu \nu} \hat{\mathcal{M}}_{JK, \ell}(\vec{p}_1 - \vec{p}_2) \, . 
                    \end{align*}
                    \vspace{-1em}
            } \\
            & \vtop{
                \begin{align*}
                    & \left[ f_{VA,1} \right]^{00}  = \frac{e d_M}{2m_e^2} (\vec p_1 \cdot \vec p_2) ~~~,~~~ \left[ f_{VA,1} \right]^{0i} = \frac{e d_M}{2m_e^2} \omega_2 p_1^i ~~~,~~~ \left[ f_{VA,4} \right]^{0i,j} = \frac{i e d_M}{m_e} \epsilon^{ijk} p_1^k ~~~,~~~ \left[ f_{VA,1} \right]^{i0} = \frac{e d_M}{2m_e^2} \omega_1 p_2^i 
                    \\
                    &\left[ f_{VA,1} \right]^{ij} = \frac{e d_M}{4m_e^2} \left( p_1^i p_1^j - \delta^{ij} (\vec p_1^2 - 2 \omega_1 \omega_2) \right) ~~~,~~~ \left[ f_{VA,4} \right]^{ij,k} = \frac{ie d_M}{4 m_e^2} \left( \epsilon^{ijm} p_1^m (p_2^k - p_1^k) + \delta^{jk} \epsilon^{iml} p_2^m p_1^l + 4 \omega_1 \epsilon^{ijk} m_e  \right)
                    \\
                    & \left[ f_{VA,5} \right]^{ij,kl} = - \frac{e d_M}{2m_e} \left( \delta^{kl} \epsilon^{ijm} + \delta^{jk} \epsilon^{ilm} \right) p_1^m
                \end{align*}
            } \\
        \bottomrule
    \end{tblr}
    \caption{ Summary of the Feynman rules generated by the UV interaction Lagrangian $\LUV_\text{int} = d_M V_{\mu\nu} \bar{\Psi} \sigma^{\mu\nu} \Psi / 2$. See App.~\ref{app:summary_tables} for the NR EFT interaction Lagrangian.}
    \label{tab:dipole_summary_table}
\end{table}

\clearpage

\section{In-Medium Photon Propagator}
\label{app:in_medium_photon_propagator}

In the absence of interactions, the photon propagator is
\begin{align}
    G_{\mu \nu}^0 = -\frac{i \, \eta_{\mu \nu}}{q^2},
    \label{eq:photon_prop_0}
\end{align}
where $q^\mu = (\omega, \bm{q})$ is a momentum four-vector, and we have neglected gauge-dependent contributions which vanish from physical processes due the Ward identity (WI). In the presence of interactions the photon propagator is modified. The modification to the propagator due to virtual insertions of  1PI diagrams, can be written as the infinite sum
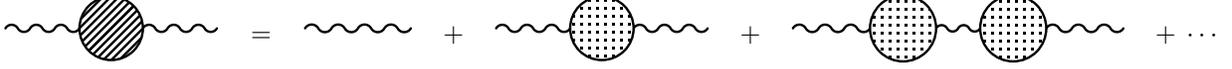
\begin{fmffile}{1PI_photon_propagator_1}
    \begin{align*}
        \begin{gathered}
            \begin{fmfgraph*}(80,75)
                \fmfleft{i} \fmfright{f}
                \fmf{photon}{i,v}
                \fmf{photon}{v,f}
                \fmfblob{.3w}{v}
            \end{fmfgraph*}
        \end{gathered} 
        \quad = \quad 
        \begin{gathered}
            \begin{fmfgraph*}(40,75)
                \fmfleft{i} \fmfright{f}
                \fmf{photon}{i,f}
            \end{fmfgraph*}
        \end{gathered}
        \quad + \quad 
        \begin{gathered}
            \begin{fmfgraph*}(80,75)
                \fmfleft{i} \fmfright{f}
                \fmf{photon}{i,v}
                \fmf{photon}{v,f}
                \fmfv{decor.shape=circle,decor.filled=gray10,decor.size=.3w}{v}
            \end{fmfgraph*}
        \end{gathered}
        \quad + \quad 
         \begin{gathered}
            \begin{fmfgraph*}(125,75)
                \fmfleft{i} \fmfright{f}
                \fmf{photon}{i,v1}
                \fmf{photon}{v1,v2}
                \fmf{photon}{v2,f}
                \fmfv{decor.shape=circle,decor.filled=gray10,decor.size=.2w}{v1}
                \fmfv{decor.shape=circle,decor.filled=gray10,decor.size=.2w}{v2}
            \end{fmfgraph*}
        \end{gathered}
        \quad + \,  \cdots  
    \end{align*}
\end{fmffile}

\noindent where the left-hand side is the resummed propagator, and the right-hand side is contribution from many 1PI diagrams (represented by dotted blobs). 

Computing the resummed propagator is aided with a different representation of the photon propagator
\begin{align}
    G_{\mu \nu}^0 = i \sum_{\lambda} \frac{\epsilon^\lambda_\mu \epsilon^\lambda_\nu}{q^2}~,
    \label{eq:photon_prop_1}
\end{align}
where $\epsilon_\lambda^\mu$ are three polarization vectors satisfying
\begin{align}
    \epsilon_{\pm}^\mu = (0, \hat{\bm{q}}_{\pm})~~~,~~~\epsilon_L^\mu = \frac{1}{\sqrt{q^2}} (|\bm{q}|, \omega \hat{\bm{q}})~~~,~~~q_\mu \epsilon^\mu_{\lambda} = 0~~~,~~~\epsilon_\lambda^\mu \epsilon^{\lambda'}_{\mu} = - \delta_{\lambda \lambda'}~~~,~~~\sum_{\lambda} \epsilon^\mu_{\lambda} \epsilon^\nu_{\lambda} = - \eta^{\mu \nu} + \frac{q^\mu q^\nu}{q^2} \, ,
    \label{eq:pol_properties}
\end{align}
where $\hat{\bm{q}}_{\pm}$ are any two vectors mutually orthonormal to $\hat{\bm{q}}$. The propagator in Eq.~\eqref{eq:photon_prop_1} is physically equivalent to that in Eq.~\eqref{eq:photon_prop_0}, since the terms proportional to $q^\mu$ vanish due to the WI. The 1PI diagram is the photon self-energy, $\Pi_{AA}$, and can be approximately decomposed in to components along each polarization vector\footnote{In complicated materials the polarization vectors in Eq.~\eqref{eq:pol_properties} may not allow for $\Pi_{AA}^{\mu\nu}$ to be decomposed directly as in Eq.~\eqref{eq:Pi_decompose}; there may be terms mixing the transverse and longitudinal polarization components. However, there is always a basis of polarization vectors (a linear combination of those in Eq.~\eqref{eq:pol_properties}) which allow for a polarization decomposition as in Eq.~\eqref{eq:Pi_decompose}. The rest of the derivation in more complicated materials follows analogously once the diagonalizing basis of polarization vectors is used.}
\begin{align}
    \Pi^{\mu \nu}_{AA} = - \sum_{\lambda} \Pi_{AA}^\lambda \, \epsilon^{\mu}_{\lambda} \epsilon^{\nu}_{\lambda}~~~,~~~\Pi_{AA}^\lambda = - \epsilon^\lambda_{\mu} \, \Pi^{\mu \nu}_{AA} \, \epsilon^\lambda_{\nu} \, .
    \label{eq:Pi_decompose}
\end{align}
Note that this $\Pi_{AA}$ is referred to as $\Pi_{AA}^\text{UV}$ in the main text. The resummed propagator is then computed as
\be
    G_{\mu \nu}  = G^0_{\mu \nu} + G^0_{\mu \alpha} \left( i \Pi^{\alpha \beta}_{AA} \right) G^0_{\beta \nu} + \! \cdots  
     = \left(i \sum_{\lambda} \frac{\epsilon_\mu^\lambda \epsilon_\nu^\lambda}{q^2} \right) + \left( i \sum_{\lambda} \frac{\epsilon_\mu^\lambda \epsilon_\alpha^\lambda}{q^2} \right) \left( i \Pi_{AA}^{\alpha \beta} \right) \left( i \sum_{\lambda} \frac{\epsilon_{\beta}^\lambda \epsilon_\nu^\lambda}{q^2} \right) + \! \cdots 
     = \, i \sum_\lambda \frac{\epsilon^\lambda_\mu \epsilon^\lambda_
    \nu}{q^2 - \Pi^\lambda_{AA}}  .~~~~~~~
    \label{eq:G_AA}
\ee

As discussed in Sec.~\ref{subsec:nr_matrix_element_def} in a more general context momentum conservation need not apply, and therefore the resummed propagator may be a function of both the incoming and outgoing momentum, $\vec{q}, \vec{q}'$, respectively. However the non-interacting propagator should be unchanged and we can write it as a function of the incoming and outgoing momentum as
\begin{align}
    G^0_{\mu \nu}(\vec{q}, \vec{q}') = G^0_{\mu \nu}(\vec{q}) \, \delta_{\vec{q}\vec{q}'} = G^0_{\mu \nu}(\vec{q}) \, \frac{(2 \pi)^3 \delta^3(\vec{q} - \vec{q}')}{\vol} 
\end{align}
Following the Feynman rules discussed in Sec.~\ref{sec:feynman_rules} we compute the NR resummed propagator as
\begin{align}
    G_{\mu \nu}^\text{NR}(\vec{q}, \vec{q}') & = G^{\text{NR}, 0}_{\mu \nu}(\vec{q}, \vec{q}') + \int \frac{d^3 \vec{q}_1}{(2 \pi)^3} \frac{d^3 \vec{q}_2}{(2 \pi)^3} \left( G_{\mu \alpha}^{\text{NR}, 0}(\vec{q}, \vec{q}_1) \right) \left( i \Pi_{AA}^{\text{NR}, \alpha \beta}(\vec{q}_1, \vec{q}_2) \right) \left( G^{\text{NR},0}_{\beta \nu}(\vec{q}_2, \vec{q}') \right) + \cdots \nonumber \\ 
    & = (2 \pi)^3 \delta^3(\vec{q} - \vec{q}') G^0_{\mu \nu}(\vec{q}) +  G^0_{\mu \alpha}(\vec{q}) \left( i \Pi^{\text{NR}, \alpha \beta}_{AA}(\vec{q},\vec{q}') \right)  G^0_{\beta \nu}(\vec{q}') + \cdots \, ,
    \label{eq:GAA_2}
\end{align}
where $G^{\text{NR},0}_{\mu \nu}(\vec{q},\vec{q}') = (2 \pi)^3 \delta^3(\vec{q} - \vec{q}') G^0_{\mu \nu}(\vec{q})$ connects the NR propagator diagram to the usual relativistic quantity. If the 1PI diagram can be approximated as
\begin{align}
    \Pi_{AA}^{\text{NR},\mu \nu}(\bm{q}, \bm{q}') \approx \Pi_{AA}^{\text{NR},\mu \nu}(\bm{q}) \, \delta_{\bm{q}\bm{q}'} = \Pi_{AA}^{\text{NR},\mu \nu}(\bm{q}) \frac{(2 \pi)^3 \delta^3(\vec{q} - \vec{q}')}{\vol} \, ,
\end{align}
as discussed in Sec.~\ref{subsec:loop_diagram}, then the propagator in Eq.~\eqref{eq:GAA_2} can be resummed as it was previously, using $\Pi^{\text{NR},\mu \nu}_{AA}(\vec{q}) = \vol \Pi^{\mu \nu}_{AA}(\vec{q})$,
\begin{align}
    G_{\mu \nu}^\text{NR}(\vec{q}, \vec{q}') & \approx (2 \pi)^3 \delta^3(\vec{q} - \vec{q}') \left( G^0_{\mu \nu}(\vec{q}) +  G^0_{\mu \alpha}(\vec{q}) \left( i \Pi^{\alpha \beta}_{AA}(\vec{q}) \right) G^0_{\beta \nu}(\vec{q}) + \cdots \right) \nonumber \\ 
    & = (2 \pi)^3 \delta^3(\vec{q} - \vec{q}') \left( i \sum_\lambda \frac{\epsilon^\lambda_\mu \epsilon^\lambda_\nu}{q^2 - \Pi_{AA}^\lambda}\right) \, ,
    \label{eq:GAA_3}
\end{align}
where the momentum conserving delta function is usually cancelled if the Feynman rule was defined in the usual way, i.e., if the left-hand side of Eq.~\eqref{eq:GAA_3} had a momentum-conserving delta function. Note that in this approximation one can either explicitly insert $G_{\mu \nu}^\text{NR}(\vec{q}, \vec{q}')$ in to a diagram and integrate over the extra undetermined momentum, or require momentum conservation and insert only the final term in the brackets in Eq.~\eqref{eq:GAA_3} (which is $G_{\mu \nu}$ in Eq.~\eqref{eq:G_AA}), ignoring the momentum conserving delta function.

The \textit{in-medium} photon propagator is the resummed photon propagator in Eq.~\eqref{eq:GAA_3} with the 1PI diagrams evaluated in a medium (as opposed to a vacuum in the usual relativistic QFT scenario). The in-medium photon propagator is usually written in terms of the dielectric function, $\vec{\varepsilon}(\vec{q}, \omega)$. The relationship between the projected photon self-energies, $\Pi_{AA}^\lambda$ and $\vec{\varepsilon}$ is usually made using the constitutive relationships of electrodynamics and equations of motion for $A^\mu$, $\partial_\nu F^{\mu \nu} = - \Pi_{AA}^{\mu \nu} A_\nu$, from which we can identify $- \Pi_{AA}^{\mu \nu} A_\nu$ as an effective four-current, $J^\mu = (J^0, \vec{J})$
\begin{align}
    \vec{J} = \vec{\sigma} \vec{E}~~~\Longleftrightarrow~~~J^\mu = - \Pi_{AA}^{\mu \nu} A_\nu
\end{align}
where $\vec{\sigma}$ is the conductivity, related to the dielectric by $\vec{\sigma} = i \omega (1 - \vec{\varepsilon})$. Since $E^i \supset - \partial_t A^i = i \omega A^i$ we can identify
\begin{align}
    \Pi^{ij}_{AA} = - \omega^2 (1 - \vec{\varepsilon}^{ij} ) \, .
\end{align}
Therefore the transverse projected self-energies are
\begin{align}
    \Pi_{AA}^{\pm} = - \epsilon^{\pm}_{\mu} \Pi_{AA}^{\mu \nu} \epsilon^{\pm}_{\nu} = \omega^2 (1 - \hat{\vec{q}}_{\pm} \cdot \vec{\varepsilon} \cdot \hat{\vec{q}}_{\pm} ) \, .
\end{align}
The longitudinal projected self-energies can be written in terms of $\Pi^{00}_{AA}$
\begin{align}
    \Pi^L_{AA} = - \epsilon^L_{\mu} \, \Pi^{\mu \nu}_{AA} \epsilon^L_{\nu} = - \frac{q^2}{|\vec{q}|^2} \Pi_{AA}^{00} \, ,
\end{align}
which, using the WI, $q_\mu \Pi^{\mu \nu}_{AA} = \Pi^{\nu \mu}_{AA} q_\mu = 0$, can be written in terms of $\Pi_{AA}^{ij}$
\begin{align}
    \Pi_{AA}^{00} = \frac{\vec{q}^i \Pi_{AA}^{ij} \vec{q}^j}{\omega^2} \, .
\end{align}
Therefore,
\begin{align}
    \Pi^L_{AA} = q^2 (1 - \hat{\vec{q}} \cdot \vec{\varepsilon} \cdot \hat{\vec{q}}) \, .
\end{align}

The in-medium photon propagator is then written in terms of the dielectric as
\begin{align}
    G^{\mu \nu} & = i \left( \frac{\epsilon_L^\mu \epsilon_L^\nu}{q^2 \, ( \hat{\bm{q}} \cdot \bm{\varepsilon} \cdot \hat{\bm{q}}) } + \frac{\epsilon_+^\mu \epsilon_+^\nu}{ \omega^2 (\hat{\bm{q}}_+ \cdot \bm{\varepsilon} \, \cdot \hat{\bm{q}}_+) - \bm{q}^2} + \frac{\epsilon_-^\mu \epsilon_-^\nu}{ \omega^2 (\hat{\bm{q}}_- \cdot \bm{\varepsilon} \, \cdot \hat{\bm{q}}_-) - \bm{q}^2} \right) \, .
\end{align}
This can be further simplified in the limit of scattering or absorption kinematics, as considered in Secs.~\ref{sec:scattering},~\ref{sec:absorption} respectively,
\begin{align}
        G^{\mu \nu} = \begin{cases}
             \displaystyle \frac{i}{\omega^2} \left( \frac{\epsilon_{\hat{\bm{q}}}^\mu \epsilon_{\hat{\bm{q}}}^\nu}{ \hat{\bm{q}} \cdot \bm{\varepsilon} \cdot \hat{\bm{q}} } + \frac{\epsilon_{+}^\mu \epsilon_{+}^\nu}{ \hat{\bm{q}}_+ \cdot \bm{\varepsilon} \cdot \hat{\bm{q}}_+ } + \frac{\epsilon_{-}^\mu \epsilon_{-}^\nu}{ \hat{\bm{q}}_- \cdot \bm{\varepsilon} \cdot \hat{\bm{q}}_- } \right) & \omega \gg |\bm{q}| \; (\text{absorption}) \\ 
             \displaystyle \frac{i}{\bm{q}^2} \left( \frac{\epsilon_{0}^\mu \epsilon_{0}^\nu}{ \hat{\bm{q}} \cdot \bm{\varepsilon} \cdot \hat{\bm{q}} } - \epsilon_{+}^\mu \epsilon_{+}^\nu - \epsilon_{-}^\mu \epsilon_{-}^\nu \right) 
             & \omega \ll |\vec{q}| \; (\text{scatter})
        \end{cases} \label{eq:final_propagator_limits} \, ,
\end{align}
where $\epsilon^\mu_{\hat{\bm{q}}} \equiv (0, \hat{\bm{q}})$ and $\epsilon^\mu_0 \equiv (1, 0, 0, 0)$ are limits of $\epsilon_L^\mu$. In the limit of an isotropic target ($\vec{\varepsilon}^{ij} = \delta^{ij} \varepsilon$) these can be simplified even further to
\begin{align}
    G^{\mu \nu} & = \begin{cases}
         \displaystyle \frac{i}{\varepsilon \, \omega^2} P^{\mu \nu}& \omega \gg |\vec{q}| \; (\text{absorption}) \\ 
         \displaystyle \frac{i}{\bm{q}^2} \left( \frac{\epsilon_{0}^\mu \epsilon_{0}^\nu}{ \varepsilon } - \epsilon_{+}^\mu \epsilon_{+}^\nu - \epsilon_{-}^\mu \epsilon_{-}^\nu \right) & \omega \ll |\vec{q}| \; (\text{scatter}) \label{eq:final_propagator_limits_isotropic}\, ,
    \end{cases}
\end{align}
where $P^{\mu \nu} = \text{diag}(0, 1, 1, 1)$.


\bibliographystyle{utphys3}
\bibliography{bibliography}

\end{document}